\documentclass[a4paper,10pt]{article}

\pdfoutput=1

\usepackage{jheppub}

\usepackage{graphicx}         % needed for figures
\usepackage{dcolumn}         % needed for some tables
\usepackage{bm}                  % for math
\usepackage{amssymb}     % for math
\usepackage{subcaption}
\usepackage{verbatim}
\usepackage{environ}

\usepackage[english]{babel}
\usepackage{xcolor}
\usepackage{graphicx}
\usepackage{amsmath}
\usepackage{dsfont}
\usepackage[makeroom]{cancel}
\usepackage{url}
\usepackage{hyperref}
\usepackage{mathabx}
\usepackage{mathtools}

\usepackage{bbm}
\usepackage{slashed}
\usepackage{braket}

\usepackage[normalem]{ulem}

\graphicspath{{figures/}}

\newcommand{\appn}[1]{Appendix~\ref{#1}}
\newcommand{\one}{\, (\mathbf{1})}
\newcommand{\two}{\, (\mathbf{2})}
\newcommand{\otwo}{\, (\mathbf{12})}
\newcommand{\three}{\, (\mathbf{3})}

\newcommand{\othree}{\, (\mathbf{13})}
\newcommand{\tthree}{\, (\mathbf{23})}
\newcommand{\otthree}{\, (\mathbf{123})}

\newcommand{\RV}{\, (\mathbf{RV})}

\newcommand{\RRV}{\mathbf{RRV}}
\newcommand{\RVV}{\mathbf{RVV}}

\newcommand{\E}{{\mbox{\tiny{E}}}}
\newcommand{\eps}{\epsilon}
\newcommand{\vareps}{\varepsilon}
\newcommand{\al}{\alpha}
\newcommand{\nn}{\nonumber}
\newcommand{\T}{\mathbf{T}}
\newcommand{\hard}{\mathcal{H}}
\newcommand{\radState}{\{k_j,\lambda_j\}}

\newcommand{\Int}{T}
\newcommand{\kOnen}{(k_1\cdot n)}
\newcommand{\kTwon}{(k_2\cdot n)}
\newcommand{\kOnekTwo}{(k_1\cdot k_2)}
\newcommand{\s}{\mathcal{S}}
\newcommand{\J}{\mathcal{J}}
\newcommand{\beq}{\begin{eqnarray}}
\newcommand{\eeq}{\end{eqnarray}}
\newcommand{\npo}{{n+1}}
\newcommand{\npt}{{n+2}}
\newcommand{\npth}{{n+3}}
\newcommand{\pn}{\Phi_n}
\newcommand{\pnpo}{\Phi_\npo}
\newcommand{\pnpt}{\Phi_\npt}
\newcommand{\sig}{\sigma}
\newcommand{\LO}{{\mbox{\tiny{LO}}}}
\newcommand{\NLO}{{\mbox{\tiny{NLO}}}}
\newcommand{\NNLO}{{\mbox{\tiny{NNLO}}}}
\newcommand{\NNNLO}{{\mbox{\tiny{N$^3$LO}}}}

\newcommand{\bS}[1]{{\bf S}_{#1}}
\newcommand{\bC}[1]{{\bf C}_{#1}}
\newcommand{\bSC}[1]{{\bf SC}_{#1}}

\newcommand{\bbL}[2]{\overline{\bf L}^{#1}_{#2}}
\newcommand{\bbS}[1]{\overline{\bf S}_{#1}}
\newcommand{\bbC}[1]{\overline{\bf C}_{#1}}

\newcommand{\as}{\alpha_s}
\newcommand{\gs}{g_s}
\newcommand{\secn}[1]{Section~\ref{#1}}
\newcommand{\nnb}{\nonumber}
\newcommand{\CG}{{\mbox{\tiny{CG}}}}
\newcommand{\varsi}{\varsigma}

\newcommand{\RVl}{RV}
\newcommand{\Norm}{\mc{N}_1}
\newcommand{\mc}{\mathcal}
\newcommand{\Vl}{V}
\newcommand{\cg}{c_\Gamma}

\newcommand{\SL}{\mathbf{S}}
\newcommand{\CL}{\mathbf{C}}

\def\eq#1{eq.~(\ref{#1})}
\def\Eq#1{Eq.~(\ref{#1})}

\preprint{\\  \rightline{MPP-2024-46}}

\title{
Strongly-ordered infrared counterterms from factorisation}

\author[a]{Lorenzo Magnea,}
\author[a]{Calum Milloy,}
\author[b]{Chiara Signorile-Signorile,}
\author[a]{Paolo Torrielli}

\affiliation[a]{Dipartimento di Fisica, Universit\`a di Torino,
and INFN, Sezione di Torino, \\ Via P. Giuria 1, I-10125 Torino, Italy}
\affiliation[b]{Max-Planck-Institut f\"ur Physik, Boltzmannstrasse 8, 
85748 Garching, Germany}

\emailAdd{lorenzo.magnea@unito.it}
\emailAdd{calumwilliam.milloy@unito.it}
\emailAdd{signoril@mpp.mpg.de}
\emailAdd{paolo.torrielli@unito.it}

\abstract{
In the context of infrared subtraction algorithms beyond next-to-leading order, it becomes 
necessary to consider multiple infrared limits of scattering amplitudes, in which several 
particles become soft or collinear in a strongly-ordered sequence. We study these limits
from the point of view of infrared factorisation, and we provide general definitions of
strongly-ordered soft and collinear kernels in terms of gauge-invariant operator matrix 
elements. With these definitions in hand, it is possible to construct local subtraction
counterterms for strongly-ordered configurations. Because of their factorised structure,
these counterterms cancel infrared poles of real-virtual contributions by construction.
We test these ideas at tree level for multiple emissions, and at one loop for single and 
double emissions, contributing to NNLO and N$^3$LO distributions, respectively.
}

\keywords{QCD corrections, factorisation, infrared subtraction}

\begin{document}
\maketitle

%%%%%%%%%%%%%%%%%%%%%%%%%%%%%%%%%%%%%%%

\section{Introduction}
\label{Intro}

In the era of precision high-energy physics, significant efforts and excitement in the 
particle physics community focus on developing methods to explore high orders in 
perturbation theory, and improving on the precision of theoretical predictions, to meet 
the standards set by current and future experiments (refer to Refs.~\cite{Heinrich:2020ybq,
Agarwal:2021ais,Badger:2023eqz} for recent reviews). An important requirement to 
access the calculation of hard processes at lepton and hadron colliders is to have a 
detailed control on infrared singularities (arising from configurations where one or more 
particles become soft or collinear), that plague both real and virtual corrections.
While this problem is completely solved at next-to-leading order (NLO)~\cite{Giele:1993dj,
Giele:1994xd,Frixione:1995ms,Catani:1996vz,Nagy:2003qn,Bevilacqua:2013iha,
Prisco:2020kyb}, the issue of treating infrared singularities in a fully general manner 
remains a challenge at NNLO. Numerous \emph{subtraction} and \emph{slicing} schemes 
are currently available~\cite{Frixione:2004is,Gehrmann-DeRidder:2005btv,Currie:2013vh,
Somogyi:2005xz,Somogyi:2006db,DelDuca:2016csb,DelDuca:2016ily,Czakon:2010td,
Czakon:2011ve,Czakon:2014oma,Anastasiou:2003gr,Caola:2017dug,Catani:2007vq,
Grazzini:2017mhc,Boughezal:2011jf,Gaunt:2015pea,Boughezal:2015dva,Sborlini:2016hat,
Herzog:2018ily,Magnea:2018hab,Bertolotti:2022aih,Capatti:2019ypt,Devoto:2023rpv,
Gehrmann:2023dxm}, as reviewed in~\cite{TorresBobadilla:2020ekr}, and a few of them 
have already been applied to a variety of NNLO calculations (see~\cite{Chen:2014gva,
Boughezal:2015dra,Caola:2015wna,Chen:2016zka,Campbell:2019gmd,Cacciari:2015jma,
Cruz-Martinez:2018rod,Gauld:2021ule,Catani:2022mfv,Chawdhry:2019bji,Chawdhry:2021hkp,
Czakon:2020coa,Gauld:2023zlv,Currie:2017eqf,Chen:2022tpk,Badger:2023mgf,
Czakon:2021mjy,Czakon:2015owf,Catani:2019hip,Buonocore:2023ljm,Brucherseifer:2014ama,
Berger:2016oht,Campbell:2020fhf,Bronnum-Hansen:2022tmr,Alvarez:2023fhi} for 
some representative results). However, none of the existing methods have so far 
provided an explicit demonstration of the cancellation of infrared singularities for arbitrary 
hadron collider processes at NNLO. Although recent work made substantial progress 
in this direction, in the context of \emph{antenna subtraction}~\cite{Chen:2022ktf,
Gehrmann:2023dxm} and \emph{nested soft-collinear subtraction}~\cite{Devoto:2023rpv}, 
such a proof for generic final states is still beyond the current state of the art for hadronic 
scattering. For $e^+e^-$ collisions, this cancellation for arbitrary final states has only been 
displayed within the framework of the \emph{local analytic sector subtraction} 
scheme~\cite{Magnea:2018hab,Magnea:2020trj,Bertolotti:2022aih}.

Given this scenario, perfecting our understanding of the physical mechanisms 
governing infrared divergences is crucial. It is intriguing to observe an ``asymmetry" in 
the control we have on soft and collinear singularities in \emph{virtual} corrections to 
scattering amplitudes, compared to our understanding of the behaviour of \emph{real-radiation}
corrections in the same regimes. The infrared content of virtual corrections has been deeply 
scrutinised in the past~\cite{Sen:1982bt,Collins:1989bt,Magnea:1990zb,Sterman:1995fz,
Catani:1998bh,Sterman:2002qn,Dixon:2008gr,Gardi:2009qi,Gardi:2009zv,Becher:2009cu,
Becher:2009qa,Feige:2014wja}, and, in the case of massless gauge amplitudes, it has been 
shown that it is governed to all orders by a limited set of universal functions defined by 
gauge-invariant operator matrix elements. These functions, in turn, obey evolution equations 
that can be solved in terms of soft and collinear anomalous dimensions, which have been 
computed in the massless case up to three loops~\cite{Almelid:2015jia,Almelid:2017qju,
Henn:2016jdu,Duhr:2022cob}, and in some cases also beyond~\cite{Henn:2019swt,
vonManteuffel:2020vjv,Agarwal:2021zft}. Our knowledge of the behaviour of real-radiation 
matrix elements under unresolved limits is somewhat more limited, as most existing results
have been derived at fixed order. Nevertheless, it has been demonstrated that, under a 
variety of conditions, these matrix elements factorise in soft and collinear limits into 
universal kernels and lower-multiplicity amplitudes~\cite{Kosower:1999xi,Catani:1998nv,
Catani:1999ss}. The necessary kernels for NNLO calculations are fully known~\cite{Campbell:1997hg,
Catani:1998nv,Bern:1999ry,Catani:1999ss,Kosower:1999xi,Catani:2000pi}, while 
partial results are available at N$^3$LO~\cite{DelDuca:1999iql,Badger:2004uk,
Duhr:2013msa,Li:2013lsa,Banerjee:2018ozf,Bruser:2018rad,Dixon:2019lnw,
Catani:2019nqv,DelDuca:2020vst,Catani:2021kcy,DelDuca:2022noh,Czakon:2022dwk,
Czakon:2022fqi,Catani:2022hkb}.

A theoretical framework to systematically construct unresolved radiative contributions, 
serving as local subtraction counterterms, and leveraging the well-understood structure
of infrared divergences in virtual corrections, would greatly benefit the ultimate organisation 
of subtraction methods. This paper aims at filling this gap, complementing the study 
performed in Ref.~\cite{Magnea:2018ebr}. In that paper, a general method was proposed 
to identify local infrared subtraction counterterms, to any order in perturbation 
theory. The core idea of the method consists in exploiting completeness relations and 
general theorems on the cancellation of infrared singularities, to infer the expression 
of real-radiation counterterms from the factorised form of virtual corrections. In the latter, soft 
and collinear divergences are organised in gauge-invariant matrix elements of fields 
and Wilson lines. Ref.~\cite{Magnea:2018ebr} also tested the agreement between 
the expressions of the candidate counterterms and the known results at NLO, 
and presented the organisation of the relevant counterterms at NNLO. An interesting 
aspect of NNLO calculations that was not addressed in Ref.~\cite{Magnea:2018ebr}, 
and which is the focus of the present paper (see also Ref.~\cite{Magnea:2022twu}), 
is the issue of identifying and dealing with strongly-ordered soft and collinear limits, 
where groups of particles become unresolved in a hierarchical manner. For example, 
at NNLO, up to two partons can become soft and up to three partons can become collinear 
at the same time. When these double-unresolved limits are analysed, one needs to disentangle 
\emph{uniform limits}, where soft and collinear emissions become unresolved at the 
same rate, and \emph{strongly-ordered} limits, where one or more particles become 
unresolved at a faster rate than the others. These hierarchical configurations, upon 
integration, play the delicate role of cancelling singularities arising in mixed real-virtual 
contributions. In the present paper, we will tackle the problem of systematically building 
strongly-ordered subtraction counterterms, exploiting the principles of infrared 
factorisation. Through specific examples, we will illustrate how matrix elements 
involving fields and Wilson lines, that describe the factorised emission of soft and 
collinear particles, can be {\it re-factorised} in strongly-ordered configurations. We will 
then exploit this nested factorisation structure to display the required cancellation of 
singularities between strongly-ordered limits and real-virtual contributions.

The paper is organised as follows: in \secn{Archi} we present the architecture 
of infrared subtraction in full generality, starting at NLO and then illustrating the 
organisation of relevant counterterms up to N$^3$LO. The resulting structure 
smoothly generalises to any order, and we discuss the counting of necessary 
counterterms at N$^k$LO, for generic $k$. In \secn{FactoSubtra}, we discuss 
the factorisation approach to explicitly construct the counterterms introduced 
in \secn{Archi}. Given such a construction, we will focus on strongly-ordered 
limits at tree level in \secn{ReFactoTree}. We will show that ordered configurations 
can be easily modelled by products of soft and jet functions, with appropriate 
colour and spin contractions. This result paves the way for a formal, all-order 
definitions of hierarchical limits. We then recall that strongly-ordered configurations
must interplay with the real-virtual corrections. For this reason, in \secn{ReFactoLoop}, 
we analyse the single radiative contributions at one-loop order, and we obtain 
the relevant counterterms. Combining the results of \secn{ReFactoTree} and 
\secn{ReFactoLoop} will allow us to derive the form of strongly-ordered counterterms 
in terms of factorised soft and jet functions, presented in Section~\ref{dsocfrfvcr}. 
We conclude in Section~\ref{sec:summary_and_future_prospects} with a summary 
of our results and a brief discussion of future prospects. Four Appendices present
some technical aspects: in particular, Appendix \ref{scollcanc} discusses in detail how
the cancellation of singularities in terms of universal matrix elements is achieved 
at NLO, emphasising the role played by phase-space mappings in the collinear case.

%%%%%%%%%%%%%%%%%%%%%%%%%%%%%%%%%%%%%%%

\section{The architecture of infrared subtraction}
\label{Archi}

Consider renormalised scattering amplitudes $\mathcal{A}_n \! \left(p_i \right)$ 
involving $n$ massless coloured particles carrying momenta $p_i$ ($i = 1, \ldots, n$). 
For simplicity, we focus on the case of final-state particles with a total momentum 
$Q = \sum_{i = 1}^n p_i$, generated by the decay of a colour-singlet current, and 
we do not display the dependence on the strong coupling $\al_s(\mu^2)$ and the 
dimensional infrared regulator $\eps = (4 - d)/2 < 0$, where $d$ is the number of 
space-time dimensions. We expand the amplitude in perturbation theory as
\beq
\label{pertexpA}
  {\cal A}_n (p_i) \, = \, {\cal A}_n^{(0)} (p_i) \, + \, {\cal A}_n^{(1)} (p_i) 
  \, + \,
  {\cal A}_n^{(2)} (p_i) \, + \, \ldots \, , 
\eeq
where $\mathcal{A}_n^{(k)}(p_i)$ is the $k$-loop correction, including the appropriate 
power of the strong coupling constant. Given the amplitude, one can compute finite
differential distributions for a generic infrared-safe observable $X$, whose perturbative 
expansion we write as
\beq
\label{pertexpsig}
  \frac{d \sigma}{d X} \, = \, \frac{d \sigma_\LO}{d X} \, + \, 
  \frac{d \sigma_\NLO}{d X} \, + \, 
  \frac{d \sigma_\NNLO}{d X} \, + \, \ldots \, .
\eeq
For example, the leading-order distribution is given by
\beq
\label{BornDist}
  \frac{d\sig_\LO}{dX} \, = \,  \int d \pn\, B_n \,\delta_n \left( X \right) \, ,
\eeq
where $B_n \equiv | \mathcal{A}_{n}^{(0)} |^2$ is the Born-level squared amplitude, 
$\pn$ is the $n$-particle phase space, and $\delta_n (X) \equiv \delta (X - X_n)$ fixes 
$X_n$, the expression for the observable appropriate for an $n$-particle configuration, 
to the prescribed value $X$. At higher orders, infrared divergences arise from both
loop and phase-space integrations, and must be properly handled. We begin our 
discussion with a brief overview of the structure of the problem, order by order
in perturbation theory.

%%%%%%%%%%%%%%%%%%%%%

\subsection{Subtraction at NLO}
\label{ArchiNLO}

At NLO, there are infrared divergent contributions from the one-loop virtual 
correction in \eq{pertexpA}, and further divergences arising from the phase-space 
integration of unresolved real radiation. These divergences cancel in the sum, 
so that one can compute\footnote{The same formula can be found in eq.~(2.3) 
in Ref.~\cite{Bertolotti:2022aih}. However, we point out that in this paper we 
will adopt a slightly different notation. For instance, we will always specify the 
number of partons contributing to the matrix elements, for example $V_n$ 
and $R_\npo$. We will use the same convention for the various conterterms. 
We believe this choice to be beneficial for clarity, in the context of the general 
discussion pursued here.}
\beq
\label{sigNLO}
	\frac{d\sig_\NLO}{dX} \, = \, \lim_{d\to 4} \left[\int\, d\pn \,V_n \, \delta_n (X) + 
	\int\, d \pnpo\, R_{n+1} \, \delta_\npo (X) \right] \, ,
\eeq
where $V_n \equiv 2 \mathbf{Re} \big[ \mathcal{A}_n^{(0) \dagger} \mathcal{A}_n^{(1)} 
\big]$, and $R_{n+1} \equiv | \mathcal{A}_{n+1}^{(0)} |^2$. The main idea of infrared
subtraction is to introduce {\it local counterterms} mimicking the phase-space behaviour 
of real radiation in all singular regions, but simple enough to be analytically integrated 
over the unresolved degrees of freedom. At NLO, this amounts to defining a 
single function in $\Phi_\npo$ (in practice, a sum of contributions from non-overlapping 
singular regions), which we denote with $K_\npo^{\one}$, where the superscript 
emphasises the fact that only one particle can become unresolved at NLO. A proper 
definition of a local counterterm requires several steps, and a detailed construction 
was described for example in Refs.~\cite{Magnea:2018hab,Bertolotti:2022ohq}. 
For the purposes of this paper, we will not need a complete implementation, so we 
will provide here only a brief outline, which does not depend on the detailed definitions of the 
counterterms. 

The first step in our construction is to define {\it projection operators} that 
extract leading-power contributions to $R_\npo$, for all singular regions, in 
the relevant variables. Since divergences arise only in soft and collinear limits, 
at NLO the required operators are $\bS{i}$, extracting the soft limit for particle 
$i$ (where $\bS{i}\, R_{\npo}$ is singular only if $i$ is a gluon), and $\bC{ij}$, 
extracting the collinear limit for particles $i$ and $j$. In our 
approach, we then split the real-radiation phase space by introducing {\it sector 
functions} ${\cal W}_{ij}$, inspired by the method of Ref.~\cite{Frixione:1995ms}, 
forming a partition of unity, so that each radiation 
sector has at most one soft and one collinear singularity. This makes it possible 
to define a single operator collecting all NLO singular limits, without double 
counting, as
\beq
\label{L1def}
  {\bf L}^{\one} \, R_\npo \, = \, \sum_{i = 1}^\npo 
  \sum_{\substack{j=1 \\j \neq i}}^{\npo} \Big( \bS{i} + 
  \bC{ij} - \bS{i} \bC{ij} \Big) \, R_{\npo} {\cal W}_{ij} \, .
\eeq
The next step is to introduce {\it phase-space mappings}, following Ref.~\cite{Catani:1996vz}, 
in order to factor the 
radiative phase space as the product of a Born-level phase space times a 
measure for the degrees of freedom of the unresolved radiation, according 
to\footnote{Such factorisation is identical to the one presented in eq.~(2.5) 
in Ref.~\cite{Bertolotti:2022aih}, with the convention $d \Phi_{{\rm rad}, 1}^\npo 
\equiv d \Phi_{{\rm rad}}$. Analogously, we will use $d \Phi_{{\rm rad}, 2}^\npt$ 
as a synonym for $d \Phi_{{\rm rad}, 2}$ used in Ref.~\cite{Bertolotti:2022aih}.}
\beq
\label{phaspa1}
  d \Phi_{\npo} \, \equiv \frac{\varsi_{\npo}}{\varsi_n} \, d \Phi_n \, 
  d \Phi_{{\rm rad}, 1}^\npo \, ,
\eeq
where we explicitly display the appropriate symmetry factors $\varsi_p$. The 
soft and collinear operators can then be {\it improved} by expressing the limits 
in terms of mapped variables, allowing for a complete factorisation of soft and 
collinear kernels (which depend only on radiative degrees of freedom) from 
Born-level squared matrix elements (depending only on the momenta of the 
resolved particles). Note that different mappings can be implemented for different 
sectors, and even for different terms in the soft and collinear kernels, in order to 
simplify subsequent integrations. Note also that one can, if needed, adjust the 
actions of improved limits on sector functions, or include subsets of non-singular 
terms, in order to ensure the consistency of nested improved limits, or to ameliorate 
numerical stability. A detailed discussion about different mappings and strategies 
to adapt them to the various singular kernels can be found in Ref.~\cite{Bertolotti:2022aih}. 

We denote the improved operators by $\bbS{i}$, $\bbC{ij}$, and $\bbS{i} \bbC{ij}$, 
respectively, and we use them to build an improved version of the subtraction 
operator defined in \eq{L1def}, which we denote by $\bbL{\one}{}$. We can then 
finally define our NLO counterterm as
\beq
\label{NLOcount}
  K^{\one}_\npo \, = \, \bbL{\one}{} \, R_\npo \, .
\eeq
With these definitions, we are now in a position to integrate the local counterterm 
$K^{\one}_\npo$ over the unresolved degrees of freedom, for any fixed Born 
configuration, defining the {\it integrated counterterm}
\beq
\label{intcountNLO}
  I_n^{\, ({\bf 1})} \, = \, \int d \Phi_{\rm rad, 1}^\npo \, K_\npo^{\one} \, .
\eeq
The infrared poles of $I_n^{\, ({\bf 1})}$ are, by construction, the same as those 
arising from the exact integration of $R_\npo$ in \eq{sigNLO}, and thus must 
cancel the explicit poles of the virtual correction $V_n$. Similarly, subtracting 
the counterterm $K^{\one}_\npo$ from $R_\npo$ yields a function that is integrable 
in $\Phi_{\npo}$ directly in $d=4$. It is then possible to rewrite \eq{sigNLO} {\it identically} as
\beq
\label{subtNLO}
  \frac{d \sigma_\NLO}{d X} \, = \, 
  \int d \pn \Big[ V_n + I_n^{\, ({\bf 1})} \Big] \, \delta_n (X) \, + \, 
  \int d \Phi_\npo \, 
  \Big[  R_\npo \, \delta_\npo (X) - K_\npo^{\one} \,
  \delta_n (X) \Big] \, .
\eeq
Both combinations in square brackets are suitable for a direct numerical 
evaluation in $d=4$. Importantly, the infrared (IR) safety of the observable is necessary 
for the cancellation, which ensures $\delta_\npo (X)$ to turn smoothly into 
$\delta_n (X)$ in all unresolved limits.

%%%%%%%%%%%%%%%%%%%%%

\subsection{Subtraction at NNLO}
\label{ArchiNNLO}

The intricacy of the subtraction problem starts to show up at NNLO, where the 
cancellation of divergences requires mixing double-virtual corrections, to be 
integrated in the $n$-particle phase space, with real-virtual contributions, to be 
integrated in the $(\npo)$-particle phase space, and with integrated double-real 
radiation in $\pnpt$. The observable distribution is then
\beq
\label{NNLOX}
  \frac{d \sig_\NNLO}{d X} = \lim_{d \to 4} 
  \Bigg[ \! \int d \pn \, VV_n \, \delta_n (X) + \int d \pnpo \, 
  RV_\npo \, \delta_\npo (X) 
  + \int d \pnpt \, RR_\npt \, \delta_\npt (X) 
  \Bigg] , \qquad
\eeq 
where
\beq
\label{defNNLO}
  VV_n \, = \,  \left| {\cal A}_n^{(1)} \right|^2 \! + 
  2 {\bf Re}  \left[ {\cal A}_n^{(0) \dagger} {\cal A}_n^{(2)} \right] ,
  \quad
  RV_\npo \, = \, 2 {\bf Re} \left[ {\cal A}_\npo^{(0) \dagger} \, 
  {\cal A}_\npo^{(1)} \right] , 
  \quad
  RR_\npt \, = \, \left| {\cal A}_\npt^{(0)} \right|^2 \! . \quad
\eeq
The double-virtual matrix element $VV_n$ features up to quadruple poles in $\eps$, 
while the real-virtual correction $RV_{n+1}$ displays up to double poles in $\eps$, 
and up to two phase-space singularities. Finally, the double-real matrix element is 
finite in $d=4$, but is affected by up to four phase-space singularities. Following 
the procedure outlined at NLO, at this order we need to consider all possible 
double-unresolved limits of $RR_\npt$, to be added to the single-unresolved ones 
that we have already introduced. The relevant configurations involve: $i)$ double-soft 
emission of two partons $i$ and $j$, described by the projection operator 
$\bS{ij}$, $ii)$ triple-collinear splitting of a single Born-level parton, described 
by the operator $\bC{ijk}$, $iii)$ double-collinear splitting of two different Born-level 
partons, given by the operator $\bC{ij,kl}$, and, finally, $iv)$ joint emission of a soft
parton $i$ and a collinear pair $jk$, given by the operator $\bSC{i,jk}$. Since up to 
four particles are involved in these limits, the partition of the double-radiative phase 
space will involve sector functions ${\cal W}_{ijkl}$, with up to four different partonic 
labels. Each sector is designed to contain up to two soft singularities and up to two 
collinear singularities. Importantly, the relevant limits of $RR_\npt$ can be organised 
into three sets, which, in analogy with \eq{L1def}, we denote by
\beq
\label{L2def}
  {\bf L}^{\one} \, RR_\npt \, , \qquad \, {\bf L}^{\two} \, RR_\npt \, ,\qquad \,
  {\bf L}^{\otwo} \, RR_\npt \, \equiv \, {\bf L}^{\one} {\bf L}^{\two} \, RR_\npt \, .
\eeq
As was the case at NLO, ${\bf L}^{\one} RR_\npt$ collects (without double-counting)
all singular contributions associated with {\it single-unresolved} radiation: it is given
by \eq{L1def} with the replacement $R_\npo \to RR_\npt$, and with NNLO sector functions.
Similarly, ${\bf L}^{\two} RR_\npt$ collects all singularities due to {\it double-unresolved} 
radiation. Finally, ${\bf L}^{\otwo} \, RR_\npt$ collects all {\it strongly-ordered} limits, 
in which two particles become unresolved at different rates. These singular limits 
lie in the overlap of single-unresolved and double-unresolved radiation, and must 
be subtracted from their sum in order to avoid double counting. The operators 
${\bf L}^{\one}$, ${\bf L}^{\two}$ and ${\bf L}^{\otwo}$ are given by iterations
of the fundamental soft and collinear limits, and each term in these sums must 
be {\it improved}, in order to construct a fully subtracted form of the distribution in 
\eq{NNLOX}. To this end, we need to factorise the $(\npt)$-particle phase space
as in \eq{phaspa1}. In this case we have
\beq
\label{phaspafac2}
  d \Phi_{\npt} = \frac{\varsi_\npt}{\varsi_\npo} \, d \Phi_\npo \, d \Phi_{{\rm rad}, 1}^\npt 
  \, , \qquad \quad
  d \Phi_{\npt} = \frac{\varsi_\npt}{\varsi_n} \, d \Phi_n \, d \Phi_{{\rm rad}, 2}^\npt \, . 
\eeq
The procedure to improve the NNLO singular limits, beginning with the choice of 
suitable mappings of the double-radiative phase space, is intricate, and highly 
constrained by consistency requirements: in the case of massless final-state
radiation it is described in Ref.~\cite{Bertolotti:2022aih}. Ultimately, improved 
limits can be used to define the required local counterterms for double-real 
radiation, as
\beq
\label{NNLOcountRR}
  K^{\one}_\npt \, = \, \bbL{\one}{} \, RR_\npt \, ,  \qquad
  K^{\two}_\npt \, = \, \bbL{\two}{} \, RR_\npt \, ,  \qquad
  K^{\otwo}_\npt \, = \, \bbL{\otwo}{} \, RR_\npt \, .  
\eeq
When a phase-space parametrisation yielding \eq{phaspafac2} is in place, one 
can define {\it integrated counterterms} for double-real radiation as
\beq
\label{intcountNNLO}
  I_\npo^{\, ({\bf 1})} \, = \, \int d \Phi_{\rm rad, 1}^\npt \, K_\npt^{\one} \, , \qquad
  I_n^{\, ({\bf 2})} \, = \, \int d \Phi_{\rm rad, 2}^\npt \, K_\npt^{\two} \, , \qquad
  I_\npo^{\otwo} \, = \, \int d \Phi_{\rm rad, 1}^\npt \, K_\npt^{\otwo} \, . \quad
\eeq
The last ingredient to construct a fully subtracted version of \eq{NNLOX} is a 
local counterterm for the phase-space singularities of the real-virtual contribution 
$RV_\npo$. Since only one particle is radiated, and can become unresolved, 
it is natural to mimic the NLO procedure, and make use of the limit ${\bf L}^{\one} 
\, RV_\npo$, defined as in \eq{L1def}. The corresponding improved real-virtual counterterm reads
\beq
\label{NNLOcountRV}
  K^{\RV}_\npo \, = \, \bbL{\one}{} \, RV_\npo \, .
\eeq
As discussed below, however, the improvement of the real-virtual local counterterm 
is non-trivial, due to the presence of explicit infrared poles in $RV_\npo$, which are not 
associated with phase-space singularities. Once the appropriate improvement 
has been identified, we can define the real-virtual integrated counterterm as
\beq
\label{intcountNNLORV}
  I_n^{\RV} \, = \, \int d \Phi_{\rm rad, 1}^\npo \, K_\npo^{\RV} \, .
\eeq
Putting together the ingredients assembled so far, we can now write a fully 
subtracted form of the generic NNLO distribution, \eq{NNLOX}, which is the 
NNLO equivalent of \eq{subtNLO}. It is given by
\beq
\label{subtNNLO} 
  \frac{d \sig_\NNLO}{dX}
  & = &
  \int d \Phi_n \, \Big[ VV_n + I^{\two}_n + I^{\RV}_n \Big] \, \delta_n (X) 
  \\
  & + & 
  \int d \Phi_\npo \, \bigg[ \Big( RV_{\npo} + I^{\one}_\npo \Big) \, 
  \delta_\npo (X)
  \, -  \, \Big( K^{\RV}_\npo + I^{\otwo}_\npo \Big) \, \delta_n (X) \bigg]
  \nn \\
  & + & 
  \int d \Phi_\npt \, \bigg[ RR_\npt \, \delta_\npt (X) - 
  K^{\one}_\npt \, \delta_\npo (X) - \Big( K^{\two}_\npt - 
  K^{\otwo}_\npt \Big) \, \delta_n (X) 
  \bigg] \, . \nn
\eeq
With the definitions we have presented above, \eq{subtNNLO} is an {\it identical} 
rewriting of \eq{NNLOX}. To analyse \eq{subtNNLO}, we begin by noting that 
the third line is integrable in $\Phi_\npt$ by construction, since all singular regions 
have been subtracted with no double counting. The analysis of the second line, 
which involves both explicit poles in $\eps$ and phase-space singularities, is 
more delicate. With a minimal definition of the local counterterms (including 
only leading-power contributions to the singular limits), standard theorems for 
the cancellation of IR divergences imply that the integral $I_\npo^{\, ({\bf 1})}$ 
must cancel the poles of $RV_\npo$. The resulting sum, however, will still be 
affected by phase-space singularities, when the radiated particle becomes 
unresolved. To take care of this problem, the counterterm $K^{\RV}_\npo$ is 
designed to cancel the phase-space singularities of $RV_\npo$. Furthermore, 
the integral $I_\npo^{\otwo}$ must share the phase-space singularities of 
$I_\npo^{\, ({\bf 1})}$, since the limit taken in defining $K_\npt^{\otwo}$ from 
$K_\npt^{\one}$ does not affect the leading-power contributions in the variables 
relative to the first unresolved particle. We conclude that the second line in \eq{subtNNLO} 
is free of phase space singularities, and the first parenthesis is finite in $d=4$. 
However, with a minimal definition of $K_\npo^{\RV}$, we are  not guaranteed that 
the second parenthesis will be finite: for example, the poles in $K_\npo^{\RV}$ 
will depend on the phase-space mappings used in \eq{NNLOcountRV}, while 
the poles in $I^{\otwo}_\npo$ will depend on the mappings and parametrisations 
used in the double-radiation phase space, and incorporated in \eq{NNLOcountRR}. 
Nevertheless, the complete cancellation of poles in the second parenthesis of the 
second line can be enforced by adjusting the definition of $K_\npo^{\RV}$ to match 
the poles of $I_\npo^{\otwo}$, without affecting phase-space singularities. Since 
this adjustment involves only terms that are not singular when the radiated 
particle becomes unresolved, we understand it here as part of the necessary
improvement of the ${\bf L}^{\one}{}$ operator, leading to the definition of 
$\bbL{\one}{}$, when acting on $RV_{\npo}$. Having established the finiteness and 
integrability of both the second and the third line in \eq{subtNNLO}, the cancellation 
of poles in the first line follows directly from the KLN theorem.

%%%%%%%%%%%%%%%%%%%%%

\subsection{Subtraction at higher orders}
\label{ArchiHigh}

It is interesting to attempt to extrapolate the patterns appearing at NLO and at 
NNLO to higher orders. At N$^3$LO the required matrix elements are
\beq
\label{defN3LO}
  VVV_n \, = \,  2 {\bf Re}  \left[ {\cal A}_n^{(0) \dagger} {\cal A}_n^{(3)}  + 
  {\cal A}_n^{(1) \dagger} {\cal A}_n^{(2)} \right] \, ,
  \quad && \quad
  RVV_\npo \, = \, \left| {\cal A}_\npo^{(1)} \right|^2 + 2 {\bf Re} 
  \left[ {\cal A}_\npo^{(0) \dagger} \, {\cal A}_\npo^{(2)} \right] \, , \nonumber \\
  RRV_\npt \, = \, 2 {\bf Re} \left[ {\cal A}_\npt^{(0) \dagger} \, 
  {\cal A}_\npt^{(1)} \right] \, , \quad && \quad
  RR_\npth \, = \, \left| {\cal A}_\npth^{(0)} \right|^2 \, .
\eeq
The uniform limits in which three particles become unresolved are
\beq
\label{unilim3}
  \bS{ijk} \, , \quad \bC{ijkl} \, , \quad \bC{ij, klm} \, , \quad \bC{ij, kl, mn}  
  \, , \quad \bSC{i, jkl} \, , \quad \bSC{i, jk, lm}  \, , \quad \bSC{ij, kl} \, ,
\eeq
where, following the conventions introduced at NNLO, we use commas to separate
clusters of particles originating from the same hard parton, and we treat all soft particles 
as a single cluster. Thus, for example, in the limit $\bC{ij, klm}$ particles $i$, $j$
and particles $k$, $l$, $m$ form two distinct collinear clusters, while in the limit
$\bSC{ij, kl}$ particles $i$ and $j$ are soft, while particles $k$ and $l$ form a collinear 
pair. These limits must of course be added on top of the single- and double-unresolved 
ones. Importantly, at N$^3$LO there are several new possibilities for strong ordering of
infrared limits. In fact, in analogy with \eq{L2def}, one must introduce triple-unresolved 
limits
\beq
\label{L3def}
  {\bf L}^{\three} \, RRR_\npth \, , \quad && \quad
  {\bf L}^{\othree} \, RRR_\npth \, \equiv \, {\bf L}^{\one} {\bf L}^{\three} \, 
  RRR_\npth \, , \\
  {\bf L}^{\tthree} \, RRR_\npth \, \equiv \, {\bf L}^{\two} {\bf L}^{\three} \, 
  RRR_\npth \, , \quad &&  \quad
  {\bf L}^{\otthree} \, RRR_\npth \, \equiv \, {\bf L}^{\one} {\bf L}^{\two} {\bf L}^{\three} \, 
  RRR_\npth \, . \nonumber 
  \qquad
\eeq
Here ${\bf L}^{\three}$ gives the uniform limit, where three particles become unresolved 
at the same rate, ${\bf L}^{\othree}$ gives the strongly-ordered limits in which one 
particle becomes unresolved much faster than the other two, ${\bf L}^{\tthree}$ 
covers the case in which two particles become unresolved at the same rate, but 
much faster than the third particle. Finally ${\bf L}^{\otthree}$ gives the completely 
ordered limits in which each particle becomes unresolved at a different rate. All of 
the limits in \eq{L3def} will be expressed as iterations of the fundamental 
limits in \eq{unilim3}, and each term in these sums will have to be improved, using 
the phase-space factorisations
\beq
\label{phaspafac3}
  d \Phi_{\npth} = \frac{\varsi_\npth}{\varsi_\npt} \, d \Phi_\npt \, 
  d \Phi_{{\rm rad}, 1}^\npth \, ,
  \quad \!
  d \Phi_{\npth} = \frac{\varsi_\npth}{\varsi_\npo} \, d \Phi_\npo \, 
  d \Phi_{{\rm rad}, 2}^\npth \, ,
  \quad \!
  d \Phi_{\npth} = \frac{\varsi_\npth}{\varsi_n} \, d \Phi_n \, 
  d \Phi_{{\rm rad}, 3}^\npth \, .  \qquad
\eeq
The required counterterms for limits involving three unresolved particles will then 
be of the form
\beq
\label{N3LOcountRRR}
  K^{\, ({\bf h})}_\npth \, = \, \bbL{\, ({\bf h})}{} \, RRR_\npth \, , 
\eeq
with ${\bf h} \in \{ {\bf 3}, {\bf 13}, {\bf 23}, {\bf 123} \}$, and the corresponding 
integrated counterterms will be given by
\beq
\label{intcountN3LO}
  I_{n + 3 - q}^{\, ({\bf h})} \, = \, \int d \Phi_{{\rm rad}, q}^\npth \, K_\npth^{\, ({\bf h})} \, ,
\eeq
where $q$ is the number of particles going unresolved at the highest rate, 
as in \eq{intcountNNLO}. In order to complete the N$^3$LO subtraction formula, 
one will further need a single local counterterm for single real radiation in 
$RVV_\npo$, which we denote by $K^{\, (\RVV)}_\npo$, and three local 
counterterms for double real radiation in $RRV_\npt$, which we denote by 
$K^{\, (\RRV, \, {\bf h})}_\npt$, with ${\bf h} \in \{ {\bf 1}, {\bf 2}, {\bf 12} \}$, 
following the pattern set in \eq{NNLOcountRR}. These $RVV$ and $RRV$ 
counterterms must be integrated in the appropriate phase spaces, according 
to the rules introduced at NLO and at NNLO. Thus we define
\beq
\label{intcountN3LO2}
  I_n^{\, \RVV} \, = \, \int d \Phi_{\rm rad, 1}^\npo \, K_\npo^{\, (\RVV)} \, ,  
  \qquad \quad
  I_{n + 2 - q}^{\, (\RRV, \, {\bf h})} \, = \, \int d \Phi_{{\rm rad}, q}^\npt \, 
  K_\npt^{\, (\RRV, \, {\bf h})} \, .
\eeq
We are finally in a position to write a {\it master formula} for (final-state) 
N$^3$LO subtraction. It takes the form
\beq
\label{N3LOfin}  
  \frac{d \sig_\NNNLO}{dX} & = & \int d \Phi_n \, 
  \Big[ VVV_n + I_n^{\three} + I_n^{\, (\RRV, \,{\mathbf 2})} + I_n^{\, (\RVV)} 
  \Big] \, \delta_n (X) \nn \\
  && \hspace{-15mm} + \, \int d \Phi_\npo \, \bigg[
  \Big( RVV_\npo + I^{\two}_\npo + I^{\, (\RRV, \, \mathbf{1})}_\npo
  \Big) \, \delta_\npo (X) \, - \,
  \Big( K^{\, (\RVV)}_\npo + I^{\, (\mathbf{2 3})}_\npo
  + I^{\, (\RRV, \, \mathbf{12})}_\npo \Big) \, \delta_n (X) \bigg] \nn \\
  && \hspace{-15mm} + \, \int d \Phi_\npt \, \bigg\{
  \Big( RRV_\npt + I^{\one}_\npt \Big) \, \delta_\npt (X) -
  \Big( K^{\, (\RRV, \mathbf{1})}_\npt + I^{\otwo}_\npt \Big) \, \delta_\npo (X)
  \nn \\
  & & \hspace{3cm} - \, \bigg[
  \Big( K^{\, (\RRV, \mathbf{2})}_\npt + I^{\, (\mathbf{13})}_\npt \Big) -
  \Big( K^{\, (\RRV, \mathbf{12})}_\npt  + I^{\, (\mathbf{123})}_\npt \Big)
  \bigg] \, \delta_n (X) \bigg\} \nn \\
  && \hspace{-15mm} + \, \int d\Phi_\npth \, \bigg[ RRR_\npth \, 
  \delta_\npth (X) - K^{\one}_\npth \, \delta_\npt (X) -
  \Big( K^{\two}_\npth - K^{\otwo}_\npth \Big) \, \delta_\npo (X) \nn \\
  && \hspace{3cm} - \,
  \Big( K^{(\mathbf{3})}_\npth - K^{\, (\mathbf{13})}_\npth
  - K^{\, (\mathbf{23})}_\npth + K^{\, (\mathbf{1 2 3})}_\npth \Big) \,
  \delta_n (X) \bigg] \, .
\eeq
In a concrete implementation, each one of the four phase space integrands in 
\eq{N3LOfin} will be free of non-integrable singularities, and finite in $d=4$. 
For example, in the $(\npth)$-particle phase space, single-unresolved and 
double-unresolved singular contributions have been subtracted, and their 
overlap added back, in the first line. Triple-unresolved singular contributions 
are subtracted in the second line, with their overlap with single- and 
double-unresolved ones removed, and the overlap of the overlaps added 
back in. In the $(\npt)$-particle phase space, infrared poles will cancel in the 
first parenthesis according to general theorems, and the phase-space
singularities of $RRV_\npt$ are fully subtracted by construction. Furthermore, 
the phase-space singularities of the three local counterterms are matched 
by the three corresponding integrated counterterms, also by construction. 
This would leave out possible uncancelled explicit poles, not associated 
with phase-space singularities, in each one of the last three parentheses: 
these can be handled in the process of improving the relevant limit operators, 
as was done at NNLO. A similar pattern applies to the integrand in the 
$(\npo)$-particle phase space.

We see that a general-purpose subtraction algorithm at N$^3$LO requires 
the construction of 11 local counterterm functions\footnote{We have described 
the structure of subtraction for final-state emissions only, however the inclusion 
of initial-state radiation does not essentially affect this counting. Rather, in a 
sector approach, some terms in the collinear parts of the counterterms, 
associated with initial-state radiation sectors, must be treated and integrated   
separately.}, out of which 5 involve various forms of strong ordering, while 6 
correspond to uniform limits. At the price of introducing some more formal 
notations, it is not too difficult to generalise this counting to N$^k$LO, for 
generic $k$. We find that the total number of local counterterms is then
\beq
\label{countcount}
  c(k) \, = \, 2^{k+1} - 2 - k \, ,
\eeq
and, of these, only $k (k +1 )/2$ correspond to uniform limits, while the 
remaining ones involve strong ordering: for example, already for $k = 4$
the number of strongly-ordered counterterms (16) exceeds the number of 
uniform ones (10). Moreover, as we have seen, the pattern of cancellations 
involving mixed real-virtual local counterterms and the integrated versions of 
counterterms with extra real radiation is especially delicate, due to the interplay 
of phase-space singularities and explicit poles. Clearly, a general understanding 
of infrared subtraction will require mastering strongly-ordered infrared limits, 
and the cancellation of their singularities. We believe that a factorisation-based 
approach will be instrumental to this understanding, and we now turn to illustrate 
this viewpoint.

%%%%%%%%%%%%%%%%%%%%%%%%%%%%%%%%%%%%%%%

\section{The factorisation approach to subtraction}
\label{FactoSubtra}

Conceptually, the approach to subtraction that was discussed in the previous 
section can be described as {\it bottom-up}, according to the ordering used to 
analyse the lines in \eq{subtNNLO} and \eq{N3LOfin}: one starts by considering 
the singular limits for multiple real-radiation matrix elements, and subsequently 
integrates the resulting counterterms, finally achieving the cancellation of virtual 
poles. However, we believe that it is interesting to consider a complementary 
{\it top-down} viewpoint, beginning with an analysis of infrared poles in the virtual 
correction, with the tools of factorisation, and then constructing infrared-finite soft 
and collinear cross sections, from which one can extract local counterterms.
This approach was pursued in Ref.~\cite{Magnea:2018ebr}, where a general
prescription was given to construct local counterterms for uniform soft and 
collinear limits, in terms of matrix elements of fields and Wilson lines. We now 
discuss how this approach can be extended to strongly-ordered local counterterms: 
this extension will ultimately guarantee the cancellation of infrared poles and 
phase-space singularities in the intermediate lines of \eq{subtNNLO} and \eq{N3LOfin}.

To begin our discussion, we recall the well-known infrared factorisation formula
for massless gauge-theory scattering amplitudes~\cite{Dixon:2008gr,Gardi:2009qi,
Becher:2009qa,Feige:2014wja} 
\beq
\label{AmpFact}
  \mathcal{A}_n \big( \{p_i\} \big) \, = \, \prod_{i = 1}^n \left[ 
  \frac{ \J_i \big( p_i, n_i \big)}{\J_{\E_i} \big( \beta_i, n_i \big)} \right] \, 
  \s_n \big( \{\beta_i\} \big) \, \hard_n \big( \{p_i\}, \{n_i\} \big) \, ,
\eeq
where we introduced four-velocities $\beta_i \equiv \sqrt2 \, p_i/\mu$, and reference 
vectors $n_i$, for each one of the external legs. The soft function $\s_n$ captures 
long-wavelength gluon exchanges between external particles: it does not 
depend on the spin of the latter, but it is a non-trivial operator 
in colour space, acting on the finite hard coefficient $\hard_n$. A jet function 
$\J_i$ is associated to each external particle, capturing collinear singularities: 
these depend only on the momentum and spin of the hard particle $i$, and 
are proportional to the identity in colour space. The eikonal jets $\J_{\E_i}$ 
encapsulates the overlap of soft and collinear singularities, that is present both 
in $\s_n$ and in $\J_i$, and would otherwise be double-counted. 

The factorisation functions $\s_n$, $\J_i$ and $\J_{\E_i}$ appearing in 
\eq{AmpFact} have explicit operator definitions involving semi-infinite Wilson lines 
aligned with the external-particle trajectories,
\beq
\label{WilsonLine}
  \Phi_{\beta_i} (\infty,0) \, \equiv \, P \exp \left\{ {\rm i} g_s \T^a \int_0^\infty dz \,
  \beta_i \cdot A_a (z) \right\} \, ,
\eeq
where the symbol $P$ identifies the path ordering, $A_a$ is the gluon field, and $g_s$ 
is the strong coupling. These Wilson lines capture soft-gluon radiation off external particles; 
further Wilson lines along the unphysical directions $n_i$ are employed in the 
definition of the jet and eikonal jet functions. In the context of subtraction, it is 
useful to consider the virtual soft and jet functions in \eq{AmpFact} as special 
cases of more general objects, which we describe as {\it radiative} soft and jet 
functions, and can be used to organise soft and collinear {\it real} radiation. 
In the soft case, we begin by defining {\it eikonal form factors}
\beq
\label{eikFF}
  \s_{n,f_1 \ldots f_m} \big( \{ \beta_i \}; \radState \big) \, \equiv \, 
  \braket{\radState | \, 
  T \left[ \prod_{i=1}^n \Phi_{\beta_i}(\infty,0) \right] |0} \, ,
\eeq
where $T$ is the time-ordering operator. The equation above describes the 
radiation of $m$ particles of flavours $f_j$, momenta $k_j$ and spin polarisations 
$\lambda_j$ ($j = 0, \ldots, m$), from the Wilson lines representing hard particles, 
including virtual corrections in the soft approximation. Similarly, we define 
{\it collinear form factors}, in order to describe collinear radiation from external 
particle $i$. These are spin-dependent quantities, involving the quantum field 
responsible for the creation or absorption of the hard particle. For final-state 
quarks we write
\beq
\label{collFFq}
  \mathcal{J}_{q, f_1 \ldots f_m}^{\alpha} \big(x; n; \{k_j, \lambda_j\} \big) \, \equiv \, 
  \braket{\radState | \, T \left[ \bar{\psi}^{\alpha}(x) \Phi_n(x,\infty) \right] |0} \, ,
\eeq
where one of the final-state particles is the quark created by the field $\bar{\psi}$ at point 
$x$. For final-state gluons, on the other hand, we choose~\cite{Becher:2010pd}
\beq
\label{collFFg}
  g_s \, \mathcal{J}_{g, f_1 \ldots f_m}^{\nu} \big( x; n; \{k_j, \lambda_j\} \big) 
  \, \equiv \, 
  \braket{\radState | \, T \Big[ \Phi_n (\infty, x) \big( {\rm i} D^{\nu} 
  \Phi_\beta (x, \infty) \big) 
  \Big] |0} \, ,
\eeq
where the $\Phi_\beta$ Wilson line is in the adjoint representation, oriented along
the direction of a final-state gluon within the set $\{k_j\}$, and $D_\nu = \partial_\nu 
- {\rm i} g_s A_\nu$. In \eq{collFFq} and \eq{collFFg}, the index $j$ ranges in 
$\{1, \dots, m\}$, where the $m=1$ case reads $\mathcal{J}_{f_i, f_i}=\mathcal{J}_{i}$.

Finally, the eikonal versions of collinear form factors are spin-independent, and only 
characterised by the colour representation of the hard emitter. We can thus use
the definition in \eq{collFFq}, and replace the field by a Wilson line along the classical
quark or gluon trajectory, which gives
\beq
\label{eikcollF}
  \mathcal{J}_{\E_i, f_1 \ldots f_m} \big(\beta_i; n_i;  \{k_j, \lambda_j\} \big) \, \equiv \, 
  \braket{\radState | \, T \Big[ \Phi_{\beta_i} (\infty, 0) \Phi_{n_i} (0, \infty) \Big] |0} \, .
\eeq
The purely virtual soft and jet functions appearing in \eq{AmpFact} are instances
of these form factors in the cases $m = 0$ (for the eikonal form factors and eikonal
jets, corresponding to no final-state radiation), and $m = 1$ (for the collinear form
factors, corresponding to the emission of the single particle created by the field).

At cross-section level, eikonal and collinear form factors must be appropriately 
squared, building up radiative soft and jet functions, which are fully local in the 
degrees of freedom of soft and collinear real radiation. Specifically, the radiative 
soft function, responsible for the emission of $m$ soft particles by $n$ resolved emitters, 
is defined by
\beq
\label{RadSoftFunc}
  S_{n, f_1 \ldots f_m} \big( \{\beta_i\}; k_1, \dots, k_m) \, = \,  
  \sum_{ \{ \lambda_j \} } \s^\dagger_{n, f_1 \ldots f_m} 
  \big( \{\beta_i\}; \radState \big) \, \s_{n, f_1\ldots f_m} \big( \{\beta_i\}; \radState \big) \, ,
\eeq
where we summed over the polarisation of the final-state soft particles, regardless of 
their flavour\footnote{Although Wilson lines only radiate gluons, those gluons can in 
turn produce $q \bar{q}$ pairs.}. Radiative jet functions must take into account the fact 
that hard-collinear emissions carry non-negligible momentum. At cross-section level
one will therefore need a convolution rather than a simple product, with one of the two
collinear form factors evaluated at a displaced location $x$, Fourier-conjugate to 
the total momentum $\ell$ carried by final-state particles. We thus define
\beq
\label{RadJetFuncs}
  J_{f, f_1 \ldots f_m}^{\al \beta} \big(\ell; n; k_1, \ldots, k_m \big) =
  \sum_{ \{ \lambda_j \} } \, \int d^d x \, {\rm e}^{ {\rm i} \ell \cdot x} \, 
  \mathcal{J}_{f, f_1 \ldots f_m}^{\al, \dagger} \big( 0; n; \{k_j, \lambda_j\} \big) 
  \mathcal{J}_{f, f_1 \ldots f_m}^{\beta} \big( x; n; \{k_j,\lambda_j\} \big) \, . \quad
\eeq
In \eq{RadJetFuncs}, $f$ denotes the flavour of the parent particle, which can be
either a quark or a gluon, and we have adopted a unified notation, where $\al$ and 
$\beta$ are indices in the spin-1/2 or the spin-1 representations of the Lorentz group, 
depending on the flavour $f$. Performing the $x$ integral will fix $\ell = \sum_j k_j$. 
To exemplify, for the discussion of NLO corrections we will need $J_{f, f_1 f_2}^{(0)}$,
describing the tree-level splitting of type $f \rightarrow f_1+f_2$; similarly, for
tree-level triple-collinear emissions, relevant for NNLO corrections, we will need 
$J_{f, f_1 f_2 f_3}^{(0)}$, corresponding to the branching $f \rightarrow f_1+f_2+f_3$;
finally, real-virtual corrections involving  one-loop splittings will be described by the 
one-loop jet function  $J_{f, f_1 f_2}^{(1)}$. We tested the definition of the radiative 
jet function at tree-level in the case of single and double radiation, and found agreement 
with known results (see for instance Ref.~\cite{Catani:1999ss} for a comprehensive 
list of splitting functions). The single radiative jet functions are given explicitly in 
Appendix~\ref{app:jetFuncs}. For cross-section-level radiative eikonal jets, the 
Fourier transform is not necessary, since they can be computed directly in the soft
approximation\footnote{Note that this convention is different from the one adopted 
in Ref.~\cite{Magnea:2018ebr}.}. We have
\beq
\label{EikRadJetFunc}
  J_{\E_i, f_1 \ldots f_m} \big( \beta_i; n_i; k_1, \ldots, k_m \big) \, = \, 
  \sum_{ \{ \lambda_j \} } \mathcal{J}_{\E_i, f_1 \ldots f_m}^{\dagger} 
  \big( \beta_i; n_i; \{k_j, \lambda_j\} \big) \, 
  \mathcal{J}_{\E_i, f_1 \ldots f_m} \big( \beta_i; n_i; \{k_j, \lambda_j\} \big) \, .
\eeq
As argued in Ref.~\cite{Magnea:2018ebr}, the cross-section-level soft and jet 
functions thus defined provide natural candidates to build local soft and collinear
counterterms for subtraction algorithms, to any order in perturbation theory. Indeed, 
integrating \eq{RadSoftFunc} over the radiative $m$-particle phase space, summing 
over the number of particles, and using {\it completeness}, one finds
\beq
  \sum_{m = 0}^\infty \sum_{\{f_i\}}  \int d \Phi_m \, S_{n, f_1 \ldots f_m} 
  \big(\{ \beta_i \}, k_1, \dots, k_m \big) \, = \, 
  \bra{0} \,  \overline{T} \left[ \prod_{i = 1}^n 
  \Phi_{\beta_i} (0, \infty) \right] T \left[ \prod_{i = 1}^n 
  \Phi_{\beta_i} (\infty, 0) \right] \ket{0} \, , \qquad 
\label{CompleteSoft}
\eeq
where $\overline{T}$ is the anti-time ordering.
The {\it r.h.s.} of \eq{CompleteSoft} represents a total cross section in the presence
of Wilson-line sources, and it is infrared finite order by order in perturbation theory.
This implies that the phase-space integrals of the $m$-particle radiative soft functions 
defined in \eq{RadSoftFunc} do indeed cancel the virtual poles arising from soft 
virtual corrections. Similarly, integrating \eq{RadJetFuncs} over phase space, summing
over the number of radiated particles, and using completeness, one finds (for quarks)
\beq
&&
  \sum_{m = 1}^\infty \! \sum_{\{f_i\}} \int d \Phi_m
  J_{q, f_1 \ldots f_m}^{\alpha \beta} \! \big( \ell; n; k_1, \ldots , k_m \big)
\nnb\\
&&
\hspace{25mm}
 = \,
  {\rm Disc} \bigg[ \! \int \! d^d x \, {\rm e}^{{\rm i} \ell \cdot x}
  \bra{0} T \Big[ \Phi_n (\infty, x) \psi^\beta (x) \bar{\psi}^\alpha (0) 
  \Phi_n (0, \infty) \Big] \! \ket{0}
  \!  \bigg] \, ,
\label{CompleteColl}
\eeq
and similarly for gluons. The {\it r.h.s.} of \eq{CompleteColl} represents the discontinuity 
of a two-point function in the presence of Wilson lines, and it is infrared finite order by 
order, as was the case for \eq{CompleteSoft}. This shows that the radiative jet functions
in \eq{RadJetFuncs} provide candidate local collinear counterterms to cancel virtual
collinear singularities.

%%%%%%%%%%%%%%%%%%%%%

\subsection{A top-down approach to subtraction at NLO}
\label{NLOSubtra}

Before moving on to strongly-ordered soft and collinear limits, we illustrate how one 
can build and integrate local infrared counterterms within the framework presented in 
\secn{FactoSubtra}, and previously discussed in some detail in Ref.~\cite{Magnea:2018ebr}. 
We begin at NLO, where some of the technical issues can be easily clarified. For 
the sake of notational simplicity, from now on we drop the $n$ dependence from the 
argument of jet functions, and we expand all functions in perturbation theory using 
the same conventions as in \eq{pertexpA}. With these definitions, the NLO virtual 
correction $V_n$ can be obtained by expanding \eq{AmpFact} to NLO, and can be 
written as
\beq
\label{VirtualNLO}
   V_n & = & {\hard_n^{(0)}}^\dagger \, S_{n}^{(1)} \big(\{ \beta_i \} \big) \, \hard_n^{(0)} \, - \,
   {\hard_n^{(0)}}^\dagger \, S_{n}^{(0)} \, \hard_n^{(0)} \, \sum_{i=1}^n J_{\E_i}^{(1)} 
   \big( \beta_i \big)
   \nn \\ &&
  + \sum_{i=1}^n \, \sum_{f_j} \, \int \frac{d^d \ell_i}{(2 \pi)^d} \, 
  {\hard_{n, \, \alpha_i}^{(0)}}^{\!\!\!\! \dagger} \! \left( \{p\}_{\slashed{i}}, \ell_i \right) \, 
  J_{f_i, f_j}^{(1) \, \alpha_i \beta_i} \big(\ell_i; p_i \big) \, 
  S_{n}^{(0)} \, \hard_{n, \, \beta_i}^{(0)} \! \left( \{ p \}_{\slashed{i}}, \ell_i \right) 
  \, + \, \text{finite} 
  \nn \\ & \equiv &
  V_n^{\, ({\rm s})} \, + \, \sum_{i=1}^n \, V_{n, \, i}^{\, ({\rm hc})} \, + \, \text{finite}
  \, , \qquad
\eeq
where we distinguish the soft singular contribution $V_n^{\, ({\rm s})}$, given by the 
first term in \eq{VirtualNLO}, from hard-collinear contributions associated with each 
external particle, denoted by $V_{n, \, i}^{\, ({\rm hc})}$ and given by the remaining two 
terms. The tree-level hard function $\hard_n^{(0)}$ depends on the momentum and 
spin (as well  as colour) of the partons in the Born process, while the tree-level soft 
function $S_{n}^{(0)}$ is just a colour tensor connecting the two hard functions. 
The notation $\hard_{n, \alpha_i}^{(0)} \big( \{p\}_{\slashed{i}},\ell_i \big)$ means 
that one needs to replace $p_i$ with $\ell_i$ in the hard function, as well as remove 
the wave function of parton $i$ (the spinor for quarks and the polarisation vector for 
gluons), uncovering the spin index $\al_i$. Since $J_{f_i, f_j}^{(1)}(\ell_i;p_i) \sim 
\delta(\ell_i - p_i) \, \delta_{f_i f_j}$, the $\ell_i$ integral and the flavour sum are 
trivial: we include them here in order to match the necessary notation when dealing 
with the case of real radiation, where the corresponding integrals will identify $\ell_i$ 
as the parent particle for collinear splittings, and the flavour sums will be non trivial. 

We can now exploit the finiteness of \eq{CompleteSoft} and of \eq{CompleteColl} (as 
well as the analogous relation for eikonal jets) to write the NLO \emph{completeness 
relations}\footnote{We defer to Appendix \ref{scollcanc} a proper treatment of UV 
divergences associated with these relations.}
\beq
\label{FinCondNLO1}
  S_n^{(1)} \big( \{ \beta_i \} \big) + \int d \Phi(k) \, S_{n, g}^{(0)} \big( \{ \beta_i \}; k \big) 
  & = & \, \text{finite} \, , \\
\label{FinCondNLO2}
  J_{\E_i}^{(1)} (\beta_i) +  \int d \Phi(k) \, J_{\E_i, g}^{(0)} (\beta_i; k) 
  & = & \, \text{finite} \, , \\
\label{FinCondNLO3}  
  \sum_{f_1} \int d \Phi(k_1) \, J_{f, f_1}^{(1) \, \alpha \beta} (\ell; k_1) + \sum_{f_1, f_2} 
  \varsi_{f_1 f_2} \int d \Phi(k_1) d \Phi(k_2) \, 
  J_{f, f_1 f_2}^{(0) \, \alpha \beta} (\ell; k_1, k_2) & = & \text{finite} \, , 
\eeq
where the flavour sum extends to all final-state flavour combinations compatible 
with the Feynman rules (with each combination counted only once: for example, 
if $\{ f_1, f_2 \} = \{q,g\}$ is included, then $\{ f_1, f_2 \} = \{g,q\}$ is excluded).
Furthermore, $\varsi_{f_1 f_2}$ is a phase-space symmetry factor, equal to $1/2$ 
when $f_1 = f_2 = g$ and equal to one in all other cases at NLO. For single soft 
emissions, we have used the fact that only the radiation of a single gluon is allowed, 
which removes the need for a flavour sum in \eq{FinCondNLO1} and in \eq{FinCondNLO2}. 
The finiteness conditions in eqs.~(\ref{FinCondNLO1}-\ref{FinCondNLO3}) immediately 
suggest expressions for candidate soft and hard-collinear local infrared counterterms 
at NLO, which are given by 
\beq
\label{CandCountNLO1}
  K^{({\bf 1}, \, \rm{s})}_\npo \big( \{ p_i \}, k \big) & = & 
  {\hard_n^{(0)}}^\dagger \big( \{ p_i \} \big) \, S_{n, g}^{(0)} \big( \{ \beta_i \}; k \big) \,
  \hard_n^{(0)} \big( \{ p_i \} \big) \, , \\
\label{CandCountNLO2}
  K^{({\bf 1}, \, \rm{hc})}_{\npo, \, i} \big( \{ p_i \}, k_1, k_2 \big) 
  & = & 
  \sum_{f_1, f_2}
  \Bigg[ \int \frac{d^d \ell_i}{(2\pi)^d} \,\, {\hard_{n, \alpha_i}^{(0)}}^{\!\!\!\dagger} 
  \big( \{ p \}_{\slashed{i}}, \ell_i \big) J_{f_i, f_1f_2}^{(0) \, \alpha_i \beta_i} 
  \big( \ell_i; k_1, k_2 \big)
  \, S_{n}^{(0)} \, \hard_{n, \beta_i}^{(0)} \big( \{ p \}_{\slashed{i}}, \ell_i \big) \nn \\
  & & \hspace{1cm} - \,\, {\hard_n^{(0)}}^\dagger \big( \{ p_i \} \big) \,
  \bigg( \sum_{j = 1}^2 J_{\E_i, f_j}^{(0)} \big( \beta_i; k_j \big) \bigg) 
  S_{n}^{(0)} \, \hard_n^{(0)} \big( \{ p_i \} \big) 
  \Bigg] \\
  & \equiv & {\hard_n^{(0)}}^\dagger  
  \sum_{f_1, f_2} \bigg( J_{f_i, f_1f_2}^{(0)} -
  \sum_{j = 1}^2 J_{\E_i, f_j}^{(0)} \bigg)  
  S_{n}^{(0)} \, \hard_n^{(0)} \, , \nn
\eeq
where in the last line we have introduced a shorthand notation to denote the
convolution of the hard function with the jet function over the total collinear 
momentum $\ell_i$, including the spin sum. Note again that eikonal jets with 
a single final-state quark emission, $J_{\E_i, q}$, vanish: thus the sum on the 
second line of \eq{CandCountNLO2} includes only one term, except in the case 
$f_1 = f_2 = g$. For all flavour combinations it is straightforward to check, using the 
results of Appendix~\ref{app:jetFuncs}, that the jet combination in round brackets 
in the latter equation is completely free of soft phase-space singularities ({\it i.e.}~both 
those associated to $f_1$ and to $f_2$). Hence we define the 
{\it hard-collinear} single-radiative jet at $k$ loops as
\beq
\label{eq:hc_jet_double_real}
  J^{(k), \, \text{hc}}_{f_i, f_1 f_2}
  \, \equiv \,
  J^{(k)}_{f_i, f_1 f_2} -
  \sum_{j = 1}^2 J_{\E_i, f_j}^{(k)}
  \, .
\eeq
With these definitions, \eq{FinCondNLO1} implies that the phase-space integral 
(over the gluon momentum $k$) of \eq{CandCountNLO1} will cancel the explicit 
infrared poles of the first term in \eq{VirtualNLO}; similarly, \eq{FinCondNLO2} 
shows that the phase-space integration of momentum $k_j$ in the last term in 
\eq{CandCountNLO2} will cancel the explicit infrared poles of the second term 
in \eq{VirtualNLO}; finally, \eq{FinCondNLO3} shows that the phase-space 
integration of the first term in \eq{CandCountNLO2} over momenta $k_1$ and 
$k_2$ will cancel the explicit infrared poles of the last term in \eq{VirtualNLO}, 
after the (trivial) integration in $d \Phi(k_1)$. 

It is important, at this stage, to emphasise that the candidate counterterms in 
(\ref{CandCountNLO1}) and (\ref{CandCountNLO2}) are not quite ready for 
implementation in a subtraction scheme. The first problem is that the 
phase-space integrals in \eq{FinCondNLO1} and \eq{FinCondNLO2} are affected 
by ultraviolet divergences, since the respective integrands correctly reproduce 
the amplitude only in the limit of soft $k$. Physically, we expect that infrared 
singularities will be independent of the choice of the ultraviolet regulator, but still 
a specific scheme needs to be devised. A second problem is that \eq{FinCondNLO3} 
is valid as distribution identity, since the first term has support when $\ell$ is on-shell, 
while in the second term $\ell$ is off-shell, and divergences arise in the on-shell limit. 
In order to implement \eq{FinCondNLO3} locally in the Born phase-space we will 
therefore need to introduce appropriate phase-space mappings, expressing $\ell$ 
in terms of an on-shell momentum in the real-radiation contribution. A detailed 
implementation of the soft and collinear cancellations in a form useful for subtraction 
is presented in \appn{scollcanc}.

%%%%%%%%%%%%%%%%%%%%%

\subsection{Factorisation structure at NNLO}
\label{NNLOSubtra}

Following the same procedure as at NLO, one can extract the poles of the two-loop 
amplitude by expanding all factors of \eq{AmpFact} at the proper order, and then 
organising them in terms of cross-section-level soft and jet functions. The final 
result for $VV_{n}$ can be written as~\cite{Magnea:2018ebr}
\beq
\label{virtwolo}
  VV_{n} & = & VV_{n}^{(2 {\rm s})} +  VV_{n}^{(1 {\rm s})} + \sum_{i = 1}^n 
  \bigg[ VV_{n, \, i}^{(2 {\rm hc})} + \sum_{j= i+1}^n \! VV_{n, \, ij}^{(2 {\rm hc})} 
  +  VV_{n, \, i}^{(1 {\rm hc},\, 1 {\rm s})} 
  + VV_{n, \, i}^{(1 {\rm hc})} \bigg]
  + \, {\rm finite} \, , \quad \,\,\,\, \quad
\eeq
where the superscripts identify the soft or hard-collinear nature of the poles collected 
in the different terms. Explicitly,
\beq
\label{virtwolopieces}
  VV_{n}^{(2 {\rm s})} & = & {{\cal H}_n^{(0)}}^\dagger \, S_n^{(2)} \big(\{ \beta_i \} \big)
  \, {\cal H}^{(0)}_n \, , 
  \nn \\
  VV_{n}^{(1 {\rm s})} & = & {{\cal H}_n^{(0)}}^\dagger S_n^{(1)} \big(\{ \beta_i \} \big)
  {\cal H}^{(1)}_n \, + \, {{\cal H}_n^{(1)}}^\dagger S_n^{(1)} \big(\{ \beta_i \} \big) \, 
  {\cal H}^{(0)}_n \, ,  
  \nn \\
  VV_{n, \, i}^{(2 {\rm hc})} & = & {{\cal H}_n^{(0)}}^\dagger \bigg[ 
  J^{(2)}_{f_i,f_i} \big( p_i \big) - J_{\E_i}^{(2)} \big( \beta_i \big) \, - \, 
  J_{\E_i}^{(1)} \big( \beta_i \big) \, J^{(1), \, \text{hc}}_{f_i, f_i} \bigg] \, 
  S_n^{(0)} \, {\cal H}^{(0)}_n \, , 
  \nn \\
  VV_{n, \, ij}^{(2 {\rm hc})} & = & {{\cal H}_n^{(0)}}^\dagger
  J^{(1), \, \text{hc}}_{f_i, f_i} \, 
  J^{(1), \, \text{hc}}_{f_j, f_j} \, 
  S_n^{(0)} \, {\cal H}^{(0)}_n \, , 
  \\
  VV_{n, \, i}^{({1\rm hc} ,\,  {1\rm s})} & = & {{\cal H}_n^{(0)}}^\dagger 
  J^{(1), \, \text{hc}}_{f_i, f_i} \, 
  S_n^{(1)}  \big(\{ \beta_j \} \big) \, {\cal H}^{(0)}_n \, , 
  \nn \\
  VV_{n, \, i}^{(1{\rm hc})} & = & {{\cal H}_n^{(0)}}^\dagger 
  J^{(1), \, \text{hc}}_{f_i, f_i} \, S_n^{(0)} \, 
  {\cal H}^{(1)}_n \, + \, {{\cal H}_n^{(1)}}^\dagger 
  J^{(1), \, \text{hc}}_{f_i, f_i} \, S_n^{(0)} \, 
  {\cal H}^{(0)}_n \, ,  \quad \nn 
\eeq
where we slightly simplified the notation (as compared to \eq{VirtualNLO}), by performing,
where needed, the trivial integration over the total jet momenta $\ell_i$.
We also introduced a symbol for the hard-collinear one-loop non-radiative jet
\beq
\label{hccomb}
  J^{(1), \, \text{hc}}_{f_i, f_i} \, \equiv \, J^{(1)}_{f_i, f_i}(p_i) - J_{\E_i}^{(1)}(\beta_i) \, , 
\eeq
which is free of $\epsilon$ poles of soft origin\footnote{For gluon jets, note that the subtraction 
of soft poles involves a single eikonal jet for virtual corrections, as in \eq{hccomb}, while for
the real radiation of two gluons one needs to subtract two eikonal contributions, as in 
\eq{eq:hc_jet_double_real}, since both gluons can independently become soft. The 
symmetry factor $\varsi_{f_1f_2}$ multiplying the phase-space integral in \eq{FinCondNLO3} 
is then crucial to compensate for this extra factor of two, ensuring the consistency of the 
two definitions in \eq{hccomb} and in \eq{eq:hc_jet_double_real}.}. 

Similarly to the double-virtual case, the explicit infrared poles of the real-virtual correction 
will be given by
\beq
\label{realvirtpol}
  RV_\npo & = & 
  {{\cal H}_\npo^{(0)}}^{\!\! \dagger} \, S_\npo^{(1)} \big(\{ \beta_i \} \big) 
  \, {\cal H}^{(0)}_\npo
  \, + \,  \sum_{i = 1}^\npo {{\cal H}_\npo^{(0)}}^{\!\! \dagger}
  J^{(1), \, \text{hc}}_{f_i, f_i} \, 
  S_\npo^{(0)} \, {\cal H}^{(0)}_\npo
  \, + \, {\rm finite} \nonumber \\
  & \equiv & 
  RV_\npo^{({\rm s})} + \sum_{i = 1}^n RV_{\npo, \, i}^{({\rm hc})} 
  \, + \, {\rm finite} \, , 
\eeq
where, however, one must keep in mind that the finite contributions that we are not 
displaying in \eq{realvirtpol} will be affected by phase-space singularities, when the 
radiated particle becomes soft or collinear. These singular contributions, as well as those 
stemming from the other terms in $RV_{\npo}$, are subtracted by introducing the local 
real-virtual counterterm $K^{\RV}_\npo$.

Finding an explicit expression for $K^{\RV}_\npo$ and $K^{\two}_\npt$ requires 
extending the conditions given in eqs.~(\ref{FinCondNLO1}-\ref{FinCondNLO3}) 
to two loops. They read
\beq
\label{completeness2_S}
  S_n^{(2)}  \big( \{ \beta_i \} \big) 
  \, + \, 
  \int d \Phi (k_1) \, S_{n, g}^{(1)}  \big( \{ \beta_i \}; k_1 \big)  
  \hspace{4cm} && \\   
  \, + \sum_{f_1,f_2} \varsi_{f_1 f_2}
  \int d \Phi (k_1) \, d \Phi (k_2) \, S_{n, f_1 f_2}^{(0)} \big( \{ \beta_i \}; k_1, k_2 \big)
  & = & {\rm finite}
  \, ,  \nn \\
\label{completeness2_SC}
  J^{(2)}_{\E_i} \big( \beta_i \big)
  \, + \, 
  \int d \Phi (k_1) \, J_{\E_i, g}^{(1)} \big( \beta_i; k_1 \big)
  \hspace{4cm}  && \\ 
  \, + \,
  \sum_{f_1,f_2} \varsi_{f_1 f_2}
  \int d \Phi (k_1) \, d \Phi (k_2) \, J_{\E_i, f_1 f_2}^{(0)} \big( \beta_i; k_1, k_2 \big)
  & = & 
  {\rm finite} 
  \, , \nn
  \qquad \\
\label{completeness2_C}
  \sum_{f_1} \int d \Phi(k_1) \, J_{f, f_1}^{(2) \, \alpha \beta} (\ell; k_1) \, + \,
  \sum_{f_1, f_2} \varsi_{f_1 f_2} \int d \Phi(k_1) d \Phi(k_2) \,
  J_{f, f_1 f_2}^{(1) \, \alpha \beta} (\ell; k_1, k_2) 
  \hspace{1cm}
  && \\ 
  \, +
  \sum_{f_1, f_2, f_3} \varsi_{f_1 f_2 f_3} \int d \Phi (k_1) \, d \Phi (k_2) \,  d \Phi (k_3) \, 
  J_{f, f_1 f_2 f_3}^{(0) \, \alpha \beta}(\ell; k_1, k_2, k_3)
  & = & 
  {\rm finite} 
  \, , \nn
\eeq
where again the flavour sums extend to all final-state flavour combinations compatible 
with the Feynman rules, as discussed in \secn{NLOSubtra}, and the symmetry factors 
differ from unity when identical particles are emitted in the final state, as expected. As 
was the case at NLO, eqs.~(\ref{completeness2_S}-\ref{completeness2_C}) naturally 
provide tentative expressions for local counterterms at NNLO, following the logic outlined 
in Ref.~\cite{Magnea:2018ebr}.  More specifically, starting from \eq{virtwolopieces}, 
and using eqs.~(\ref{completeness2_S}-\ref{completeness2_C}), one can easily identify 
the contributions to the double-unresolved counterterm $K^{\two}_\npt$. Such contributions
live in the $(\npt)$-particle phase space, and are thus proportional to a double-radiative 
jet or soft function, or to the product of a single-radiative soft function and a single-radiative 
jet function, properly combined into hard-collinear corrections. An example of the first 
configuration is given by the third term in \eq{completeness2_S}, which reproduces 
the singularities of two {\it uniformly soft} emissions. Pursuing this line of attack, the 
double-unresolved counterterm $K^{\two}_\npt$ can be organised according to the 
soft or hard-collinear character of the radiated particles. We write
\beq
\label{OrgaDU}
  K_{n+2}^{\bf (2)} & = &
  K_{n+2}^{({\bf 2}, \, {\rm 2 s})} 
  \, + \, 
  \sum_{i = 1}^n
  \bigg[
  K_{n+2, \, i}^{({\bf 2}, \, \rm{2hc})} 
  \, + \, 
  \sum_{j = i+1}^n  K_{n+2, \, i j}^{({\bf 2}, \, \rm{2hc})} 
  \, + \,
  K_{\npt, \, i}^{({\bf 2}, \, \rm{1 hc, \, 1s})}  
  \bigg] \, ,
\eeq
where each term, evaluated at tree level, is defined in a phase space with $\npt$ 
particles, two of which will become unresolved. As suggested by the notation in 
\eq{OrgaDU}, the two radiations can be both soft (including soft-collinear), or both 
hard and collinear, in which case they can be associated to either one or two detected 
particles, or, finally, one of them can be soft while the other one is hard and collinear
to one of the Born-level particles\footnote{We note that the counterterm in \eq{OrgaDU}
is written, in the spirit of our top-down approach, for a fixed set of Born momenta,
and with a specific assignment of unresolved momenta. When the subtraction is 
implemented on the full double-real matrix element, it will be necessary to perform 
a further sum over the possible assignments of unresolved momenta in the set of
$\npt$ final-state particles, in order to account for all singular contributions. Note 
also that in Ref.~\cite{Magnea:2018ebr} the counterterms in \eq{OrgaDU} were 
written in compact form, keeping the flavour sums implicit, whereas here the flavour 
structure is given in detail, at the price of a slightly more cumbersome notation.}.
Individual contributions read
\beq
\label{CandCountNNLO1}
  K^{({\bf 2}, \, \rm{2s})}_\npt \big( \{ p_i \}, k_1, k_2 \big) \, = \,  
  {\hard_n^{(0)}}^\dagger  \sum_{f_1,f_2} S_{n, f_1 f_2}^{(0)} 
  \big( \{ \beta_i \}; k_1, k_2 \big) \,
  \hard_n^{(0)}  \, , 
\eeq
for double-soft emission,
\beq
\label{CandCountNNLO2}
  K^{({\bf 2}, \, {\rm 2hc})}_{\npt, \, i} \big( \{ p_i \}, k_1, k_2, k_3 \big)
  \, = \, 
 {\hard_n^{(0)}}^{\dagger}  \sum_{f_1, f_2,  f_3}
  \bigg[ 
  J^{(0), \, \text{hc}}_{f_i, f_1 f_2 f_3}
  - \! \sum_{jkl \in \{123,312,231\}} \!\!
  J_{\E_i, f_j}^{(0)}  \, J^{(0), \, \text{hc}}_{f_i, f_k f_l} 
  \bigg]  \, 
  S_{n}^{(0)} \, {\cal H}^{(0)}_n 
  \, , \qquad \, 
\eeq
for double hard-collinear emission from a fixed Born-level particle $i$,
\beq
\label{CandCountNNLO3}  
  K^{({\bf  2}, \, {\rm 2hc})}_{\npt, \, ij} \big( \{ p_i \}, k_1, k_2, k_3, k_4 \big)
  \, = \, 
  {\hard_n^{(0)}}^{\dagger}  \sum_{f_1,f_2,f_3,f_4}\!\!
  J^{(0), \, \text{hc}}_{f_i, f_1 f_2}  \,
  J^{(0), \, \text{hc}}_{f_j, f_3 f_4} \,
  S_{n}^{(0)} \, {\cal H}^{(0)}_n \,
 + (i \leftrightarrow j) \, ,
\eeq
for two hard-collinear emissions from two distinct Born-level particles $i$ and $j$,
and finally
\beq
\label{CandCountNNLO4}
  &&
  K^{({\bf 2}, \, {\rm 1hc, \, 1s})}_{\npt, \, i} \big( \{ p_i \}, k_1, k_2, k_3 \big)
  \, = \,  
  {\hard_n^{(0)}}^{\dagger}  \sum_{f_1, f_2, f_3}
  \sum_{jkl \in \{123,312,231\}} \,
  J^{(0), \, \text{hc}}_{f_i, f_k f_l} 
  \, S_{n, f_j}^{(0)} \, \hard_n^{(0)} \, ,
\eeq
for the joint emission of a soft particle and a hard-collinear particle. Once again, 
flavour sums extend to all final-state flavour configurations allowed by the Feynman 
rules, and we introduced the necessary sums over assignments of soft and collinear 
momenta within the sets of radiated particles; furthermore, we have introduced the 
definition
\beq
\label{eq:hc_jet_3rad}
  J^{(0), \, \text{hc}}_{f_i, f_1 f_2 f_3}
  \, \equiv \,
  J^{(0)}_{f_i, \, f_1 f_2 f_3} \, - \,  \sum_{jk \in \{12,13,23\}}
  J_{\E_i, f_j f_k}^{(0)} \, .
\eeq
The most intricate case is clearly the double hard-collinear emission from a single 
Born-level parton, given in \eq{CandCountNNLO2}. There, the first term, defined in 
\eq{eq:hc_jet_3rad} gives the full collinear double emission, with the subtraction of 
configurations where two of the three final-state particles are soft (the only allowed 
flavour combinations in this case are $\{f_j, f_k\} = \{ g, g\}$ and $\{f_j, f_k\} = \{ q, \bar{q}\}$); 
the second term subtracts configurations where only one final-state particle is soft, while 
the remaining two particles form a hard-collinear pair: in this case, the soft emission
factorises from the hard-collinear one, resulting in a single-radiative eikonal jet,
so that $f_j$ must be a gluon. We emphasise that symmetry factors for functions
involving the radiation of identical particles, for example a factor of $1/6$ for the
jet function $J_{g, ggg}$, must be included in the phase-space measure when the
counterterm is integrated, according to \eq{completeness2_C}.

We turn next to the real-virtual counterterm $K^{\RV}_\npo$. In order to construct a 
candidate counterterm in this case, we need to collect all contributions arising from
eqs.~(\ref{completeness2_S}-\ref{completeness2_C}), which are defined in a 
phase space with $n$ detected particles, involve a single extra unresolved radiation, 
and have one of the factor functions evaluated at one loop. Note that our goal here is to 
organise all phase-space singularities of the real-virtual contribution to the cross section, 
while retaining control of all explicit infrared poles arising in the loop. To this end, we write 
\beq
\label{twoloctRV}
  K_{n+1}^{\RV} 
  & = &
  K_{n+1}^{({\bf  RV}, \, {\rm s})} 
  \, + \, 
  \sum_{i = 1}^n 
  \bigg[
  K_{n+1, \, i}^{({\bf RV}, \, \rm{hc})} 
  \, + \, 
  \!\!\sum_{j =i +1}^n\! 
  K_{n+1, \, i j}^{({\bf RV}, \, \rm{hc})} 
  \, + \,
  K_{n+1, \, i}^{({\bf RV}, \, \rm{1hc , \, 1s})} 
  \, + \,
  K_{n+1, \, i}^{({\bf RV}, \, \rm{1hc})} 
  \bigg]
  \, . \quad
\eeq
Importantly, in \eq{twoloctRV} the distinction between soft and hard-collinear radiation
applies both to the single real radiation and to the loop correction. Thus, for example,
$K_{n+1}^{({\bf  RV}, \, {\rm s})}$ collects all terms that have a soft phase-space 
singularity, possibly accompanied by a pole of soft origin. Therefore we define
\beq
\label{twoloctRVs}
  K^{({\bf RV}, \, {\rm s})}_\npo 
  & = & 
  {\hard_n^{(0)}}^{\dagger} \, S_{n, \, g}^{(0)} \, {\cal H}^{(1)}_n 
  \, + \, 
  {\hard_n^{(1)}}^{\dagger} \, S_{n, \, g}^{(0)} \, {\cal H}^{(0)}_n 
  \, + \,
  {\hard_n^{(0)}}^{\dagger} \, S_{n, \, g}^{(1)} \, {\cal H}^{(0)}_n  
  \, .
\eeq
In \eq{twoloctRVs}, the first two terms have soft singularities arising from soft-gluon
emission, accompanied by finite one-loop corrections: thus, they have no explicit 
poles. The last term contains soft poles, accompanied by soft phase-space 
singularities. As customary in our approach, we include soft-collinear configurations
in the soft sector, both for virtual and for real contributions. Next, we consider 
hard-collinear configurations associated with the $i$-th external leg. They are 
given by
\beq
\label{twoloctRVhci}
  K^{({\bf RV}, \, {\rm hc})}_{\npo, \, i} 
  & = & 
  {\hard_n^{(0)}}^{\dagger}  \sum_{f_1, f_2}
  \bigg[ J^{(1), \, \text{hc}}_{f_i, \, f_1 f_2} - \!
  \sum_{kl = \{12, 21 \}}
  \Big( J_{f_i, \, f_k}^{(1)} - J^{(1)}_{\E_i} \Big) \, J^{(0)}_{\E_i, \, f_l}
  - 
  J_{\E_i }^{(1)} \, J^{(0), \, \text{hc}}_{f_i, f_1 f_2} \bigg] \, S_{n }^{(0)} \, {\cal H}^{(0)}_n \, ,
  \nonumber \\
  & = & 
  {\hard_n^{(0)}}^{\dagger}  \sum_{f_1, f_2}
  \bigg[ J^{(1), \, \text{hc}}_{f_i, \, f_1 f_2} - \!
  \sum_{kl = \{12, 21 \}}
   J^{(1), \, \text{hc}}_{f_i, \, f_k} \, J^{(0)}_{\E_i, \, f_l}
  - 
  J_{\E_i }^{(1)} \, J^{(0), \, \text{hc}}_{f_i, f_1 f_2} \bigg] \, S_{n }^{(0)} \, {\cal H}^{(0)}_n \, ,
  \qquad
\eeq
where the first term is defined in \eq{eq:hc_jet_double_real}, and we used the fact 
that $J_{f_i, \, f_k}^{(1)}$ carries momentum $k_k = k_i$, and is flavour-diagonal, 
$f_k=f_i$. In the second line of \eq{twoloctRVhci}, the first term in brackets contains 
the one-loop contributions to all relevant collinear splitting kernels (with a single unresolved 
radiation). This term is in fact affected by both collinear phase-space singularities and 
collinear poles, while it is free from singularities and poles of soft origin. The second 
term subtracts all hard-collinear one-loop virtual poles that are accompanied by 
soft-collinear phase-space singularities; finally, the third term subtracts hard-collinear 
phase-space singularities that are accompanied by soft-collinear one-loop poles. 

Next, we need to consider the case in which collinear poles and collinear phase-space 
singularities are associated with different external legs: these contributions are given by
\beq
\label{twoloctRV1hcij}
  K^{({\bf RV}, \, {\rm hc})}_{\npo, \, ij} 
  & = & 
  {\hard_n^{(0)}}^{\dagger}  
  J^{(1), \, \text{hc}}_{f_i, f_i}
  \sum_{f_1, f_2}
  J^{(0), \, \text{hc}}_{f_j, f_1 f_2} \,  S_{n}^{(0)} \, {\cal H}^{(0)}_n 
  \, + (i\leftrightarrow j) \, . \qquad  \,\,\,
\eeq
A further contribution accounts for soft phase-space singularities accompanied by 
hard-collinear virtual poles on leg $i$, as well as hard-collinear phase-space 
singularities in emissions from leg $i$, accompanied by soft virtual poles. It is
\beq
\label{twoloctRV1hc1s}
  K^{({\bf RV}, \, {\rm 1hc, 1s})}_{\npo, \, i} 
  & = & 
  {\hard_n^{(0)}}^{\dagger}
  \sum_{f_1,f_2} \bigg[
  \sum_{kl = \{12, 21 \}} J_{f_k, \, f_k}^{(1), \, \text{hc}} \,   
  S_{n, \, f_l}^{(0)} 
  \, + \,
  J^{(0), \, \text{hc}}_{f_i, f_1 f_2} \, 
  S_{n}^{(1)}
  \bigg] \, {\cal H}^{(0)}_n \, , 
\eeq
where we used again the fact that the virtual jet is flavour-diagonal, so that
$J_{f_k, \, f_k}^{(1), \, \text{hc}} = J_{f_i, \, f_k}^{(1), \, \text{hc}}$. 

Finally, one needs to account for hard-collinear phase-space singularities due to 
emissions from leg $i$, accompanied by the finite part of one-loop corrections.
These give
\beq
\label{twoloctRV1hc}
  K^{({\bf RV}, \, {\rm 1hc})}_{\npo, \, i} 
  & = &
  {\hard_n^{(0)}}^{\dagger}  \sum_{f_1, f_2}
  J^{(0), \, \text{hc}}_{f_i, f_1 f_2} \, 
  S_{n}^{(0)} \, {\cal H}^{(1)}_n \,
  + \, {\hard_n^{(1)}}^{\dagger}  \sum_{f_1, f_2}
  J^{(0), \, \text{hc}}_{f_i, f_1 f_2} \, S_{n}^{(0)} \, {\cal H}^{(0)}_n \, .
\eeq
The expressions given in eqs.~(\ref{twoloctRVs}-\ref{twoloctRV1hc}) reproduce all
the phase-space singularities of \eq{realvirtpol}, including those that are not accompanied 
by virtual poles. Note that only \eq{twoloctRVs} and \eq{twoloctRV1hc} contain
non-universal one-loop ingredients, which must of course occur at this stage at 
NNLO. All other contributions are given by universal soft and collinear functions.

To conclude our discussion, the simplest required ingredient at NNLO is the single-unresolved 
counterterm in the $(\npt)$-particle phase space, $K^{\one}_\npt$. This has precisely the 
same form as its NLO counterpart, given in \eq{CandCountNLO1} and in \eq{CandCountNLO2}, 
with the replacement $n \to (\npo)$. Our next goal is therefore to study the double-radiative 
local counterterms introduced in eqs.~(\ref{CandCountNNLO1}-\ref{CandCountNNLO4}), 
in the hierarchical limit in which one of the two radiated particles becomes unresolved 
at a faster rate with respect to the second one. We expect, and verify below, that in 
these limits radiative functions {\it refactorise}, and this feature allows for an easy identification 
of the  remaining local counterterm required for NNLO subtraction, namely $K^{\otwo}_\npt$. 
 
%%%%%%%%%%%%%%%%%%%%%%%%%%%%%%%%%%%%%%%

\section{Factorisation of radiative functions in strongly-ordered limits at tree level}
\label{ReFactoTree}

It is clear from our discussion in \secn{Archi} that strongly-ordered soft and 
collinear limits play an increasingly important role for subtraction at higher 
perturbative orders. For practical purposes, the construction of strongly-ordered 
counterterms, starting from the corresponding `unordered' ones, is not difficult: 
one simply needs to take suitable scaling limits on subsets of soft and collinear 
momenta. It is, however, very interesting to study these limits from an operator
point of view, expressing strongly-ordered counterterms in terms of operator 
matrix elements related to those in eqs.~(\ref{RadSoftFunc}-\ref{EikRadJetFunc}).
This would be of significant help when attempting to prove the line-by-line finiteness
of subtracted distributions, by means of completeness relations and power-counting
arguments. Furthermore, a discussion of the factorisation properties of Wilson-line 
matrix elements including radiation is of intrinsic interest, and indeed issues of
`refactorisation' of soft and collinear cross sections have already arisen
in the context of resummation for inclusive rates, for example in
Ref.~\cite{Laenen:2000ij}. In this section, we  will present a discussion of the 
factorisation properties of radiative functions, deriving general results
at tree level.

To begin our discussion, we note that the jet and soft functions defined in 
eqs.~(\ref{RadSoftFunc}-\ref{EikRadJetFunc}) reproduce the relevant multiple singular 
configurations in the absence of any hierarchy among unresolved partons. Thus, for 
example, at NNLO the counterterms derived from these functions are naturally identified 
as contributions to the unordered counterterm $K_{n+2}^{\bf (2)}$. A procedure is then 
necessary in order to extract the strongly-ordered configurations entering $K_{n+2}^{\bf (12)}$, 
and similarly for higher-order counterterms. 

%%%%%%%%%%%%%%%%%%%%%

\subsection{Tree-level radiative soft functions}
\label{RasofuTree}

Consider, as a first example, the case of double-soft radiation at tree level. In the limit in 
which one of the two radiated soft gluons is much softer than the other, say $k_2 \ll k_1$, 
the strongly-ordered double-soft current is given by \cite{Catani:1999ss}
\beq
  \left[ J_{\CG}^{(0), \, \rm s.o.} \right]_{\mu_1 \mu_2}^{a_1 a_2} 
  \left( \{\beta_i\}; k_1, k_2 \right) \, = \, 
  \left(J^{(0) \, a_2}_{\mu_2}(k_2) \, \delta^{a_1 a}+ {\rm i} g_s \mu^\eps\, 
  f^{a_1 a_2 a} \, \frac{k_{1 , \, \mu_2}}{k_1\cdot k_2} \right)
  J^{(0)}_{\mu_1, \, a} (k_1) \, ,
\label{2sCGso}
\eeq
where
\beq 
  J_\mu^{(0) \, a} (k) \, = \, g_s\mu^\eps \, \sum_{i = 1}^n \,
  \frac{\beta_{i,\, \mu}}{\beta_i \cdot k} \, \mathbf{T}^a_i  \, 
\label{treesoftcurexp}
\eeq
is the single-gluon tree-level soft current. Notice that it is a colour matrix, so 
that the ordering in \eq{2sCGso} is fixed. This expression can of course be obtained from 
factorisation by considering the tree-level double-radiative soft function ${\cal S}^{(0)}_{n, \, gg} 
\left(\{\beta_i\};k_1, \, k_2 \right)$, stripping off the two gluon polarisation vectors, rescaling 
$k_2$ by a factor $\xi_2$, and retaining only the leading power in the limit  $\xi_2 \to 0$. 
As discussed above, however, it is desirable to give a definition of strongly-ordered soft 
operators without resorting to an {\it a posteriori} limit operation on unordered configurations: 
this can be achieved by applying soft factorisation in an iterative fashion.\footnote{We emphasise 
that strongly-ordered soft emission to leading IR accuracy was already investigated 
in Ref.~\cite{Catani:1984dp,Catani:1985xt}, where both real and virtual emission contributions 
are discussed.}

\begin{figure}
  \centering
  {\includegraphics[scale=0.2]{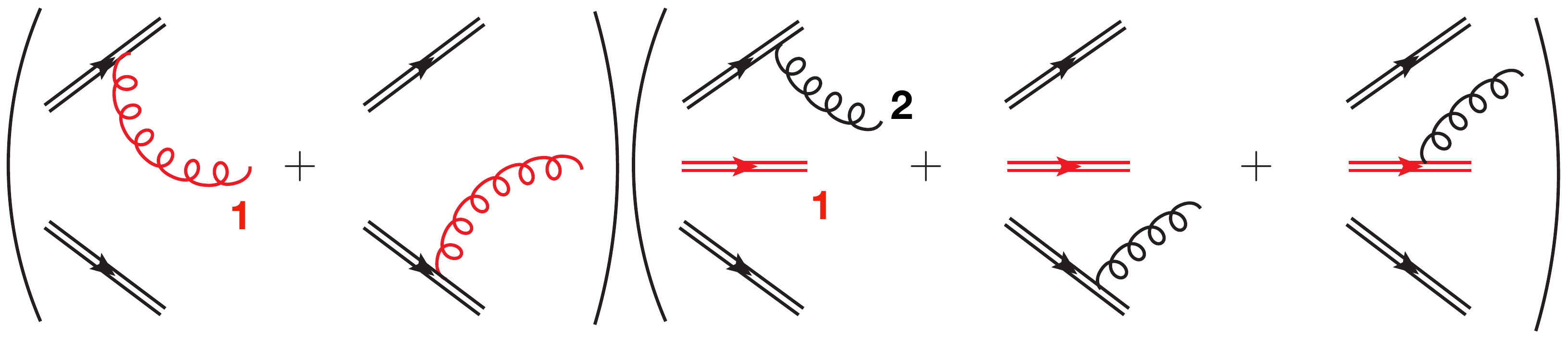}}
  \caption{The factorisation of an eikonal form factor, illustrated for the case of two 
  hard lines and strongly-ordered double-soft radiation.}
  \label{wilsonisation}
\end{figure}

The key idea is that, in the limit $k_2 \ll k_1 \ll \mu$, with $\mu$ a typical hard 
scale of the process, gluon 1 (corresponding to momentum $k_1$) is soft with 
respect to the $n$ hard Born partons, but, in turn, is seen as a hard parton if 
probed by gluon 2 (with momentum $k_2$). This implies that the soft emission 
of gluon 1 is described by a soft current featuring $n$ Wilson lines, corresponding 
to the Born partons, while the emission of gluon 2 is represented by a soft current 
featuring $n+1$ Wilson lines: one of these (in the adjoint representation) corresponds 
to gluon 1. In the language of factorisation, we may apply the standard techniques 
of the soft approximation to the matrix element ${\cal S}^{(0)}_{n, \, gg} \left(\beta_i\,;
k_1, \,  k_2 \right)$, in the limit when gluon 2 is softer than all other particles. 
In this limit, such a matrix element factorises, and the resulting soft function
is, as expected, a Wilson-line correlator, where gluon 2 is still treated as a
final-state parton, while gluon 1 has been replaced by a Wilson line in the adjoint 
representation\footnote{One could perhaps describe this factorisation by saying
that the harder gluon has `Wilsonised', becoming a classical source for all much
softer radiation.}. This factorisation leaves behind a `hard' function which, consistently, 
is given by the Wilson line correlator for the radiation of  gluon 1 from the
$n$ hard Born-level partons. This tree-level factorisation is represented pictorially
in Fig.~\ref{wilsonisation}, in the simplified case with $n=2$.

In formulae, the natural definition for a strongly-ordered tree-level double-soft 
radiative function, which we denote by ${\cal S}^{(0)}_{n; \, g, g}$, is thus
\beq
\label{softrads.o.}
  \Big[{\cal S}^{(0)}_{n; \, g, g} \Big]^{a_1 a_2}_{\{d_i e_i\}}
  \big(\{\beta_i\};k_1, k_2) & \equiv &  
  \bra{k_2, a_2} \, 
  %%%%% 
  T \bigg[
  \Phi^{\,\, a_1 b}_{\beta_{k_1}} (0,\infty) 
  \prod_{i = 1}^{n} \Phi^{\quad \,\,\,\,\, c_i }_{\beta_{i}, \, d_i}(\infty,0) \bigg] \, \ket{0} \nn\\
  & & \hspace{1cm}
  \times \,
  \bra{k_1, b} \,  
    T \bigg[
    \prod_{i=1}^{n}
  \Phi_{\beta_i, \, c_i e_i}(\infty,0)
  \bigg]  \ket{0} \Big|_{\rm tree} \nn \\
  & = & \left[ {\cal S}^{(0)}_{n+1, \, g} \right]^{a_2, \, a_1 b}_{\{d_i c_i\}}
  \left(\beta_{k_1}, \{\beta_{i}\}; k_2 \right) \, 
  \left[ {\cal S}^{(0)}_{n, \, g} \right]_{b, \, \{ c_i e_i \}} \! \left(\{\beta_{i}\};k_1 \right)\, ,
\eeq
where, in the first line, one recognises the factorised emission amplitude of 
gluon 2 off $n+1$ Wilson lines, multiplied times the radiation of gluon 1 
off $n$ Born-level Wilson lines (the second line). For clarity, in \eq{softrads.o.}
we have written explicitly all colour indices: one sees that the first eikonal 
form factor, involving $n+1$ Wilson lines, acts as a colour operator on the 
second form factor, with the adjoint index $b$ of the Wilson line associated 
with gluon 1 contracting with the index of the gluon state in the second line.
It is straightforward to verify that Eq. \eqref{softrads.o.} yields precisely
\beq
  \Big[{\cal S}^{(0)}_{n; \, g, g} \Big]^{a_1 a_2}
  \left(\{\beta_{i}\};k_1, k_2 \right)=
  \epsilon^{* \, \mu_1} (k_1) \;
  \epsilon^{* \,  \mu_2} (k_2) \,
  \left[ J_{\CG}^{(0), \, \rm s.o.} \right]_{\mu_1 \mu_2}^{a_1 a_2} 
  \left(\{\beta_i\} ; k_1, k_2 \right) \, .
\label{checksoftradso}
\eeq
The strongly-ordered double-soft counterterm $K^{({\bf 12}, \,  \rm s)}_\npt$ is then
obtained by squaring \eq{softrads.o.}, in the spirit of \eq{CandCountNLO1}.

By iterating this factorisation procedure, one may define the strongly-ordered 
triple-soft current, in the kinematic limit $k_3 \ll k_2 \ll k_1 \ll \mu$, as
\beq
  \left[ {\cal S}_{n; \, g,g,g}^{(0)} \right]^{a_1a_2a_3}_{\{ f_i e_i \}}
  \left(\{\beta_{i} \} ;k_1, k_2, k_3 \right) & \equiv &
  \left[ {\cal S}^{(0)}_{n+2, \, g} \right]^{a_3}_{\{ f_i d_i \}, \, a_1 b_1, \, a_2 b_2}
  \left[ {\cal S}^{(0)}_{n+1, \, g} \right]^{b_2}_{\{ d_i c_i \}, \, b_1 g_1}
  \left[ {\cal S}^{(0)}_{n, \, g} \right]^{g_1}_{\{ c_i e_i \}} \nn \\ 
  & = & \langle k_3, a_3 | \,
  T \bigg[
  \Phi_{\beta_{k_1}}^{a_1 b_1}(\infty,0) \,
  \Phi_{\beta_{k_2}}^{a_2b_2} (\infty,0) 
  \prod_{i = 1}^n \Phi_{\beta_{i}}^{f_i d_i} (\infty,0) \bigg] | 0 \rangle \nn \\ 
  & & \times \, \langle k_2, b_2 | \,
  T \bigg[
  \Phi_{\beta_{k_1}}^{b_1g_1}(0, \infty) 
  \prod_{i=1}^n \Phi_{\beta_{i}}^{d_ic_i}( \infty,0)  \bigg]
  | 0 \rangle \nn \\
  && \times \, \langle k_1,g_1 | T \bigg[ \prod_{i=1}^n
  \Phi_{\beta_{i}}^{c_ie_i}(\infty,0) \bigg] | 0 \rangle
  \Big|_{\rm tree} \, ,
\label{softs.o.3}
\eeq
where, on top of the double radiation already detailed in \eq{softrads.o.}, we 
recognise in second line the emission of the softest gluon 3 (with momentum $k_3$) 
from a set of $n+2$ Wilson lines, two of which are `Wilsonised' versions of gluons 1 
and 2. \Eq{softs.o.3} yields a natural generalisation to the case of triple radiation of 
the strongly-ordered two-gluon eikonal form factor given in \eq{checksoftradso},
\beq
\label{3soft}
  \left[{\cal S}_{n; \, g, g, g}^{(0)} \right]^{a_1 a_2 a_3} & = &
  \epsilon_{\mu_3}^* (k_3) \,
  \epsilon_{\mu_2}^* (k_2) \,
  \epsilon_{\mu_1}^* (k_1) \nn \\
  & & \times \,
  \Bigg[
  J^{\mu_3}_{a_3}(k_3) \, \delta^{a_1 b_1} \, \delta^{a_2 b_2}
  + \, {\rm i} g_s \mu^\eps \, f^{a_1 a_3 b_1} \, \delta^{a_2 b_2} \,
  \frac{k_1^{\mu_3}}{k_1\cdot k_3}
  + \, {\rm i} g_s\mu^\eps \, f^{a_2 a_3 b_2} \, \delta^{a_1 b_1} \, 
  \frac{k_2^{\mu_3}}{k_2\cdot k_3}
  \Bigg] \nn \\
  & & \times \,
  \Bigg[
  J^{\mu_2}_{b_2} (k_2) \, \delta^{b_1 c_1}
  + \, {\rm i} g_s\mu^\eps \, f^{b_1 b_2 c_1} \, \frac{k_1^{\mu_2}}{k_1\cdot k_2}
  \Bigg] J_{c_1}^{\mu_1}(k_1) \, ,
\eeq
in complete agreement with the strongly-ordered limit of the triple-soft current 
presented in Ref.~\cite{Catani:2019nqv}. Based on the above physically motivated 
discussion, and on the explicit form of the strongly-ordered currents for up to three 
soft radiations at tree-level, it is natural to build an ansatz for the  strongly-ordered 
eikonal form factor for the radiation of $m$ gluons with momenta $k_m \ll k_{m-1} 
\ll \dots \ll k_1\ll \mu$. We write
\beq
\label{nsoft}
  \left[ {\cal S}_{n; \, g, \ldots, g}^{(0)} \right]^{a_{1,1} \dots \,
  a_{1,m}}_{\{ b_{1, \ell} \, b_{m+1, \ell }\} } \hspace{-6pt}
  & \equiv &
  \prod_{i = 1}^m
  \bra{k_{m - i + 1} , a_{i, \, m-i+1}}
  T \bigg[
  \prod_{p = 1}^{m-i} \Phi_{\beta_{k_p}}^{a_{i, p} \, a_{i+1, p}} (\infty,0)
  \prod_{\ell = 1}^n \Phi_{\beta_\ell}^{b_{i, \ell} \, b_{i+1, \ell} } (\infty,0)
  \bigg]
  \ket{0} \Big|_{\rm tree} \nn \\
  & = & \prod_{i = 1}^m 
  \left[ {\cal S}^{(0)}_{n+m-i, \, g} \right]^{a_{i, m-i+1}}_{\{ b_{i, \ell} \, b_{i+1, \ell } \}, \, 
  a_{i, 1} \, a_{i+1, 1}, \, \dots, \,  a_{i, m-i} \, a_{i+1, m-i}} \nn \\
  & = & \prod_{i = 1}^m \epsilon_{\mu_{m-i+1}}^* (k_{m-i+1})
  \Bigg[ J^{\, \mu_{m-i+1}}_{a_{i, m-i+1}} (k_{m-i+1})
  \prod_{p = 1}^{m-i} \delta^{\, a_{i, p} \, a_{i+1, p}} \nn \\
 && \qquad  \quad + \,\, {\rm i} g_s \mu^\eps\sum_{r = 1}^{m - i}
  \frac{k_r^{\, \mu_{m-i+1}}}{k_r \cdot k_{m-i+1}} \, 
  f^{\, a_{i, r} \, a_{i, m-i+1} \, a_{i+1, r}} \, 
  \prod_{\substack{j = 1 \\ j \neq r}}^{m-i} \delta^{\, a_{i, j} \, a_{i+1, j}} \Bigg] \, ,
\eeq
which reduces to \eq{softrads.o.} and \eq{softs.o.3} for $ m= 2$ and $m = 3$, 
respectively. We point out that the factorisation arguments presented for the totally 
ordered configuration $k_m\ll k_{m-1}\ll \dots \ll k_1 \ll \mu$ generalise to less
hierarchical kinematics, in which for instance $q$ among the radiated gluons have
comparable softness: in such a case, the corresponding sequence of $q$ single-radiative 
form factors is replaced by a single form factor radiating $q$ gluons.

While the analysis above has focused on tree-level soft functions, the underlying
physics, and the fact that the discussion can be phrased in terms of operator 
matrix elements, strongly suggests that the structure of the proposed factorisations can be
extended to higher orders, upon retaining hard-collinear and finite loop contributions, 
as well as including renormalisation factors. Indeed, the fact that the softest radiation 
can be factorised from the remaining (harder) ones, in terms of a soft function, where 
harder radiated gluons have `Wilsonised', follows from the standard rules of the soft 
approximation, and Ward identities could be applied, where appropriate, since the 
original eikonal form factor is gauge invariant. These qualitative arguments can be 
verified at one-loop by examining the strongly-ordered limit of the one-loop current for 
the radiation of two soft gluons~\cite{Zhu:2020ftr,Catani:2021kcy,Czakon:2022dwk}, 
and indeed one recovers a natural generalisation of \eq{softrads.o.}, which we 
briefly discuss below in \secn{ReFactoLoop}. Notwithstanding this strong check, 
a proper generalisation to all orders of a nested factorisation such as \eq{nsoft} 
would require a thorough analysis, which is left for future work.

%%%%%%%%%%%%%%%%%%%%%

\subsection{Tree-level radiative jet functions}
\label{RaJeTree}

Strongly-ordered collinear configurations need to be analysed next. Let us consider, 
for instance, the triple-collinear configuration corresponding to a kinematic situation 
in which three partons $i$, $j$, $k$ become collinear, with relative angles $\theta_{ij}, 
\theta_{ik}, \theta_{jk} \to 0$, but with two partons displaying a dominant collinearity, 
say $\theta_{ij} \ll \theta_{ik}, \theta_{jk}$. It is known that the strongly-ordered collinear 
limit of squared scattering amplitudes factorises into products of Altarelli-Parisi kernels, which 
are matrices in spin space. For instance, the NNLO strongly-ordered collinear configuration 
for a $q \, \to \, q_1' \, \bar q'_2 \, q_3$ branching is given by
\beq
\label{FGs.o.}
  \lim_{\theta_{12} \ll \theta_{13} \to 0} RR_{\npt}
  \, = \, 
  \frac{(8 \pi \as)^{\, 2}}{s_{12} \, s_{[12]3}} \,
  \, P^{\rho\sigma}_{q \to gq} \big( z_{[12]}, q_\bot \big) \, 
  d_{\rho \mu} \big( k_{[12]} \big) \, 
  P^{\mu \nu}_{g \to q \bar q} \left(\frac{z_1}{z_{[12]}}, k_\bot \right) \, 
  d_{\sigma\nu} \big( k_{[12]} \big) \, B_n \, , \quad
\eeq
where the intermediate-particle momentum is $k_{[12]} \equiv  k_1 + k_2$, its collinear 
energy fraction is $z_{[12]} \equiv z_1 + z_2 = 1 - z_3$, and $s_{[12]3} = 2 \, k_{[12]} 
\cdot k_3$. Finally,
\beq
\label{polsum}
  d_{\mu \nu} \big( k_{[12]} \big) \, = \, - g_{\mu \nu} + \frac{k_{[12] \mu} 
  n_\nu + k_{[12] \nu} n_\mu}{k_{[12]} \cdot n} \, ,
\eeq 
with $n^2 = 0$, represents the gluon polarisation sum. The momenta $q_\perp$ 
and $k_\bot$ in the splitting kernels specify the transverse directions for the 
successive branchings, $q \rightarrow g_{[12]} q_3$ and $g_{[12]} \rightarrow q'_1 
\bar q'_2$, respectively: their definitions follow from the Sudakov parametrisation
of momenta $k_i$ ($i = 1, 2, 3$), according to
\beq
  k_i^\mu \, = \, z_i \, p^\mu + k_{\bot i} - \frac{k_{\perp  i}^2}{z_i} 
  \frac{n^\mu}{2 p \cdot n} \, , \qquad \quad
  q_\bot \, = \, k_{\bot 3} \, , \qquad \quad 
  k_\bot \, = \, z_2 k_{\bot 1} - z_1 k_{\bot 2} \, .
\label{sudakov}
\eeq
The kernel $P^{\rho\sigma}_{q\to gq}$ describes the first splitting of a parent quark 
into a quark-gluon pair, with gluon spin indices un-contracted. As such, it represents 
the spin matrix acting on the subsequent splitting of the virtual gluon in a quark-antiquark 
pair, described by $P_{g\to q\bar q}^{\mu\nu}$. The explicit form of the relevant kernels is
\beq
\label{expliker}
  P^{\mu\nu}_{g\to q\bar q}(z, k)
  \, = \,
  T_R
  \left( - g^{\mu \nu} + 4 z (1 - z) \frac{k^\mu k^\nu}{k^2} \right)
  \, ,
  \qquad
  P^{\rho\sigma}_{q\to gq}(z, k)
  \, = \,
  \frac{C_F}{2 \, T_R} \, z \, P^{\rho\sigma}_{g\to q \bar q} (1/z, k)
  \, .
\eeq
Our goal is now to express such collinear refactorisations by means of radiative
jet functions. This can be done without any loss of information, since the jet functions 
defined in \eq{RadJetFuncs} retain full dependence on gluon (as well as quark) spin.
As a first example, the strongly-ordered triple-collinear kernel in \eq{FGs.o.} can be 
rewritten in the factorisation language as
\beq
\label{eq:coll_str_ord_example1}
  \int \frac{d^d \ell}{(2\pi)^d} \,
  \left[ \lim_{\theta_{12} \ll \theta_{13} \to 0}
  J_{q,qq'\bar q'}^{(0)}(\ell;k_1,k_2,k_3) \right]
  & \equiv &
  \int \frac{d^d \ell}{(2\pi)^d} \,\,
  J_{q,gq;g,q'\bar q'}^{(0)} (\ell;k_1,k_2,k_3) \\
  & = & \int \frac{d^d \ell}{(2\pi)^d} \,\,
  J^{\, ; \rho\sigma \, (0)}_{q,gq}\big( \ell; k_{[12]}, k_3 \big) \, 
  \int \frac{d^d \ell'}{(2\pi)^d} \,\,
  J^{\rho\sigma \, (0)}_{g, q' \bar q'}(\ell';k_1,k_2) \, , \nnb
\eeq
where the first line sets up the notation for a strongly-ordered jet function, specifying
the sequential splittings involved. The integrals over $d^d\ell$ and $d^d\ell'$ just 
serve the purpose of resolving the Dirac delta-function constraints contained in the 
definition of jet functions, which in turn fix the momentum of the splitting parton to 
equal the momentum sum of its decay products. Note also that in $J^{\, ; \rho\sigma 
\, (0)}_{q,gq}$ we have left implicit the spin indices associated with the parent quark, to 
lighten the notation. The semicolon serves as a marker for this implicit dependence, 
and Lorentz spin indices after semicolon are associated to the daughter gluon created 
by the splitting. Conversely, in the secondary branching described by $J^{\rho\sigma \, 
(0)}_{g, q' \bar q'}$ the Lorentz indices are related to the splitting gluon. It is not difficult 
to verify that, upon retaining the leading-power contribution in the transverse momenta, the 
radiative quark jet $J^{\, ; \rho\sigma \, (0)}_{q,gq}$ yields the $P^{\rho\sigma}_{q \to gq}$ 
kernel of \eqref{FGs.o.}, while the factor $J^{\rho \sigma \, (0)}_{g, q' \bar q'}$ reproduces 
the product $d^\rho_{\mu} \, P^{\mu \nu}_{g\to q \bar q} \, d^\sigma_{\nu}$.

The factorisation of other flavour combinations in strongly-ordered limits follows 
the same lines, upon keeping proper track of the relevant Dirac or Lorentz indices. 
As an example, we consider the case of the strongly-ordered splitting $q \to qgg$,
which we display in the abelian limit for simplicity (non-abelian terms follow the 
same kinematic pattern). This limit is described by the factorised formula
\beq
\label{eq:coll_str_ord_example2}
  \int \frac{d^d \ell}{(2\pi)^d} \,
  \left[
  \lim_{\theta_{12} \ll \theta_{13} \to 0}
  J^{(\rm ab) \, (0)}_{q,qgg}(\ell;k_1,k_2,k_3)
  \right]
  & \equiv &
  \int \frac{d^d \ell}{(2\pi)^d} \,\,
  J_{q,qg;q,qg}^{(0)}(\ell;k_1,k_2,k_3) \\
  & = &
  \int \frac{d^d \ell}{(2\pi)^d} \,
  J^{\, ; \alpha \beta \, (0)}_{q,qg} \big( \ell; k_{[12]}, k_3 \big) \, 
  \int \frac{d^d \ell'}{(2\pi)^d} \,
  J^{\alpha\beta \, (0)}_{q,qg}(\ell';k_1,k_2) , \nnb
\eeq
where $\alpha$ and $\beta$ are now Dirac indices. The only case featuring a further 
complication involves a gluon splitting into gluons: in fact, this requires keeping track 
of the Lorentz indices of both the parent and the sibling gluons. For instance we find
\beq
\label{eq:coll_str_ord_example3}
  \int \frac{d^d \ell}{(2 \pi)^d} 
  \left[ \lim_{\theta_{12} \ll \theta_{13} \to 0} 
  J^{\mu\nu \, (0)}_{g,ggg} (\ell; k_1,k_2,k_3 ) \right]   
  & \equiv &
  \int \frac{d^d \ell}{(2\pi)^d} \,\,
  J_{g,gg;g,gg}^{\mu\nu \, (0)} (\ell;k_1,k_2,k_3) \\
  & = &
  \int \frac{d^d \ell}{(2 \pi)^d} \,
  J^{\, \mu \nu; \rho \sigma \, (0)}_{g,gg} \big( \ell; k_{[12]}, k_3 \big) \, 
  \int \frac{d^d \ell'}{(2 \pi)^d} \,
  J^{\rho \sigma \, (0)}_{g, gg} (\ell';k_1,k_2) , \nnb
\eeq
where the four-index jet $J^{\mu\nu;\rho\sigma \, (0)}_{g,gg}$ is defined by
\beq
\label{jetfourind}
  J_{g, gg}^{\, \mu \nu; \rho \sigma \, (0)} \big( \ell; k_a, k_b \big)
  \, = \,
  \sum_{\lambda_a, \lambda_b} \,
  \int d^d x \, {\rm e}^{ {\rm i} \ell \cdot x} \, 
 \left( \mathcal{J}_{g,gg}^{\mu;\rho \, (0)} \big(0; \{k_j, \lambda_j \} \big) \right)^\dag \,
  \mathcal{J}_{g,gg}^{\nu;\sigma \, (0)} \big(x; \{k_j ,\lambda_j \} \big) \, ,
\eeq
and $\mathcal{J}_{g,gg}^{\nu;\sigma \, (0)}$ satisfies
\beq
\label{jetfourind2}
  \mathcal{J}_{g,gg}^{\nu \, (0)}\big(x; k_a, k_b; \lambda_a, \lambda_b \big)
  \, = \,
  \mathcal{J}_{g,gg}^{\nu;\sigma \, (0)} \big(x; k_a, k_b; \lambda_a, \lambda_b \big)
  \, \eps^{*(\lambda_a)}_\sigma(k_a)
  \, .
\eeq
We emphasise that eqs.~(\ref{eq:coll_str_ord_example1})-(\ref{eq:coll_str_ord_example3})
have been found by explicitly computing the limits of the two-radiative jet functions.

This factorised structure can be generalised to the case of an arbitrary number of 
collinear emissions, all strongly ordered. Introducing appropriate notation, such
limits can be described by the expression
\beq
\label{genfactcollso}
  &&
  \int \frac{d^d \ell}{(2\pi)^d} \,
  \left[
  \lim_{\theta_{1} \ll \theta_{2} \ll \cdots \ll \theta_{m-1} \to 0}
  J^{IJ \, (0)}_{f, f_1\ldots f_m} \big( \ell;k_1,\ldots,k_m \big)
  \right]
  \nnb \\
  && \hspace{1cm}
  \equiv \, \, 
  \int \frac{d^d \ell}{(2\pi)^d} \,
  J^{IJ \, (0)}_{f, a_1b_1, \ldots, a_m b_m} \big( \ell;k_1,\ldots,k_m \big)
  \nnb \\
  && \hspace{1cm} 
  = \, \, 
  \prod_{j=1}^{m-1}
  \int \frac{d^d \ell_j}{(2\pi)^d} \,
  J^{I_{p_j}J_{p_j};I_{a_j}J_{a_j} \, (0)}_{p_j,a_jb_j}(\ell_j;k_{a_j},k_{b_j}) \, .
\eeq
The labels in the previous equation are constructed as follows. Partons $a_1,b_1\in \{1,\dots,m\}$,
stemming from the splitting of parent particle $p_1$, are those emitted with the smallest relative 
angle among all, $\theta_1$. If the next-to-smallest independent relative angle, $\theta_2$, is the
one connecting partons $c, d\in \{1,\dots,\slashed{a}_1,\dots,\slashed{b}_1,\dots,m\}$, then 
$a_2=c$ and $b_2=d$. Otherwise, if $\theta_2$ connects $c\in \{1,\dots,\slashed{a}_1,\dots,
\slashed{b}_1,\dots,m\}$ with either $a_1$ or $b_1$, then $a_2=c$ and $b_2=[a_1b_1]=p_1$. 
Iteratively, proceeding by larger and larger relative angles $\theta_j$, with $j\leq m-1$, $a_j$ is
assigned in the set $\{1,\dots,m\}$ deprived of all $a_k, b_k$ with $k<j$; if $\theta_j$ connects 
$a_j$ to a yet unassigned parton, the latter then gets labelled as $b_j$, otherwise $b_j$ is 
labelled as the ancestor of the cluster of partons to which $a_j$ is linked by $\theta_j$. 
Parent $p_j$ is $[a_jb_j]$, understanding the iterative rule $[a[bc]] = [abc]$ and so on. 
Finally, indices $I_{p_j} J_{p_j}$ ($I_{a_j}J_{a_j}$) are the Lorentz or Dirac indices of the 
$j$-th parent (first sibling), with the constraint $I_{p_1} J_{p_1} = IJ$. Once again, situations 
in which not all collinear splittings are strongly ordered, but rather successive clusters of $k>2$ 
particles are produced with parametrically similar collinearities, can be described by analogous 
factorisations. This is not difficult to achieve case by case, while writing down a general formula 
for this would involve  rather cumbersome notations.

%%%%%%%%%%%%%%%%%%%%%%%%%%%%%%%%%%%%%%%

\section{Factorisation of single-radiative functions at one loop}
\label{ReFactoLoop}

Extending the discussion of \secn{ReFactoTree} to loop level is non-trivial. 
Virtual corrections involve loop momenta which are unconstrained: thus they 
may, and do, carry soft, collinear and ultraviolet enhancements, which will
eventually survive strongly-ordered limits. These enhancements will have 
to be properly identified and analysed, as they will impact the connection 
between strongly-ordered and real-virtual counterterms. Our strategy will 
be to treat the matrix elements defining radiative soft and jet functions as 
generalised scattering amplitudes: we will then conjecture natural expressions 
for their factorisations, and verify that they hold at one loop. 

%%%%%%%%%%%%%%%%%%%%%

\subsection{Single-radiative soft function at one loop}
\label{RaSoFuLoop}

We consider first the single-radiative eikonal form factor defined by \eq{eikFF} with $m=1$. 
This amplitude-level radiative soft function is in fact, on its own accord, a scattering amplitude 
in the presence of Wilson lines acting as sources. As such, the following factorisation ansatz, 
along the lines set in \eq{AmpFact}, is expected to hold:\footnote{The jet functions, defined 
in eqs.(\ref{collFFq}-\ref{collFFg}), can be evaluated at $x=0$ in this case, without loss of 
generality.}
\beq
\label{eq:soft_fact_nlo}
  \s_{n,g} \big( \{\beta_i\}; k \big)
  \, = \,
  \s_{n+1} \big( \{\beta_i\} , \beta_k) \,
  \frac{\J^{\mu}_{g,g}(0;k)}{\J_{\E_g}(\beta_k)} \,
  \s_{n,g}^{\mathcal{H},\mu} \big( \{\beta_i\}; k \big) \, .
\eeq
Note that, at variance with previous sections, in \eq{eq:soft_fact_nlo} we 
have dropped the dependence on the polarisation of the emitted gluon, since it 
is not relevant for the present discussion. This dependence is encoded on the 
{\it l.h.s.} in the definition of $\s_{n,g}$, and on the {\it r.h.s.} in the definition of 
$\J^{\al}_{g,g}$. The first factor in \eq{eq:soft_fact_nlo} contains the virtual soft poles 
of an $(\npo)$-point amplitude, while the jet ratio contains hard-collinear 
virtual poles associated with the radiated gluon. The factor 
$\s_{n,g}^{\mathcal {H}, \mu} \big( \{\beta_i\}; k \big)$ encodes all loop contributions 
to the radiative soft function that are finite as $\eps \to 0$: thus, it describes hard 
wide-angle virtual contribution to the soft real radiation of gluon $g$ off a set 
of $n$ hard legs. Finally, observe that, starting from \eq{AmpFact}, one might 
have expected further hard collinear factors in \eq{eq:soft_fact_nlo}, associated
with the $n$ Wilson lines along the directions $\beta_i$: these factors however
are all equal to one, since the jet functions in the numerators coincide with
their soft approximations, given by the eikonal jet functions in the denominators.

At NLO, the factorisation in \eq{eq:soft_fact_nlo} leads to
\beq
\label{eq:Sng1}
  \s_{n,g}^{(1)} \big( \{\beta_i\}; k \big)
  \, = \, &
  \left[ \s_{n+1}^{(1)} \big( \{\beta_i\}, \beta_k \big) -
  \J_{\E_g}^{(1)} (\beta_k) \right] \s_{n,g}^{(0)} \big( \{\beta_i\}; k \big) 
  \nn \\ &
  + \, 
  \J^{(1) \mu}_{g,g} (0;k) \, \s_{n,g}^{(0) \mu} \big( \{\beta_i\}; k \big) \,
  + \, \text{finite} 
  \, , \quad
\eeq
where we used the fact that at tree level $\s_{n,g}^{\mathcal{H},\mu}$ coincides
with the full tree level radiative soft function $\s_{n,g}^{(0)}$ after contracting 
with the appropriate polarisation vector. 

We can now proceed to verify \eq{eq:Sng1}. We start with the known expressions 
for the relevant soft functions. At tree level, the radiative soft function gives
\beq
\label{eq:Sng0}
  \Big[
  \s_{n,g}^{(0)} \big( \{\beta_i\}; k \big)
  \Big]^{a}
  & = &
  \gs \mu^{\eps}
  \sum_{i=1}^n
  \frac{\beta_i \cdot \epsilon_\lambda(k)}{\beta_i \cdot k} \,
  \T^a_i \, ,
\eeq
where $\mu$ is the $\overline{\rm{MS}}$ renormalisation scale, including the 
appropriate correction for contributions proportional to $\gamma_E$ and 
$\ln 4 \pi$, and $ \gs$ is the bare strong coupling. The one-loop contribution 
to the virtual soft function for $(\npo)$ particles, on the other hand, reads (see 
for instance~\cite{Dixon:2008gr})
\beq
\label{eq:Snp1}
  \Big[
  \s_{n+1}^{(1)} \big( \{\beta_i\}, \beta_k \big)
  \Big]^{ab}
  & = &
  \frac{\as}{4\pi} \,
  \sum_{\substack{i,j = 1 \\ i \neq j}}^{n}
  {\bf T}_i \cdot  {\bf T}_{j} \,
  \delta^{ab}
  \bigg[
  \frac1{\eps^2}-\frac1{\eps} \ln \frac{2 \, p_i \cdot p_j}{\mu^2}
  \bigg]
  \nnb \\
  & + &
  \frac{\as}{2\pi} \,
  \sum_{i=1}^{n}
  {\bf T}_i \cdot  ({\bf T}_{k})^{ab}
  \bigg[
  \frac1{\eps^2}-\frac1{\eps} \ln \frac{2 \, p_i \cdot k}{\mu^2}
  \bigg]
  \, ,
\eeq
where $p_i = \mu/\sqrt{2} \,  \beta_i$, and we have understood the gluon with momentum 
$k$ to be at position $n+1$, {\it i.e.} $p_\npo = k$, we wrote explicitly the colour indices of 
the Wilson line associated with that gluon, and all poles in $\eps$ are of purely soft origin. 
Since \eq{eq:Snp1} is the leading-order term for the virtual soft function, the coupling $\as$ 
can equivalently be taken to be bare or renormalised, without affecting the argument below.

The first term of \eq{eq:Sng1} is found by multiplying \eq{eq:Sng0} to the right 
of \eq{eq:Snp1}. Importantly, one can show that
\beq
\label{eq:JCGun}
  \s_{n+1}^{(1)} \big( \{\beta_i\}, \beta_k \big) \, \s_{n,g}^{(0)} \big( \{\beta_i\}; k \big) \, = \,
  \s_{n,g}^{(0)} \big( \{\beta_i\}; k \big) \, \s_n^{(1)} \big(\{\beta_i\}\big) + 
  \vareps_\lambda(k) \cdot J_{\text{CG}}^{(1), \, \text{b}} \big( \{\beta_i\}; k \big) \, ,
\eeq
where $J_{\text{CG}}^{(1), \, \text{b}}$ is the bare one-loop Catani-Grazzini (CG) 
soft-gluon current for final-state radiation~\cite{Catani:2000pi}, given by
\beq
  J_{\text{CG}}^{(1), \, \text{b}, \, \mu}\big(\{\beta_i\};k\big)
  \, = \, 
  - \, \frac{\al_s}{4\pi} \, g_s \mu^{\eps} \, {\rm i} f^{abc} 
  \sum_{i,j = 1}^n \T_i^b \T_j^c \, \frac1{\eps^2}
  \left( \frac{\beta_i^\mu}{\beta_i\cdot k} - \frac{\beta_j^\mu}{\beta_j\cdot k} \right)
  \left(\frac{\mu^2 \, p_i \cdot p_j}{2 \, p_i \cdot k \, p_j \cdot k}
  \right)^\eps \, .
\eeq
Eq.~(\ref{eq:Snp1}) is useful because the combination $\s_{n,g}^{(0)}\,\s_n^{(1)}$ appears
in the one-loop renormalised radiative amplitude $\mathcal{A}_{n,1}^{(1)}$
\beq
  \mathcal{A}_{n,1}^{(1)} \big( \{p_i\}; k \big) & = & 
  \s_{n,g}^{(0)} \big( \{\beta_i\}; k \big)
  \sum_{i=1}^n \Big[  \J_i^{(1)} \big( \{p_i\} \big)- \J_{\E_i}^{(1)} \big( \{\beta_i\}\big) \Big] \,  
  \hard_n^{(0)} \big( \{p_i\} \big) 
  \nnb\\
 &&
  +\s_{n,g}^{(0)} \big( \{\beta_i\}; k \big) \hard_n^{(1)} \big( \{p_i\} \big) 
  + \s_{n,g}^{(1)} \big( \{\beta_i\}; k \big) \hard_n^{(0)} 
  \big( \{p_i\} \big) 
  \nn \\ 
  & = & \s_{n,g}^{(0)} \big( \{\beta_i\}; k \big) \, \mathcal{A}_{n}^{(1)} \big( \{p_i\} \big)
 \nn \\
  && + 
  \left[ \s_{n,g}^{(1)} \big( \{\beta_i\}; k \big) - \s_{n,g}^{(0)} \big( \{\beta_i\}; k \big) \,
  \s_n^{(1)} \big( \{\beta_i\} \big) \right] \mathcal{A}_n^{(0)} \big( \{p_i\} \big) \, , \qquad
  \label{eq:An11FactNew}
\eeq
where $\mathcal{A}_n^{(0)} = \hard_n^{(0)}$. We can now compare \eq{eq:An11FactNew} 
to the equivalent formulation from~\cite{Catani:2000pi}: in that approach one writes
\beq
\label{eq:An11CG}
  \mathcal{A}_{n,1}^{(1)} \big( \{p_i\}; k \big)
  \, = \, \vareps_\lambda(k) \cdot \left[ J_{\text{CG}}^{(0)} \big( \{\beta_i\}; k \big) \,
  \mathcal{A}_n^{(1)} \big( \{p_i\}\big) + J_{\text{CG}}^{(1), \, \text{r}} \big( \{\beta_i\}; k \big)
  \mathcal{A}_n^{(0)} \big( \{p_i\} \big) \right] \, ,
\eeq
where $J_{\text{CG}}^{(0)}$ is equal to the expression in \eq{eq:Sng0}.
We note that all quantities in \eq{eq:An11CG} are renormalised, and in particular we now 
feature the \emph{renormalised} CG current $J_{\text{CG}}^{(1), \, \text{r}}$. Thus we can 
identify
\beq
  \vareps_\lambda(k) \cdot J_{\text{CG}}^{(1), \, \text{r}} \big( \{\beta_i\}; k \big) \, = \, 
  \s_{n,g}^{(1)} \big( \{\beta_i\}; k \big) - \s_{n,g}^{(0)} \big( \{\beta_i\}; k \big) \,
  \s_n^{(1)} \big( \{\beta_i\} \big) \, .
\eeq
Then, using \eqref{eq:JCGun}, we have
\beq
\label{eq:Sng1b0}
  \s_{n,g}^{(1)} \big( \{\beta_i\}; k \big) & = & \s_{n+1}^{(1)} \big( \{\beta_i\}, \beta_k \big) \,
  \s_{n,g}^{(0)} \big( \{\beta_i\}; k \big) + \left[ J_{\text{CG}}^{(1), \, \text{r}} 
  \big( \{\beta_i\}; k \big) - J_{\text{CG}}^{(1), \, \text{b}} \big( \{\beta_i\}; k \big) \right]
  \cdot \vareps_\lambda(k) \nnb
  \\ & = &
  \s_{n+1}^{(1)} \big( \{\beta_i\}, \beta_k \big) \, \s_{n,g}^{(0)} \big( \{\beta_i\}; k \big) - 
  \frac{\al_s}{4\pi} \frac{b_0}{2 \eps} J_{\text{CG}}^{(0)} \big( \{\beta_i\}; k \big) \cdot
  \vareps_\lambda(k) \nnb
  \\ & = & 
  \left[ \s_{n+1}^{(1)} \big( \{\beta_i\}, \beta_k \big) - \frac{\al_s}{4 \pi} 
  \frac{b_0}{2\eps} \right] \s_{n,g}^{(0)} \big( \{\beta_i\}; k \big) \, ,
\eeq
where $b_0 = (11C_A - 4 T_R n_f)/3$ in our normalisation, with $n_f$ being the 
number of active flavours in the process. Eq.~(\ref{eq:Sng1b0}) is actually equivalent to 
\eq{eq:Sng1}, which we set out to prove: indeed, the second term in the last line is the 
hard-collinear contribution from the radiated gluon, {\it i.e.} the contribution arising from 
the ratio of the two jet functions in \eq{eq:Sng1}. Such a ratio is computed for instance 
in Ref.~\cite{Falcioni:2019nxk} (see eqs.~(2.21) and (2.26) there). The $\beta$-function 
coefficient $b_0$ arises in this context as the anomalous dimension of the gluon jet 
function.\footnote{Note that in Ref.~\cite{Falcioni:2019nxk} different conventions are 
used for the normalisation of the $\beta$-function coefficients.} We note that the result 
in \eq{eq:Sng1b0} is compatible with the soft factorisation displayed in \eq{softrads.o.} 
upon converting the softer real gluon into a soft virtual radiation, namely selecting soft 
loops only.

In order to apply these ideas to subtraction, we need to explore what happens at 
cross-section level. In order to do this, we begin by examining the general expression 
of the real-virtual contribution to the cross section, $RV_\npo$, in the limit in which the 
emitted gluon $k$ becomes soft. The corresponding kernel features explicit soft and 
hard-collinear poles relevant to all $\npo$ external legs. When comparing it with the 
cross-section-level version of \eq{eq:Sng1}, we then expect the two expressions to 
differ by the hard-collinear poles associated with all particles but the radiated gluon. 
This expectation is verified in the following.

The soft limit of $RV_{\npo}$ in the $\overline{\rm{MS}}$ scheme can be found in 
Refs.~\cite{Bern:1998sc,Bern:1999ry,Kosower:1999rx,Catani:2000pi,Somogyi:2006db},
and reads\footnote{This is the $\bS{k}$ limit defined in \secn{Archi}. Note 
that, with a slight abuse of notation, we are using $k$ both for the ordering number
of the gluon in the set of radiated particles, and for its momentum.}
\beq
\label{eq:soft_RV}
  \bS{k} \, RV_\npo & = & - \,
  \Norm \, \delta_{f_k g}
  \sum_{\substack{i \neq k \\ j\neq i,k}} \, 
  \mc I^{(k)}_{ij} \, 
  \Bigg[ \Vl_{n, ij} - \bigg( \Norm \, \cg \, \frac{C_A}{\eps} \, 
  \pi \cot(\pi \eps) \Big( \mc I^{(k)}_{ij} \Big)^{\eps}
  + \frac{\as}{4\pi} \, \frac{b_0}{\eps} \bigg) B_{n, ij}
  \nnb \\ [-3mm] && \hspace{35mm}
  + \, \Norm \, \cg \,  \frac{2 \pi}\eps
  \sum_{p \neq i,j,k} \Big( \mc I^{(k)}_{jp} \Big)^{\eps} B_{n, ijp}
  \Bigg] \, ,
\eeq
where the colour-correlated Born and virtual contributions are defined by $B_{n, ij} 
= \mc{A}_n^{(0) \dag}  ({\bf T}_i \cdot {\bf T}_j) \mc{A}_n^{(0)}$ and by $V_{n, ij} = 
2 \text{Re}\big[\mc{A}_n^{(0) \dag}  ({\bf T}_i \cdot {\bf T}_j) \mc{A}_n^{(1)} \big]$,
respectively. Furthermore, we introduced the notations
\begin{align}
    \mc I^{(i)}_{ab} \, &= \, \frac{s_{ab}}{s_{ai}s_{bi}} \, , \hspace{-5mm}
    & B_{n, ijp} \, = \, f_{abc} \, 
    \mc{A}_n^{(0)\dag} \, {\bf T}^a_i {\bf T}^b_j {\bf T}^c_p \, \mc{A}_n^{(0)} \, , \nnb\\
    \Norm \, &= \,  8 \pi \as \mu^{2 \eps}  \, ,  \hspace{-5mm}
    & \cg \, = \, \frac1{(4\pi)^{2-\eps}} 
    \frac{\Gamma(1+\eps)\Gamma^2(1-\eps)}{\Gamma(1-2\eps)} \, ,
\end{align}
where $s_{ab}=2 p_a\cdot p_b$. In order to make the pole content of \eq{eq:soft_RV} 
explicit, we need to extract the divergent contributions to the colour-correlated virtual 
matrix element $V_{n, ij}$. This can be written as
\beq
  \Vl_{n, ij} \Big|_{\rm poles} & = & - \frac{\as}{2 \pi} \,
  \Bigg\{
  B_{n, ij} \sum_{m \neq k} \bigg[ \delta_{f_m g} \left(
  \frac{C_A}{\eps^2} + \frac{b_0}{2\eps} \right) + 
  \delta_{f_m \{q, \bar q\}} \, C_F \left(
  \frac{1}{\eps^2} + \frac32 \frac{1}{\eps} \right)
  \bigg]
  \nnb \\ && \hspace{15mm}
  + \, \frac1{2\eps} \sum_{\substack{r \neq k \\ s\neq k,r}}
  B_{n, ijrs} \, \ln \frac{s_{rs}}{\mu^2}
  \Bigg\} \, ,
\eeq
where $B_{n, ijrs} = \mc{A}_n^{(0)\dag}   \big\{ {\bf T}_i \cdot {\bf T}_j, {\bf T}_r \cdot 
{\bf T}_s \big\} \mc{A}_n^{(0)}$. The flavour Kronecker delta functions are defined 
as follows: if $f_i$ is the flavour of parton $i$, then $\delta_{f_i g}=1$ if parton $i$ 
is a gluon, and  $\delta_{f_i g}=0$ otherwise. Similarly, we define $\delta_{f_i \{q, \bar q\}} 
\equiv \delta_{f_iq}+\delta_{f_i \bar{q}}$. Next, we expand in $\eps$ the curly bracket 
in \eq{eq:soft_RV}: one can then use colour conservation to show that $\sum_{p \neq i,j,k}
B_{ijp} = 0$. The explicit pole content of \eq{eq:soft_RV} can then be presented
as follows
\beq
  \bS{k}\,\RVl_\npo \Big|_{\rm poles} & = & \Norm \,\, \frac{\as}{2 \pi} \,
  \sum_{\substack{i \neq k \\ j \neq i,k}} \mc I^{(k)}_{ij}
  \Bigg\{ \, B_{n, ij} \sum_{m \neq k}
  \bigg[ \delta_{f_m g} \left( \frac{C_A}{\eps^2} +
  \frac{b_0}{2}\,\frac{1}{\eps} \right) + 
  \delta_{f_m \{q, \bar q\}}\,C_F \left( \frac{1}{\eps^2} +
  \frac32 \, \frac{1}{\eps} \right) \bigg]
  \nnb \\ && \hspace{8mm}
  + \, \frac{1}{2 \eps} \sum_{\substack{r\neq k \\ s\neq r,k}} B_{n, ijrs} \, 
  \ln\frac{s_{rs}}{\mu^2} +  \bigg[
  C_A \left( \frac{1}{\eps^2} + \frac{1}{\eps} \, \ln \frac{\mu^2\,s_{ij}}{s_{ik}s_{jk}}
  \right) + \frac{b_0}{2\eps} \bigg] \, B_{n, ij} \Bigg\}
  \, . \qquad 
\eeq
One can then express the equation above in the language of factorisation, 
and find
\beq
\label{eq:SkRV}
  \bS{k} \, RV_\npo & = & \as^2 \, \mu^{2 \eps}
  \sum_{\substack{i,j = 1 \\ j \neq i}}^n
  \frac{2 \beta_i \cdot \beta_j}{\beta_i \cdot k \, \beta_j \cdot k} \,\,
  {\cal H}^{(0) \,  \dagger}_n \Bigg\{
  {\bf T}_i \cdot {\bf T}_j \sum_{m=1}^n \bigg(
  \frac{C_m}{\eps^2} + \frac{\gamma_m^{(1)}}{\eps} \bigg)
  \nnb \\ && \hspace{1cm}
  + \, {\bf T}_i \cdot {\bf T}_j \, \bigg[ C_A \bigg( \,
  \frac1{\eps^2} + \frac1\eps
  \ln \frac{\mu^2 \, \beta_i \cdot \beta_j}{2 \, \beta_i \cdot k \, \beta_j \cdot k}
  \bigg) + \frac{b_0}{2\eps} \bigg]
  \nnb \\ && \hspace{1cm}
  + \, \frac1{2 \eps}
  \sum_{\substack{r,s = 1 \\ r \neq s}}^n
  \Big\{ {\bf T}_i \cdot {\bf T}_j, {\bf T}_r \cdot {\bf T}_s \Big\}
  \ln \frac{2 \, p_r \cdot p_s}{\mu^2}
  \Bigg\} \, {\cal H}^{(0)}_n \, + \, {\rm finite} \, ,
\eeq
where $C_m = C_F$ and $\gamma_m^{(1)} = 3 \, C_F/2$ for (anti-)quarks, while
$C_m = C_A$ and $\gamma_m^{(1)} = b_0/2$ for gluons. The contributions 
proportional to quadratic Casimir eigenvalues are of soft origin, while the ones
proportional to $\gamma_m^{(1)}$ are hard-collinear single poles, and the $b_0$ 
term is due to renormalisation. After performing the relevant colour algebra, 
\eq{eq:SkRV} can be shown to satisfy
\beq
\label{eq:Sk_RV}
  \bS{k} \, RV_\npo & = & 
  {{\cal H}^{(0)}_n}^{\dagger} \,
  \left( {\s^{(0)}_{n,g}} \big( \{\beta_i\}; k \big) \right)^{\dagger}\,
  S_{n+1}^{(1)} \big( \{\beta_i\}, \beta_k \big) \,
  \s_{n,g}^{(0)} \big( \{\beta_i\}; k \big) \,
  {\cal H}^{(0)}_n
  \nnb \\ & + &
  \as^2 \, \mu^{2\eps}
  \sum_{\substack{i,j = 1 \\ j \neq i}}^n
  \frac{2 \beta_i \cdot \beta_j}{\beta_i \cdot k \, \beta_j \cdot k} \,
  {{\cal H}^{(0)}_n}^\dagger \,
  \T_i \cdot \T_j \,
  {\cal H}^{(0)}_n \,
  \left[
  \sum_{m = 1}^n \frac{\gamma_m^{(1)}}{\eps} + \frac{b_0}{2 \eps}
  \right] \, ,
\eeq
where $S_{n+1}^{(1)} = 2 \, \text{Re} \big[\s_{n+1}^{(1)} {\s_{\npo}^{(0)}}^{\!\!\dagger} \big]$, 
and, for brevity, we are henceforth omitting the mention of finite contributions. 
Eq.~(\ref{eq:Sk_RV}) is an example of the issues outlined in the introduction of this 
Section. $RV_\npo$ is a cross section involving $(\npo)$ outgoing particles, one of 
which is gluon $k$; upon taking the soft limit for $k$, the first term on the {\it r.h.s.} 
provides a natural generalisation of the tree-level soft refactorisation discussed in 
\secn{ReFactoTree}. At loop level, as expected, we find poles in $\eps$ of hard-collinear
origin, that correct a purely soft refactorisation picture: this is given by the second term on the 
right-hand side. In particular, $\gamma_m^{(1)}$ comes from the loop momentum 
being collinear to leg $m$, while $b_0$ comes from the loop momentum being 
collinear to the extra gluon, $k = n+1$. This can be further clarified by rewriting the 
second term using the cross-section-level version of \eq{eq:Sng1b0}, found by 
multiplying it times the tree-level complex conjugate and summing over the polarisations 
of the radiated gluon. Writing $\bS{k} RV_\npo$ in terms of $S_{n,g}^{(1)}$ cancels 
the term involving $b_0$ in \eq{eq:Sk_RV}, with the result
\beq
  \bS{k} \, RV_\npo & = & {{\cal H}_n^{(0)}}^{ \dagger} \, \left[
  S_{n,g}^{(1)} \big( \{\beta_i\}; k \big) + S_{n,g}^{(0)} \big( \{\beta_i\}; k \big) \, 
  \frac{\al_s}{2 \pi} \sum_{m = 1}^n \frac{\gamma_m^{(1)}}{\eps}
  \right]
  {\cal H}^{(0)}_n \, .
\eeq
One can then further rewrite the remaining hard-collinear term as a difference of 
jet functions for each external particle of the Born process, in the form
\beq
\label{eq:SkRVInFuncs}
  \bS{k} \, RV_\npo & = & {{\cal H}^{(0)}_n}^{\dagger} \,
  \left[ S_{n,g}^{(1)} \big( \{\beta_i\}; k \big) +
  \sum_{i=1}^n J^{(1), \, \text{hc}}_{f_i, f_i} \,
  S_{n,g}^{(0)} \big( \{ \beta_i \}; k \big)
  \right]
  {\cal H}^{(0)}_n \, ,
\eeq
where we have adopted the notation of \eq{virtwolopieces} for the jet function multiplying 
the hard function $\hard_n^{(0)}$. Eq.~(\ref{eq:SkRVInFuncs}) thus confirms the argument 
presented above \eq{eq:soft_RV}: the one-loop radiative soft function $S_{n,g}^{(1)}$ captures 
all the soft-virtual poles of $\bS{k} RV$, while there are still collinear poles from hard 
momenta circulating in the loop, which are correctly reproduced by the appropriate 
combination of jet functions.

We emphasise that the expressions presented in \eq{eq:SkRVInFuncs} are particularly 
suitable for verifying explicitly that the pole content of the soft limit $\bS{k} \, RV_\npo$ 
coincides (up to a sign) with the pole structure of the soft component of $I^{\otwo}_\npo$. 
This is a crucial validation of the arguments presented below \eq{subtNNLO}, and will 
be discussed in \secn{dsocfrfvcr}.  We now proceed to the analysis of the collinear limit 
of the real-virtual correction.

%%%%%%%%%%%%%%%%%%%%%

\subsection{Single-radiative jet functions at one loop}
\label{RaJeLoop}

By the same reasoning that was applied to the radiative soft function in 
\secn{RaSoFuLoop}, the radiative jet function is also expected to factorise 
as a scattering amplitude in the presence of Wilson lines, in the form\footnote{As 
before, we evaluate the jet functions at $x=0$, and we do not display the dependence 
on the polarisation vectors.}
\beq
\label{eq:ampJFact}
  \mathcal{J}_{f_i, f_1 f_2} \big(0; k_1, k_2 \big) \, = \, 
  \mathcal{S}_{3} \big( \beta_1, \beta_2, n \big) \, 
  \frac{\mathcal{J}_{f_1,f_1} \big( 0; k_1 \big)}{\mathcal{J}_{\E_1} (\beta_1)} \, 
  \frac{\mathcal{J}_{f_2,f_2} \big( 0; k_2 \big)}{\mathcal{J}_{\E_2} (\beta_2)} \, 
  \mathcal{J}_{f_i, f_1f_2}^{\mathcal{H}} \big( 0; k_1, k_2 \big) \, ,
\eeq
with one hard-collinear jet combination, $\mathcal{J}_{f_m, f_m}(0; k_m)/\mathcal{J}_
{\E_m}(\beta_m)$ for each outgoing parton, $m = 1, 2$, a three-line soft function 
$\mathcal{S}_{3} (\beta_1, \beta_2, n)$ defined by Wilson lines in the directions 
of $k_1$, $k_2$ and $n$, and finally a hard function $\mathcal{J}_{f_i, f_1 f_2}^
{\mathcal{H}} (0; k_1,k_2)$ responsible for finite corrections, due to hard virtual
particles not collinear to either $f_1$ or $f_2$. Note that there is no hard-collinear 
behaviour associated to the Wilson-line in direction $n$ (provided $n^2 \neq 0$), 
only soft singularities. Spin indices connecting the two outgoing jet functions to the 
hard jet function are implicit in \eq{eq:ampJFact}.

Squaring the factorised expression in \eq{eq:ampJFact} and evaluating the result 
at one loop gives the cross-section level jet function $J_{f_i, f_1 f_2}^{(1)}$, which 
appears in the collinear sector of $K^{\RV}_\npo$. This factorisation reads
\beq
\label{eq:J11Fact}
  J_{f_i,f_1f_2}^{(1) \al_i \beta_i} \big(\ell_i; k_1, k_2 \big) 
  & = & \\
  && \hspace{-20mm} \, = \, \int d^dx \,{\rm e}^{ {\rm i} \ell_i \cdot x}
  \Bigg\{ \Big(\mathcal{J}_{f_i, f_1 f_2}^{(0) \, \al_i} \big( 0; k_1, k_2 \big) \Big)^\dagger
  \bigg[ S_{3}^{(1)} \big( \beta_1, \beta_2, n \big) - 
  \sum_{j \in \{1,2\}} J_{\E_j}^{(1)} (\beta_j) \bigg] \,
  \mathcal{J}_{f_i, f_1 f_2}^{(0) \beta_i} \big( x ; k_1, k_2 \big) \nn \\ 
  & & + \, \int \frac{d^d q}{(2 \pi)^d} \, 
  \Big(\mathcal{J}_{f_i, f_1 f_2}^{(0) \, \al_i;  \bar{\al}_1} \big( 0; q, k_2 \big) \Big)^\dagger \, 
  J_{f_1, f_1}^{(1) \, \bar{\al}_1 \bar{\beta}_1} \big(q; k_1 \big) 
  \mathcal{J}_{f_i, f_1 f_2}^{(0) \, \beta_i; \bar{\beta}_1 } \big( x; q, k_2 \big) \nn \\ 
  & & + \, \int \frac{d^d q}{(2 \pi)^d} \,
  \Big(\mathcal{J}_{f_i, f_1 f_2}^{(0) \, \al_i ; \bar{\al}_2} \big( 0; k_1, q \big) \Big)^\dagger \, 
  J_{f_2, f_2}^{(1) \, \bar{\al}_2 \bar{\beta}_2} \big(q; k_2 \big) 
  \mathcal{J}_{f_i, f_1 f_2}^{(0) \,  \beta_i; \bar{\beta}_2} \big( x; k_1, q \big) \Bigg\} 
  + \, \text{finite} \, , \nn
\eeq
where the spin indices
undergo the conventions explained below \eq{eq:coll_str_ord_example1}.
The integrals over the parent momentum $q$ are trivial in \eq{eq:J11Fact}, since 
the cross-section-level jet functions $J_{f_m, f_m}^{(1)}$ are proportional to 
$\delta(q-k_m)$. However we will see later, in \secn{dsocfrfvcr}, that in the strongly-ordered 
limit for double-radiative jet function the corresponding momentum $q$ will play 
an important role. The cross-section-level soft function $S_{3}$ in \eq{eq:J11Fact} is 
defined as usual by squaring the vacuum expectation value of the appropriate Wilson lines. 
In this case
\beq
\label{eq:Snij}
  S_{3} \big( \beta_1, \beta_2, n \big) \, = \, \Big| \braket{0| \, T \Big[\Phi_n (0, \infty) 
  \Phi_{\beta_1} (0, \infty) \Phi_{\beta_2} (0, \infty) \Big]|0} \Big|^2 \, .
\eeq
In order for $S_{3}$ to be gauge invariant, we require colour conservation, in the 
form $\T_n + \T_1 + \T_2 = 0$, whenever the soft function operates within an 
on-shell amplitude. This implies that the Wilson line in direction $n$ carries the 
total colour charge of all the Born-level particles, excluding the colour charge of 
the parent momentum $\ell_i$.

We note that an equivalent factorisation exists for the eikonal radiative jet. At cross-section
level we write
\beq
\label{eq:JE11Fact} 
  J_{\E_i,g}^{(1)} \big( \beta_i; k \big) & = & \left[ S_{3}^{(1)} \big( \beta_i, \beta_k, n \big) - 
  J_{\E_g}^{(1)} \big( \beta_k \big) \right] \, J_{\E_i, g}^{(0)} \big( \beta_i; k  \big) \nn \\
  && \qquad 
  + \, \int \frac{d^d q}{(2 \pi)^d} \, \mathcal{J}_{\E_i, g}^{(0) \, \bar \al} 
  \big( \beta_i; q \big) \, J_{g; g}^{(1) \, \bar{\al} \bar{\beta}} \big( q; k \big) \,
  \mathcal{J}_{\E_i, g}^{(0) \, \bar \beta} \big ( \beta_i; q \big) \, .
\eeq
Here we can specify that the radiated particle is a gluon since one cannot radiate a 
single quark from a Wilson line. We will now show that eqs.~(\ref{eq:J11Fact}) 
and~(\ref{eq:JE11Fact}) hold, by explicit computation of the one-loop radiative jet.

%%%%%%%%%%%%%%%%%%%%%

\subsection{Computing radiative jet functions at one loop}
\label{sub:radiative_jet_function_calc}

In \secn{RaJeLoop} we have introduced factorisation formulas for jet and eikonal-jet 
functions at the one-loop level. In order to test such formulas, it is necessary to compute 
these functions at order $\alpha_s$. This is non-trivial, since different scales enter the 
relevant diagrams, and the corresponding Feynman rules involve denominators that 
are linear in the loop momentum.

Often, an axial gauge ($n \cdot A= 0$) is chosen when computing jet functions, since 
the Wilson line $\Phi_n$ in eqs.~(\ref{collFFq}),~(\ref{collFFg}) and~(\ref{eikcollF}) 
becomes unity in that gauge. This choice prevents the proliferation of diagrams 
involving the interaction of gluons with the auxiliary Wilson line, and also avoids 
diagrams with ghosts. At loop-level, however, there are complications that arise 
from the denominators of the axial-gauge gluon propagator~\cite{Leibbrandt:1987qv}: 
it is therefore practical to work in Feynman gauge. Another convenient choice would
be to set $n^2 = 0$, which makes the integrals easier to compute, as there is no 
additional variable, proportional to $n^2$, that they could depend upon. One 
difficulty in this selection, however, is that it introduces a set of unphysical collinear 
divergences associated with the Wilson line in the direction $n$, which would need 
to be removed. To be on the safe side and to avoid confusion, we choose $n^2 \neq 0$.

The computation of the (bare) jet functions proceeds in a standard way. The necessary 
Dirac and colour algebra for the Feynman diagrams can be performed with the software
{\tt FeynCalc}~\cite{MERTIG1991345,Shtabovenko:2016sxi,Shtabovenko:2020gxv}. 
The results are then passed to the program {\tt LiteRed}~\cite{Lee:2012cn,Lee:2013mka} 
to reduce them to a set of basis integrals using integration-by-parts identities. Some
details of the calculation are presented in \appn{app:olojetcomp}. In particular, given
the results for the relevant  master integrals, presented in eqs.~(\ref{eq:bubbleInts}),
(\ref{eq:J1111}) and~(\ref{eq:eikJ1111}), it is easy to obtain compact expressions for the 
bare jet functions at one loop. The quark radiative jet function with a single gluon 
emission reads
\beq
\label{radjqtoqgolo}
  J_{q,qg}^{(1), \, \text{b}} \big( \ell; k_1, k_2) & = & - \, \frac{\as}{2 \pi} \,
  J_{q,qg}^{(0)} \big( \ell; k_1, k_2 \big) \\
  &&\hspace{-1.5cm} \times \, \Bigg[ C_A \left( \frac{1}{\eps^2} + 
  \frac{1}{\eps} \log\frac{\mu^2 \, k_1 \cdot n}{2 k_2 \cdot n \, k_1\cdot k_2} \right) + 
  C_F \left( \frac{1}{\eps^2} + \frac{1}{\eps} \left( 1 + \log \frac{\mu^2 \, n^2}{(2 k_1
  \cdot n)^2} \right) \right) \Bigg] + \text{finite} \, , \nn
\eeq
where the superscript $\text{b}$ denotes a bare function. The gluon radiative jet function 
with the emission of a quark-antiquark pair reads
\beq
\label{radjgtoqqolo}
  J_{g,q\bar q}^{(1), \, \text{b}, \, \mu \nu} \big( \ell; k_1, k_2) & = & - \, \frac{\as}{2 \pi} \,
  J_{g,q\bar q}^{(0), \, \mu\nu} \big( \ell; k_1, k_2 \big) \\
  && \hspace{- 2cm} \times \, \Bigg[ \frac{C_A}{\eps} \left( \log\frac{n^2 \, k_1 
  \cdot k_2}{2 k_1 \cdot n \, k_2 \cdot n} - \frac{8}{3} \right) + 
  C_F \left( \frac{2}{\eps^2} + \frac{1}{\eps} \left( 3 + 2 \log \frac{\mu^2}{2 k_1\cdot k_2}
  \right) \right) + \frac{2 n_f }{3 \eps} \Bigg] + \, \text{finite} \, . \nn
\eeq
Similarly, the gluon radiative jet function with the emission of a gluon pair reads
\beq
\label{radjgtoggolo}
  J_{g,gg}^{(1), \, \text{b}, \, \mu \nu} \big( \ell; k_1, k_2 \big) & = & - \, \frac{\as}{2\pi} \, 
  J_{g,gg}^{(0), \, \mu\nu} \big( \ell; k_1, k_2 \big) \nn \\
  && \hspace{-1.5cm} \times \, C_A \, \Bigg[ \frac{2}{\eps^2} + \frac{1}{\eps}
  \left( 1 + \log \frac{\mu^2 \, n^2}{2 k_1 \cdot n \, k_2 \cdot n}
  + \log \frac{\mu^2}{ 2 k_1 \cdot k_2} \right) \Bigg] 
  + \, \text{finite} \, .
\eeq
Finally, the radiative eikonal jet function with the emission of a gluon reads
\beq
\label{radeikjgolo}
  J_{\E, g}^{(1), \, \text{b}} \big( \beta; k \big) \, = \, - \, \frac{\as}{2\pi} \, 
  J_{\E, g}^{(0)} \big( \beta; k \big) \, C_A \Bigg[ \frac{1}{\eps^2} + 
  \frac{1}{\eps} \log \left( \frac{\mu^2 \, \beta \cdot n}{2 k 
  \cdot n \, k \cdot \beta} \right) \Bigg] + \, \text{finite} \, .
\eeq
In eqs.~(\ref{radjqtoqgolo})-(\ref{radeikjgolo}) we extracted the tree-level results, given in 
\appn{app:jetFuncs}, and $\as$ is the bare coupling.
\begin{figure}
    \centering
\begin{subfigure}[b]{0.25\textwidth}
\centering
\includegraphics[width=\textwidth]{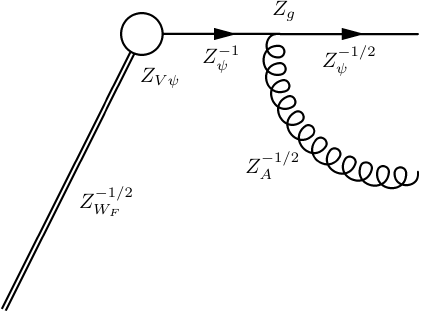}
\end{subfigure}
\begin{subfigure}[b]{0.25\textwidth}
\centering
\includegraphics[width=\textwidth]{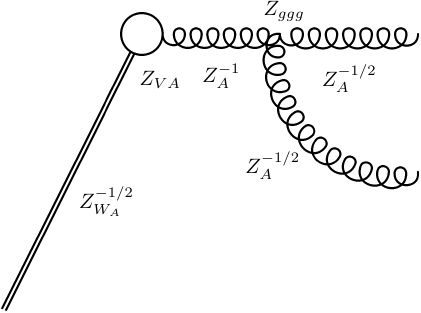}
\end{subfigure}
\begin{subfigure}[b]{0.25\textwidth}
\centering
\includegraphics[width=\textwidth]{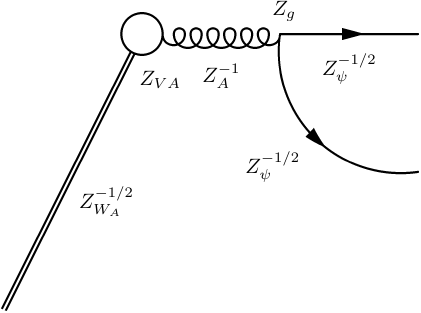}
\end{subfigure}
    \caption{Assigning renormalisation constants to different factors in the radiative 
    jet functions: radiation from the field line.}
    \label{fig:jetFunctionRenorm}
\end{figure}
It is worthwhile to discuss the renormalisation of the jet functions in detail. As we 
have chosen to regulate the infrared and ultraviolet both in dimensional regularisation, 
we cannot readily distinguish between them in the calculation (note that radiative
functions are not pure counterterms even in the eikonal limit, since the radiated
momentum provides a scale). As a consequence, the renormalisation factors need 
to be determined separately. We will do so by examining the tree-level diagrams and 
dressing each component with a renormalisation factor. All relevant factors are known
or can be easily computed. Upon renormalisation we will find a universal and transparent
expression for single-radiative jet functions at one loop.

The tree-level diagrams where the radiated particle is emitted from the field line are 
displayed in Fig.~\ref{fig:jetFunctionRenorm}. Each internal vertex carries a $Z$ factor 
for the corresponding coupling (these are $Z_{V \psi}, Z_{VA}, Z_{g}, Z_{ggg}$).
Similarly, each internal propagator carries a factor $Z^{-1}$ for the correponding field 
(in the case at hand, $Z_{\psi}^{-1}$ and $Z_{A}^{-1}$). Finally, each external leg 
carries a factor of $Z^{-1/2}$ for the corresponding wave function (here $Z_{W_F}^{-1/2},
Z_{W_A}^{-1/2}, 
Z_{A}^{-1/2}, Z_{\psi}^{-1/2}$). Altogether, the renormalisation takes the form
\beq
\label{jetrenorm}
    J_{q,qg} \big( \ell; k_1, k_2 \big) & = & J_{q,qg}^{\text{b}} \big( \ell; k_1, k_2 \big)
    \left( Z_{V \psi} Z_g Z_{\psi}^{-3/2} Z_A^{-1/2} Z_{W_F}^{-1/2} \right)^2 \, , \nn \\
    J_{g,gg}^{\mu \nu} \big( \ell; k_1, k_2 \big) & = & J_{g,gg}^{\text{b}, \, \mu \nu}
    \big( \ell; k_1, k_2 \big) \left( Z_{VA} Z_{ggg} Z_A^{-2} Z_{W_A}^{-1/2} \right)^2 \, , \\
    J_{g,qq}^{\mu \nu} \big( \ell; k_1, k_2 \big) & = & J_{g,qq}^{ \text{b}, \, \mu \nu}
    \big( \ell; k_1, k_2 \big) \left( Z_{VA} Z_g Z_\psi^{-1} Z_A^{-1} Z_{W_A}^{-1/2} 
    \right)^2 \, , \nn
\eeq
and the relevant renormalisation factors $Z_i$ are collected in \appn{app:olojetcomp}.

\begin{figure}
    \centering
\begin{subfigure}[b]{0.25\textwidth}
\centering
\includegraphics[width=\textwidth]{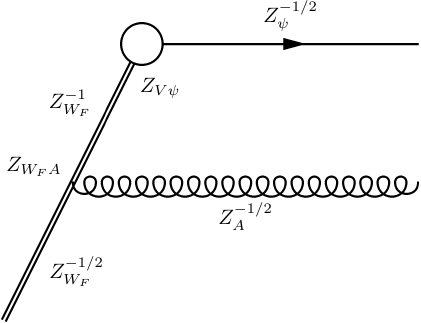}
\end{subfigure}
\hspace{1cm}
\begin{subfigure}[b]{0.25\textwidth}
\centering
\includegraphics[width=\textwidth]{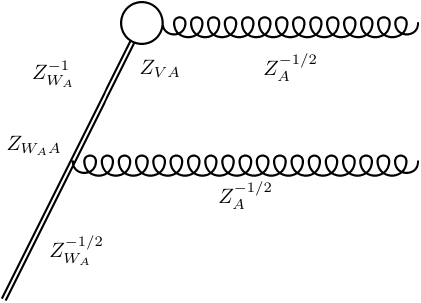}
\end{subfigure}
    \caption{Assigning renormalisation constants to different factors in the radiative 
    jet functions: radiation from the Wilson line.}
    \label{fig:jetFunctionRenorm2}
\end{figure}
Furthermore, there is another type of graph at tree-level, where a gluon is radiated from 
the Wilson line, as shown in Fig.~\ref{fig:jetFunctionRenorm2}. The corresponding 
renormalisation factors can be consistently determined by demanding that these diagrams
renormalise in the same way as those in Fig.~\ref{fig:jetFunctionRenorm}. The results
are also given in \appn{app:olojetcomp}.
\begin{figure}
    \centering
    \includegraphics[width=0.25\textwidth]{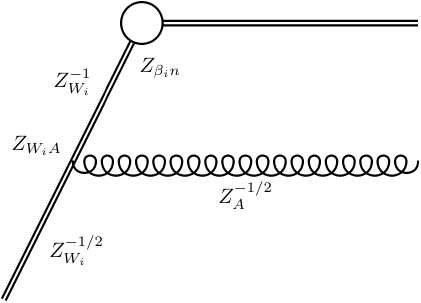}
    \caption{Assigning renormalisation constants for the eikonal radiative jet function.}
    \label{fig:eikjetFunctionRenorm}
\end{figure}
Finally, the single-radiative eikonal jet renormalisation factors are presented in 
Fig.~\ref{fig:eikjetFunctionRenorm}, showing one of the two tree-level diagrams. The 
eikonal jet thus renormalises as
\beq
\label{eikjetren}
  J_{\E_i} \big( \beta_i; k \big) \, = \, J_{\E_i}^{\text{b}} \big( \beta_i; k) 
  \left(Z_{\beta_i n} Z_{W_i A} Z_{W_i}^{- 3/2} Z_A^{-1/2} \right)^2 \, ,
\eeq
where $Z_{\beta_in}$ is the renormalisation of the vertex connecting the Wilson lines 
in directions $\beta_i$ and $n$ (see \eq{eq:Z_beta_n}). On the other hand, light-like Wilson 
lines do not carry a `wave function renormalisation', as all self-energy diagrams vanish, 
since they are proportional to $\beta_i^2 = 0$. Once renormalised in the $\overline{\rm{MS}}$ 
scheme, single-radiative jet functions can be written universally in the form
\beq
\label{eq:jetFunctionUniversal}
  && J_{f_i, f_1 f_2}^{(1)} \big( \ell_i; k_1, k_2 \big) \, = \, - \, \frac{\al_s}{2 \pi} \, 
  J_{f_i, f_1 f_2}^{(0)} \big( \ell_i; k_1, k_2 \big)
  \Bigg[
  \frac{\gamma_1^{(1)}}{\eps} + \frac{\gamma_2^{(1)}}{\eps}
  + \frac{C_1 + C_2}{\eps^2}  \\
  && \hskip18pt + \, \frac{1}{\eps} \left(
  C_i + 2 \T_1 \cdot \T_2 \, \log \frac{(k_1 \cdot k_2) n^2}{2 k_1 \cdot n \, k_2\cdot n} + 
  C_1 \log\frac{n^2 \mu^2}{4 (k_1 \cdot n)^2} + C_2 \log\frac{n^2 \mu^2}{4 (k_2\cdot n)^2}
  \right) + \mathcal{O}(\eps^0)
  \Bigg] \, , \nn
\eeq
where $\gamma_i^{(1)}$ is the one-loop collinear anomalous dimension for parton $i$, 
given below \eq{eq:SkRV}: the terms involving these anomalous dimensions are of hard-collinear
 origin and feature in the difference between radiative jet and eikonal jet functions 
arising from \eq{eq:J11Fact}. 

To complete the proof of factorisation at one-loop, we need to show that the remaining 
terms build the soft function $S_{3}^{(1)} (\beta_1, \beta_2, n)$. To this end, recall that 
$S_3^{(1)} (\beta_1, \beta_2, n)$ is a Wilson-line correlator defined with lines $\beta_1$, 
$\beta_2$ and $n$, as in \eq{eq:Snij}. In particular, at one-loop, it is the sum of three 
terms, given by
\beq
\label{eq:S3}
  S_3^{(1)} \big( \beta_1, \beta_2, n \big) & = &
  \frac{\al_s}{\pi} \, \T_1 \cdot \T_2 \left( \frac{1}{\eps^2} - \frac{1}{\eps} \log \beta_1 \cdot
  \beta_2 \right) + \frac{\al_s}{2 \pi} \, \T_1 \cdot \T_n \left[ \frac{1}{\eps^2} + 
  \frac{1}{\eps} \left( 1 + \log \frac{n^2}{2(\beta_1\cdot n)^2} \right) \right] \nn \\
  && + \, \frac{\al_s}{2 \pi} \, \T_2 \cdot \T_n \left[ \frac{1}{\eps^2} + \frac{1}{\eps}
  \left(1 + \log \frac{n^2}{2(\beta_ 2\cdot n)^2} \right) \right] \, , 
\eeq
where the first term captures the correlation between the two light-like Wilson lines 
in directions $\beta_1$ and $\beta_2$, and can be borrowed directly from eq.~(3.9) in 
Ref.~\cite{Dixon:2008gr} up to taking twice its real part. The next two terms in \eq{eq:S3}
connect one of the 
light-like lines to the line along $n$, which is off the light cone. The corresponding 
expression can be derived from eq.~({3.11}) in Ref.~\cite{Dixon:2008gr}. Using colour conservation 
in the form $\T_n = - \, \T_1 - \T_2$ we find then 
\beq
\label{eq:S31eval}
  S_3^{(1)} \big( \beta_1, \beta_2, n \big) & = & - \, \frac{\al_s}{2 \pi} \, \Bigg[
  \frac{C_1 + C_2}{\eps^2} \\
  && \hspace{-15mm} + \, \frac{1}{\eps} \left( C_{[12]} + 2 \T_1 \cdot \T_2 
  \log \frac{(\beta_1 \cdot \beta_2) \, n^2}{2\beta_1 \cdot n \, \beta_2 \cdot n} + 
  C_1 \log\frac{n^2}{ 2(\beta_1 \cdot n)^2 } + C_2 \log \frac{n^2}{2(\beta_2 \cdot n)^2}
  \right) \Bigg] \, ,  \nn
\eeq 
where $C_{[12]} \equiv (\T_1 + \T_2)^2$ is the quadratic Casimir of the parent particle
radiating 1 and 2. Using \eq{eq:S31eval} in \eq{eq:jetFunctionUniversal}, 
with $\beta_i = \sqrt{2} \, k_i/\mu$, we finally find
\beq
\label{finjetfun}
  J_{f_i, f_1 f_2}^{(1)} \big( \ell_i; k_1, k_2 \big) \, = \, 
  J_{f_i, f_1 f_2}^{(0)} \big( \ell_i; k_1, k_2 \big)
  \bigg[
   - \, \frac{\al_s}{2 \pi} \frac1{\eps} \left(\gamma_1^{(1)} + \gamma_2^{(1)} \right) + \, 
   S_3^{(1)} \big( \beta_1, \beta_2, n \big) + \mathcal{O}(\eps^0)
  \bigg] \, . \quad
\eeq
Thus, identifiying, as before, $\gamma_i^{(1)}$ as the difference between the one-loop 
jet function and its eikonal counterpart, we have indeed obtained the factorisation of the 
radiative one-loop jet function, as given in \eq{eq:J11Fact}. One can similarly verify that
\eq{eq:JE11Fact} also holds.

%%%%%%%%%%%%%%%%%%%%%%%%%%%%%%%%%%%%%%%

\section{A top-down approach to strongly-ordered counterterms} 
\label{dsocfrfvcr}

Before proceeding, we find it useful to summarise what we have achieved so far.
In \secn{ReFactoTree} we obtained expressions for strongly-ordered radiative 
functions at tree level, by applying soft and collinear factorisation in an iterative 
fashion: we conjectured the form of such hierarchical configurations to all orders,
and we verified {\it a posteriori} their correspondence with know results. On the 
other hand, in \secn{ReFactoLoop} we derived the form of soft and collinear limits 
of squared matrix elements at one loop, starting from factorisation concepts: in doing so, 
we treated soft and jet functions as generalised scattering amplitudes featuring 
Wilson lines as sources. These two constructions may seem unrelated, and based 
on different principles. However, soft and collinear limits of real-virtual contributions, 
and strongly-ordered limits of double-real contributions, must be intertwined. Indeed, 
the corresponding counterterms $K^{\RV}_\npo$ and $K_\npt^{(\mathbf{12})}$ 
(or, more precisely, its integral $I_{n+1}^{(\mathbf{12})}$) have to combine appropriately 
in order to ensure the cancellation of explicit poles in the second line of \eq{subtNNLO},
similarly to what happens with the combination of $V_{n}$ and $I_{n}^{(\mathbf{1})}$
at NLO. One can thus expect $K^{\RV}_\npo$ and $K_\npt^{(\mathbf{12})}$ to be 
connected by NLO-like completeness relations, analogous to those presented in
eqs.~(\ref{FinCondNLO1}-\ref{FinCondNLO3}). This reasoning suggests that, 
by following the {\it top-down approach} introduced in \secn{NLOSubtra}, one can 
obtain explicit expression for double-real strongly-ordered counterterms, directly 
from the factorised form of real-virtual counterterms: this is the goal of this Section.

To proceed, we insert the factorised real-virtual functions found in \secn{RaSoFuLoop} 
and in \secn{RaJeLoop} in the expression for $K^{\RV}_{n+1}$. We then apply the 
NLO completeness relations in eqs.~(\ref{FinCondNLO1}-\ref{FinCondNLO3}), and 
we arrive at the factorised strongly-ordered terms discussed in \secn{ReFactoTree}. 
From these expressions, after a preliminary analysis in \secn{sec:softRV} and in 
\secn{sec:collRV}, we proceed to derive, in \secn{sub:finding_k_12}, the expression 
for the complete counterterm $K_{n+2}^{(\mathbf{12})}$ responsible for strongly-ordered 
configurations. In Appendix~\ref{sub:checks}, we will verify that indeed our result 
coincides with the single-unresolved limit ${\bf L}^{\one}$ of $K^{(\mathbf{2})}_\npt$, 
defined in \eq{L2def}.

%%%%%%%%%%%%%%%%%%%%%

\subsection{Soft sector}
\label{sec:softRV}

In this section we show that, by applying the NLO-like completeness relations in 
eqs.~(\ref{FinCondNLO1}-\ref{FinCondNLO3}) to the factorised one-loop radiative 
soft function $S_{n,g}^{(1)}$, we arrive at NNLO finiteness conditions linking 
the real-virtual contribution to double-real strongly-ordered configurations.

We start by considering the factorised one-loop soft function given in \eq{eq:Sng1}.
Constructing the corresponding cross-section level expression, we obtain
\beq
\label{sigmalevS}
  S_{n,g}^{(1)} \big( \{\beta_i\}; k \big) & = &
  {\s_{n,g}^{(0)}}^{\dagger} \big( \{\beta_i\}; k \big)
  \left[
  S_{n+1}^{(1)} \big( \{\beta_i\}, \beta_k \big) - J_{\E_g}^{(1)} (\beta_k)
  \right]
  \s_{n,g}^{(0)} \big( \{\beta_i\}; k \big)
  \nnb\\ &&
  + \, \int\frac{d^d\ell}{(2\pi)^d} \,
  \Big( \s_{n,g}^{(0) \mu} \big( \{\beta_i\}; \ell \big) \Big)^{\dagger} \,
  J_{g,g}^{(1) \mu \nu} (\ell; k) \,\,
  \s_{n,g}^{(0)\nu} \big( \{\beta_i\}; \ell \big) \, ,
\eeq
where $\beta_{k}$ is the velocity of the radiated gluon with momentum $k$. In this form, 
we can readily apply the relations in eqs.~(\ref{FinCondNLO1}-\ref{FinCondNLO3}) 
to the inner one-loop functions on the right-hand side. Effectively, this means exchanging 
a loop function with an integrated radiative function. This gives
\beq
\label{eq:RV12cancelSoftv1}
  S_{n,g}^{(1)} \big( \{\beta_i\}; k_1 \big) & + &
  \int d \Phi(k_2) \Bigg\{
  \Big( \mathcal{S}^{(0)}_{n,g} \big( \{\beta_i\}; k_1 \big) \Big)^{\dagger}
  \left[ S_{n+1,g}^{(0)} \big( \{\beta_i\}, \beta_{k_1}; k_2 \big) - 
  J_{\E_g,g}^{(0)} (\beta_{k_1};k_2) \right]
  \mathcal{S}_{n,g}^{(0)} \big( \{\beta_i\}; k_1 \big)
  \nn \\ & + &
  \int \frac{d^d\ell}{(2\pi)^d} \, 
  \Big( \mathcal{S}^{(0) \mu}_{n,g} \big( \{\beta_i\}; \ell \big) \Big)^{\dagger} \!
  \sum_{f_1, f_2}  J_{g, f_1f_2}^{(0) \, \mu\nu} (\ell;k_1,k_2) \, \,
  \mathcal{S}_{n,g}^{(0) \nu} \big( \{\beta_i\}; \ell \big)
  \Bigg\} \, = \, \text{finite} \, .
\eeq
Eq.~(\ref{eq:RV12cancelSoftv1}) can be slightly refined by replacing momentum $k_1$ 
with the combination $k_1 + k_2$ in the tree-level soft function responsible for the harder
emission. This makes no difference in the soft limit for $k_2$, but it proves useful to 
maintain consistency in collinear limits, as discussed in Appendix \ref{sub:checks}.
With this understanding, we can now identify the strongly-ordered soft function, defined 
in \eq{softrads.o.}, as the first term in the integrand of \eq{eq:RV12cancelSoftv1}. 
We can then write
\beq
\label{eq:RV12cancelSoft}
  S_{n,g}^{(1)} \big( \{\beta_i\}; k_1 \big)  \! & + & \!
  \int d \Phi(k_2) \,
  \bigg\{ S_{n;g,g}^{(0)} \big( \{\beta_i\}; k_{[12]}; k_2 \big) - 
  \Big( \! \hspace{1pt} \mathcal{S}_{n,g}^{(0)} \big( \{\beta_i\}; k_{[12]} \big) 
  \! \Big)^{\! \dagger} J_{\E_g,g}^{(0)} (\beta_{k_1}; k_2) \, 
  \mathcal{S}_{n,g}^{(0)} \big( \{\beta_i\}; k_{[12]} \big)
  \nn \\ & + & \!
  \int \frac{d^d\ell}{(2\pi)^d} \, 
  \Big( \! \hspace{1pt} \mathcal{S}^{(0) \, \mu }_{n,g} \big( \{\beta_i\}; \ell \big) 
  \! \Big)^{\! \dagger}
  \! \sum_{f_1, f_2} J_{g, f_1f_2}^{(0) \, \mu \nu} (\ell; k_1, k_2) \, \,
  \mathcal{S}_{n,g}^{(0) \, \nu} \big( \{\beta_i\}; \ell \big)
  \Bigg\} \, = \, \text{finite} \, .
\eeq
As usual, the $k_2$ phase-space integration in the first line cancels the poles 
of $S_{n,g}^{(1)}$ originating from soft radiation at wide angles from the directions 
$\{\beta_i, k_1\}$, while the convolution on the second line cancels collinear 
poles (including soft-collinear ones, which were subtracted in the first line) 
associated with the emitted gluon. As discussed below \eq{eq:soft_fact_nlo}, 
there are no hard-collinear poles associated with the directions of the $n$ 
Wilson lines.

%%%%%%%%%%%%%%%%%%%%%

\subsection{Collinear sector}
\label{sec:collRV}

We can similarly apply the NLO finiteness conditions in eqs.~(\ref{FinCondNLO1}-\ref{FinCondNLO3}) 
to the factorised one-loop single-radiative jet function $J_{f_i,f_1f_2}^{(1)}$, which will 
lead to a finite relation between the latter and the strongly-ordered double-radiative jet 
function. We start with the factorised one-loop radiative jet in \eq{eq:J11Fact}. Applying 
the NLO completeness relations to the inner one-loop functions of that equation we find
\beq
\label{eq:compl_jet_RV}
  J_{f_i,f_1f_2}^{(1) \al_i\beta_i} (\ell_i;k_1,k_2) & + &
  \int d\Phi(k_3) \int d^dx \, e^{ i \ell_i \cdot x} \\
  && \hspace{-2cm}
  \Bigg\{
  \Big(\mathcal{J}_{f_i,f_1f_2}^{(0) \al_i} (0;k_1,k_2) \Big)^{\dagger}
  \bigg[ S_{3, f_3}^{(0)} (\beta_1,\beta_2,n;k_3) - \sum_{j \in \{1,2\} } 
  J_{\E_j, f_3}^{(0)} (\beta_j;k_3) \bigg] 
  \mathcal{J}_{f_i,f_1f_2}^{(0)\beta_i} (x; k_1, k_2) 
  \nn \\ && \hspace{-17mm} + \,
  \int \frac{d^dq}{(2\pi)^d} \, 
  \Big(\mathcal{J}_{f_i,f_1f_2}^{(0) \al_i \bar{\al}_1} (0;q,k_2)  \Big)^{\dagger}
  J_{f_q,f_1f_3}^{(0) \bar{\al}_1\bar{\beta}_1} (q;k_1,k_3) 
  \mathcal{J}_{f_i,f_1f_2}^{(0) \bar{\beta}_1 \beta_i} (x;q,k_2) 
  \nn \\ && \hspace{-17mm} + \, 
  \int \frac{d^dq}{(2\pi)^d} \, 
  \Big( \mathcal{J}_{f_i,f_1f_2}^{(0) \al_i \bar{\al}_2} (0;k_1,q) \Big)^{\dagger}
  J_{f_q,f_2f_3}^{(0) \bar{\al}_2 \bar{\beta}_2} (q;k_2,k_3)
  \mathcal{J}_{f_i,f_1f_2}^{(0) \bar{\beta}_2 \beta_i}(x;k_1,q) 
  \Bigg\} \, = \, \text{finite} \, . \nn
\eeq
In the last two lines of the integrand of \eq{eq:compl_jet_RV} one recognises the
expressions for strongly-ordered jet functions discussed in \secn{RaJeTree}, 
specifically $J^{(0)\al_i \beta_i}_{f_i; f_q f_2; f_1 f_3, f_2}$ and $J^{(0) \al_i
\beta_i}_{f_i; f_1 f_q; f_1, f_2 f_3}$. In the same way as for the soft function in 
\secn{sec:softRV}, collinear poles of $J_{f_i, f_1 f_2}^{(1)}$ are cancelled by integrating 
the strongly-ordered jet functions, while soft non-collinear poles are cancelled by
integrating the radiative soft function in the second line, where the soft-collinear 
region has been subtracted to avoid double counting.

%%%%%%%%%%%%%%%%%%%%%

\subsection{Extracting strongly-ordered counterterms}
\label{sub:finding_k_12}

After working out soft and collinear finiteness conditions, we are ready to 
derive in this section a complete, explicit expression for the strongly-ordered 
counterterm $K_{n+2}^{\otwo}$, in a top-down approach. In order to do so, we 
first write $K_{n+1}^{\RV}$ in \eq{twoloctRV} using the factorised expressions 
we have derived. Then, applying the finiteness relations derived in the previous 
sections, we can arrive at the integrated counterterm $I_{n+1}^{\otwo}$, whose 
integrand is $K_{n+2}^{\otwo}$.

We begin by focusing on the explicit poles of the soft component 
$K_{n+1}^{(\mathbf{RV}, \, \text{s})}$. They are entirely encoded in the third and last
term on the right-hand side of \eq{twoloctRVs}, and, in particular, in the radiative, 
one-loop soft function $S_{n,g}^{(1)}$. Therefore we can write
\beq
  K_{n+1}^{(\mathbf{RV}, \, \text{s})} \, = \,  
  {\mathcal{H}_n^{(0)}}^{\dagger} \, S_{n,g}^{(1)} \, \mathcal{H}_n^{(0)} \, + \, \text{finite} \, .
\eeq
Using the finiteness condition in \eq{eq:RV12cancelSoftv1}, or equivalently in 
\eq{eq:RV12cancelSoft}, we can identify the integrals on the left-hand sides of these
equations as the integrated strongly-ordered counterterm, $I_{n+1}^{(\mathbf{12}, \, 
\text{s})}$. This provides an ansatz for the corresponding integrand functions as 
$K_{n+2}^{(\mathbf{12}, \, \text{s})}$. In Appendix~\ref{sub:checks} we will verify 
that this ansatz indeed corresponds to the single-soft limit of the double-unresolved 
counterterm $K_{n+2}^{(\mathbf{2}, \, \text{s})}$. We get
\beq
\label{eq:K12s_2}
  K_{n+2}^{(\mathbf{12}, \, \text{s})}
  \, = \,
  {\mathcal{H}_n^{(0)}}^{\dagger} 
  \sum_{f_1,f_2}
  \bigg[
  S^{(0)}_{n;f_{[12]},f_2} \big( k_{[12]}, k_2 \big)
  +
  S^{(0)}_{n,f_{[12]}}
  \Big(
  J_{f_{[12]},f_1f_2}^{(0)}
  -
  J_{\E_{[12]},f_2}^{(0)}
  \Big)
  \bigg] \,
  \mathcal{H}_{n}^{(0)}
  \, ,
\eeq
where we have understood spin indices and, as discussed in \secn{sec:softRV}, we
have included a set of non-singular contributions by attributing the total soft momentum 
to the harder gluon emission. In writing \eq{eq:K12s_2}, starting from \eq{eq:RV12cancelSoftv1}, 
we have used the fact that jet functions are colour singlets, and thus they commute with 
soft functions, which carry colour structure; furthermore, for simplicity, we are understanding 
the convolution in the parent gluon momentum $k_{[12]}$ that was explicitly written
in \eq{eq:RV12cancelSoftv1}. 

We now turn to the hard-collinear component, and examine $K^{({\bf RV}, \, 
{\rm hc})}_{\npo, \, i}$ in \eq{twoloctRVhci}. To manipulate this term, it is convenient
to use the factorised expressions for jet and eikonal jet functions derived in
\secn{RaJeLoop}. In particular, we can consider \eq{eq:J11Fact}, and then obtain 
the corresponding result at cross-section level.  For the one-loop radiative eikonal 
jet function we can directly use the result in \eq{eq:JE11Fact}. Exploiting the definition 
in \eq{eq:hc_jet_double_real}, we can write the real-virtual counterterm for the 
hard-collinear sector associated with particle $i$ as
\beq
\label{KRVhci1}
  K^{(\mathbf{RV}, \, \text{hc})}_{n+1, \,  i}
  & = &  
  {\hard_n^{(0)}}^{\dagger}
  \sum_{f_1, f_2}
  J_{f_i, f_1 f_2}^{(0), \, \text{hc}} \,
  \Big[
  S_3^{(1)} - J_{\E_i}^{(1)} +
  \sum_{k=1}^2 J_{f_k, f_k}^{(1), \, \text{hc}}
  \Big] \,
  {\cal H}^{(0)}_n
  \, ,
\eeq
where for simplicity, we have understood the necessary convolution, as before, 
as well as the spin structure, which was displayed in detail in \eq{eq:JE11Fact} 
and in \eq{eq:compl_jet_RV}. Recall that \eq{KRVhci1} describes the one-loop 
correction to the hard-collinear splitting of Born-level particle $i$ into two particles 
of flavours $f_1$ and $f_2$, summed over the consistent flavour channels: thus 
the soft function $S_3$ involves the Wilson lines associated with external legs $1$ 
and $2$, and the Wilson line for the reference vector $n_i$ of the jets along the 
direction of leg $i$. Now, using the finiteness conditions, we can derive the 
corresponding counterterm for the strongly-ordered hard-collinear configuration. 
It can be written as
\begin{eqnarray}
\label{eq:K12hci}
  K_{n+2,i}^{(\mathbf{12}, \, \text{hc})}
  & = &
  {\hard_n^{(0)}}^{\dagger} \!\!\!
  \sum_{f_1,f_2,f_3}
  \bigg[
  J_{f_i,f_1f_2}^{(0), \, \text{hc}}(\bar k_1,\bar k_2) \, S_{3,f_3}^{(0)} -
  J_{f_i,f_1f_2}^{(0), \, \text{hc}}(k_1,k_2) \, J_{\E_i,f_3}^{(0)}
  \nnb \\
  &&
  \qquad\qquad
  + \!\!\!
  \sum_{kl=\{12,21\}}
  J_{f_i,f_{[k3]} f_l}^{(0), \, \text{hc}} \,
  \Big(
  J_{f_{[k3]},f_kf_3}^{(0)} - J_{\E_k,f_3}^{(0)}
  \Big)
  \bigg] \,
  \hard_n^{(0)}
  \, .
\end{eqnarray}
In analogy to what was done in the soft sector, we have introduced a shift
of the collinear momenta, so that the hardest splitting carries the total radiated 
momentum, according to $\bar k_1 + \bar k_2 = k_1+k_2+k_3$, while the soft
function $S_3$ is built with $\bar \beta_1$, $\bar \beta_2$, and $n_i$. This 
reparametrisation does not affect singular contributions.

We now turn to the hard-collinear real-virtual counterterm $K_{n+1,ij}^{(\mathbf{RV}, 
\, \text{hc})}$ in \eq{twoloctRV1hcij}, featuring the hard-collinear radiation from 
two distinct Born-level partons. In this case, loop corrections and real radiation 
affect separately parton $i$ and $j$, and they are not intertwined. Then we can 
simply apply the completeness relations to the one-loop jet function following 
\eq{FinCondNLO3}, to obtain
\beq
\label{eq:K12hcij}
  K_{n+2,ij}^{(\mathbf{12}, \, \text{hc})} \, = \, 
  {\hard_n^{(0)}}^{\dagger}
  \sum_{f_1,f_2,f_3,f_4}
  J_{f_i,f_1f_2}^{(0), \, \text{hc}} \,
  J_{f_j,f_3f_4}^{(0), \, \text{hc}} \,
  \hard_n^{(0)}
  \, +  \, (i \leftrightarrow j)
  \, .
\eeq
which is the same as \eq{CandCountNNLO3}. The final expression is simple, since
for two uncorrelated limits `democratic' and strongly-ordered counterterms coincide. 
Note however that, in a concrete implementation, one needs to complement \eq{eq:K12hcij} 
with appropriate phase-space mappings, as discussed in Appendix \ref{scollcanc}. For example, 
when checking that collinear limits of $K^{\two}$ coincide with those of $K^{\otwo}$, 
one must assume the corresponding mappings to behave in the same way. A similar
structure emerges in the case of $K_{n+1,i}^{(\mathbf{RV},\text{1hc},\text{1s})}$,
presented in \eq{twoloctRV1hc1s}, describing singular configurations involving one 
soft and one collinear limit. Also in this case, the contributions of virtual and real 
origin are factorised, leading to a particularly simple expression, which is expected 
as a consequence of QCD colour coherence: soft emissions do not resolve individual 
colour charges of the decay products, just the total colour charge of the emitters. 
Applying completeness to \eq{twoloctRV1hc1s} we obtain
\beq
\label{eq: k12 1hc 1s}
  K_{n+2,i}^{(\mathbf{12},\text{1hc},\text{1s})} \, = \,
  {\hard_n^{(0)}}^{\dagger}  \sum_{f_1, f_2, f_3}
  \sum_{jkl \in \{123,312,231\}} \,
  J^{(0), \, \text{hc}}_{f_i, f_k f_l} 
  \, S_{n, f_j}^{(0)} \, \hard_n^{(0)} 
  \, ,
\eeq
reproducing \eq{CandCountNNLO4}, which is factorised, and thus effectively strongly 
ordered. The last term of \eq{twoloctRV}, $K^{({\bf RV}, \, {\rm 1hc})}_{\npo, \, i}$, is
presented in \eq{twoloctRV1hc}: its loop content is entirely carried by the one-loop
resolved amplitude ${\cal H}^{(1)}_n$, which is finite in $d=4$ dimensions. As a 
consequence, there is no $K_{n+2}^{(\mathbf{12})}$ contribution stemming from 
this term.

We are now in a position to assemble the different contributions and arrive at the 
final expression of the strongly-ordered double-unresolved counterterm. This reads
\beq
\label{eq:K12full}
  K_{n+2}^{(\mathbf{12})}
  \,= \,
  K_{n+2}^{(\mathbf{12}, \, \text{s})}
  +
  \sum_{i=1}^n
  \bigg[
    K_{n+2,i}^{(\mathbf{12}, \, \text{hc})} 
    + \sum_{j=i+1}^n K_{n+2,ij}^{(\mathbf{12}, \, \text{hc})}
    + K_{n+1,i}^{(\mathbf{12}, \, \text{1hc}, \, \text{1s})} 
  \bigg] \, ,
\eeq
which is structurally analogous to \eq{OrgaDU} and \eq{twoloctRV}. The results 
given above for individual contributions to \eq{eq:K12full} are obtained by exploiting 
completeness relations, starting from the factorised form of the real-virtual 
counterterm. This represents an alternative approach with respect to the one in 
\secn{ReFactoTree}, which is based on the iterative factorisation of subsequent 
emissions. As a crucial validation of our arguments, we need to verify that the two 
results agree. In particular, we have to check that \eq{eq:K12full} reproduces the 
iterative limits of double-real matrix elements, constructed by taking single-unresolved 
soft and collinear limits of the double-real counterterm. In formulae, we have to 
prove the relation $\mathbf{L}_1\big[K_{n+2}^{\two} \big] = \mathbf{L}_1\big[ 
K_{n+2}^{\otwo} \big]$, given the definition of $\mathbf{L}_1$ in \eq{L1def}, the 
expression for $K_{n+2}^{\two}$ reported in \eq{OrgaDU}, and the results in 
\eq{eq:K12full}. The required calculation is successfully carried out in 
Appendix~\ref{sub:checks}. 

%%%%%%%%%%%%%%%%%%%%%

\section{Summary and future prospects}
\label{sec:summary_and_future_prospects}

In this paper we have outlined a general procedure to identify local counterterms
capturing the singular behaviour of real-radiation squared matrix elements, including the
case of strongly-ordered and nested limits. The organisation of the relevant counterterms 
has been presented at NNLO, and sketched at N$^3$LO in Section~\ref{Archi}, following 
the spirit of previous studies performed in the context of the {\it local analytic sector 
subtraction}~\cite{Magnea:2018hab, Magnea:2018ebr, Bertolotti:2022aih}. The general
structure of this subtraction algorithm turns out to be remarkably transparent, and allows 
for a straightforward generalisation to higher orders, at least so far as a formal analysis 
is concerned. We note also that the number and complexity of independent counterterms grow 
exponentially with the perturbative order (see \eq{countcount}), but remain reasonably 
limited for the perturbative orders that are expected to be of phenomenological interest.

Having established the subtraction framework, in \secn{FactoSubtra} we went on to construct 
all the local counterterms that are needed for the NNLO subtraction procedure, relying on 
the established factorised structure of infrared poles of virtual corrections to scattering 
amplitudes. This approach was first exploited in Ref.~\cite{Magnea:2018ebr}, and it builds 
on the cancellation of infrared singularities between real and virtual corrections, ensured 
by general cancellation theorems. At cross-section level, soft and collinear functions
responsible for virtual poles are used as building blocks for inclusive cross-section-like
quantities that can be shown to be infrared finite by power counting: throughout the paper, 
we have referred to these constructions as {\it completeness relations}, since finiteness is
achieved by performing an inclusive sum over a complete set of radiation states. These 
soft and collinear cross sections then contain integrated real-radiation contributions that 
cancel virtual poles by construction: their integrands are thus readily understood as local 
soft and collinear counterterms.

We note that our approach essentially reverses the standard logic guiding the construction of 
most established infrared subtraction schemes. Typically, these schemes are based on three main 
steps: {\it i)} the identification of the real-emission subtraction terms, {\it ii)} the analytic integration 
of the counterterms over the unresolved phase space, leading to explicit poles, and {\it iii)} 
the proof of pole cancellation between integrated counterterms and virtual corrections. 
The idea of starting from the infrared structure of virtual corrections to infer the form of the 
counterterms has been recently considered by other groups, in the context of {\it nested 
soft-collinear subtraction}~\cite{Devoto:2023rpv} and {\it antenna subtraction}~\cite{Gehrmann:2023dxm}. 
We stress that in this paper we have moved a further step forward with respect to simply 
using the singularity structure of virtual amplitudes as a guideline. We have, indeed, taken 
advantage of the detailed infrared factorisation properties of fixed-angle massless amplitudes, 
and we have derived expressions for all relevant real-radiation counterterms, in a fully general 
fashion. As explained in \secn{Archi}, and noted above, we derive expressions for local counterterms 
as matrix elements of fields and Wilson lines by ``completing" the corresponding soft, collinear 
and soft-collinear building blocks that appear in the factorisation formula for virtual amplitudes. 
This approach to the definition of counterterms is, by construction, valid to all orders in perturbation 
theory and can accomodate an arbitrary number of real emissions. 

Starting from \secn{ReFactoTree}, we focused on the issue of disentangling and modelling 
strongly-ordered configurations, that are not directly manifest in the expressions for multiple 
real-radiation counterterms, which are based on uniform infrared limits. Strongly-ordered
counterterms are very important, not only because they proliferate as the perturbative order 
increases, as noted in \secn{Archi}, but also because their integrals play the delicate role of
cancelling the poles of mixed real-virtual contributions, without upsetting the balance of singular 
phase space limits for the  remaining unintegrated radiation. In all existing approaches, this
subtle cancellation has been engineered on a case-by-case basis, fine-tuning the structure 
of individual counterterms. Our aim in this paper has been to provide a systematic method
to secure this cancellation in full generality. To this end, we first made educated guesses to 
conjecture expressions for the relevant hierarchical configurations, to all orders, in terms of 
expectation values of appropriate combinations of fields and Wilson lines. Then we checked 
for agreement with results known from the literature. Next, in Section~\ref{ReFactoLoop}, 
we derived factorised expressions for soft and collinear limits of squared matrix elements
involving both real radiation and virtual corrections. With these expressions in hand, as
explained in \secn{dsocfrfvcr}, one can exploit NLO-like completeness relations to 
connect real-virtual counterterms to integrals involving multiple real radiations. These
integrals are, in turn, identified with contributions to the integrated strongly-ordered 
counterterm $I_{n+1}^{(\mathbf{12})}$. Completeness relations thus link the real-virtual
counterterm $K^{\RV}_{n+1}$ with the strongly-ordered double-real counterterm 
$K_{n+2}^{(\mathbf{12})}$, in a way that automatically  ensures the cancellations foreseen 
in our approach. We have explicitly checked that the results obtain by means of the 
``completeness procedure" (applied to the real-virtual counterterm) agree with direct 
iterated limits of double-real matrix elements. 

We emphasise that our results do not immediately translate into a concrete subtraction
algorithm, because we have assumed, but not concretely implemented, the phase-space
mappings that are necessary for all singular limits involving collinear splittings, and that
must be chosen in a consistent way when multiple nested limits are considered. These 
issues are discussed in detail in Ref.~\cite{Bertolotti:2022aih}: we believe that the 
approach presented in this paper can lead to a significant simplification of the intricate
structure of the mappings employed there.

The method we have introduced naturally lends itself to various extensions. First of all, 
the inclusion of initial-state radiation can be devised without posing, in principle, new 
theoretical challenges: indeed, there are no significant conceptual differences between
soft and jet functions involving initial and final state particles. Similarly, the extension
of the method to massive quarks (which is of considerable interest in phenomena related 
to top-quark observables, and in connection with $b$-quark mass effects) does not pose 
new conceptual issues at the level of the definition of local counterterms. However, 
in the case of massive partons, defining phase-space mappings related to their 
branchings~\cite{Catani:2002hc} and performing the corresponding integrations, 
will necessitate further work. Finally, we emphasise that the approach we have presented 
is likely to significantly influence the structuring of general N$^3$LO subtraction algorithms. 
Work is underway to explore these generalisations.

%%%%%%%%%%%%%%%%%%%%%%%%%%%%%%%%%%%%%%%

\section*{Acknowledgements}

\noindent 
We are grateful to Sandro Uccirati for participating in the initial stages of this project and 
for many useful discussions. PT has been partially supported by the Italian Ministry of 
University and Research (MUR) through grant PRIN 2022BCXSW9, and by Compagnia 
di San Paolo through grant TORP\_S1921\_EX-POST\_21\_01.

%%%%%%%%%%%%%%%%%%%%%%%%%%%%%%%%%%%%%%%

\vspace{1cm}

%%%%%%%%%%%%%%%%%%%%%%%%%%%%%%%%%%%%%%%

\appendix

%%%%%%%%%%%%%%%%%%%%%%%%%%%%%%%%%%%%%%%

\section{Soft and collinear cancellations at NLO}
\label{scollcanc}

In this appendix we present a detailed calculation verifying the cancellations implied 
by the NLO completeness relation in eqs.~(\ref{FinCondNLO1}-\ref{FinCondNLO3}). 
As we will see, this requires the introduction of an appropriate regulator to cure the 
(unphysical) UV divergences of the relevant matrix elements, and a choice of phase-space
mapping in the collinear case. We discuss soft and collinear cancellations in 
Appendices~\ref{sec:softcanc} and~\ref{sec:collcanc} respectively.

%%%%%%%%%%%%%%%%%%%%%

\subsection{Soft cancellation}
\label{sec:softcanc}

The cancellation stated in \eq{FinCondNLO1} occurs, as usual, between a virtual term, 
the one-loop soft function $S_{n}^{(1)}$, and a real-radiation term, the integrated radiative 
soft function $\int d\Phi(k) S_{n,g}^{(0)}$. The virtual contribution is understood to be 
renormalised, so it is affected only by infrared (soft and collinear) divergences (indeed,
the bare Wilson-line correlator vanishes beyond tree level, since it is constructed out of 
scale-less integrals). The phase-space integral of the real-radiation contribution, on the 
other hand, diverges at large values of the radiated momentum, since the soft gluon 
energy is not constrained by momentum conservation in the eikonal approximation.

In order to regulate this UV singularity, we introduce a soft cross section at {\it fixed total 
final-state energy}, following~\cite{Laenen:2000ij}, and then we integrate over this energy 
up to a finite cutoff, matching the scale of the virtual correction. Concretely, we displace one 
of the two Wilson line correlators defining the soft cross section by a finite timelike vector 
$y^\mu = \{y, {\bf 0} \}$, and we introduce the finite quantity
\beq
\label{eq:app_compl_soft}
  \int_0^\lambda d \xi \, 
  \int\frac{dy}{2\pi} \, 
  e^{{\rm i} \xi y} \, 
  \bra{0} \,  \overline{T} \bigg[ \prod_{i = 1}^n 
  \Phi_{\beta_i} (0, \infty) \bigg] T \bigg[ \prod_{i = 1}^n 
  \Phi_{\beta_i} (\infty, y) \bigg] \ket{0}
  \, = \,
  \text{finite} \, .
\eeq
The timelike displacement $y$ corresponds to the Fourier conjugate of a cutoff on the 
energy of the final state, represented here by $\lambda$. By letting $y \to 0$ in the 
correlator, or by letting the $\xi$ integration run unconstrained, one recovers the 
{\it r.h.s.} of \eq{CompleteSoft}. Expanding \eq{eq:app_compl_soft} to NLO (see 
also the \emph{l.h.s.} of \eq{CompleteSoft}), we obtain the completeness relation
\beq
\label{eq:softCompletenessRelation}
  && \int_0^\lambda d \xi
  \int \frac{dy}{2\pi} \, 
  e^{{\rm i} \xi y} \, \Bigg\{
  \braket{0| \,  \overline{T} \bigg[ \prod_{i = 1}^n \Phi_{\beta_i} (0, \infty) \bigg] |0}
  \braket{0| \, T \bigg[ \prod_{i = 1}^n \Phi_{\beta_i} (\infty, y) \bigg] |0} 
  \bigg \rvert_{\text{one-loop}}
  \nn \\ & & \hspace{1cm} + \, 
  \int d \Phi(k)
  \braket{0| \,  \overline{T} \bigg[ \prod_{i = 1}^n \Phi_{\beta_i} (0, \infty) \bigg] |k}
  \braket{k| \, T \bigg[ \prod_{i = 1}^n \Phi_{\beta_i} (\infty, y) \bigg] |0}
  \bigg\rvert_{\text{tree}}
  \Bigg\} \, = \, \text{finite} \, , 
\eeq
where $d\Phi(k)$ is the phase-space measure of the radiated gluon. 
Using translational invariance of the vacuum, which implies
\beq
\label{traslinv}
  \braket{0 |\, T \bigg[ \prod_{i = 1}^n \Phi_{\beta_i} (\infty, y) \bigg] |0} \, = \, 
  \braket{0 |\, T \bigg[ \prod_{i = 1}^n \Phi_{\beta_i} (\infty, 0) \bigg] |0} \, ,
\eeq 
the first term can be shown to equal $S_{n}^{(1)}$. The second term, on the other hand, 
evaluates to $\int d \Phi(k)\Theta(\lambda- k ^0) S_{n,g}^{(0)}$, where $k^0$ is the 
energy component of the total final-state momentum. This modified version of the 
completeness relation in \eq{FinCondNLO1} makes the dependence on the energy 
cutoff $\lambda$ manifest, as we can write
\beq
\label{eq:softCompleteModify}
  S_{n}^{(1)} + \int d \Phi(k) \, \Theta(\lambda - k^0) \,S_{n,g}^{(0)} \, = \, \text{finite} \, .
\eeq
We can now proceed with the explicitly validation of \eq{eq:softCompleteModify}. First 
we need to obtain a cross-section-level expression for the single-radiative soft function.
This can be easily achieved by squaring the result in \eq{eq:Sng0}, and summing over 
the polarisation and colour degrees of freedom relevant for the emission of momentum 
$k$. We obtain, as expected, the well-known result
\beq
\label{wellknown}
  S_{n,g}^{(0)} \, = \, - \, g_s^2 \mu^{2 \eps} \sum_{i \neq j = 1}^n
  {\bf T}_i \cdot  {\bf T}_j \, \frac{\beta_i \cdot \beta_j}{(\beta_i \cdot k)(\beta_j \cdot k)} \, .
\eeq
Next, we perform the phase-space integration over $d \Phi(k)$ by parametrising the 
momentum of the radiated gluon as $k^\mu = x \beta_i^\mu + y \beta_j^\mu + k_\perp^\mu$. 
The integrand and the measure in \eq{eq:softCompleteModify} become then
\beq
  d \Phi(k) \, \Theta(\lambda - k^0) \, S_{n,g}^{(0)} & = & 
  - \, 2 \al_s \mu^{2 \eps} \sum_{i \neq j = 1}^n {\bf T}_i \cdot  {\bf T}_{j} \, 
  \frac{d x}{x} \, \frac{d y}{y} \, \frac{d^{d - 2} k_\perp}{(2 \pi)^{d - 2}} \, 
  \delta( 2 \, x y \, \beta_i \cdot \beta_j + k_\perp \cdot k_\perp)
  \nn \\
  & & \times \, \Theta( x \beta_i^0 + y \beta_j^0) \, 
  \Theta( \lambda - x \beta_i^0 - y \beta_j^0) \, .
\eeq
The integrals over $x$, $y$ and $k_\perp$ can be performed to arrive at
\beq
  \int d \Phi(k) \, \Theta( \lambda - k^0) \, S_{n,g}^{(0)} \, = \, - \frac{\al_s}{2 \pi}
  \sum_{i \neq j = 1}^n {\bf T}_i \cdot  {\bf T}_j \left[ \frac1{\eps^2} - 
  \frac1{\eps} \, \ln \frac{4 \lambda^2}{\mu^2} + \mathcal{O}(\eps^0) \right] \, .
\eeq
The virtual soft function $S_{n}^{(1)}$ is given by the first line on the \emph{r.h.s.} 
of \eq{eq:Snp1},
\beq
\label{softvirtagain}
  S_{n}^{(1)}
  & = &
  \frac{\as}{2\pi} \,
  \sum_{i \neq j = 1}^n {\bf T}_i \cdot  {\bf T}_j \,
  \bigg[
  \frac1{\eps^2}-\frac1{\eps} \ln \frac{2 \, p_i \cdot p_j}{\mu^2}
  \bigg] \, .
\eeq
After identifying $\lambda^2 = p_i \cdot p_j/2$, we have satisfied \eq{eq:softCompleteModify}. 
Note that the choice of the cutoff $\lambda$ is arbitrary: indeed, the numerical factor in 
the argument of the logarithm in \eq{softvirtagain} can be freely chosen by rescaling the
momenta $p_i$ and $p_j$. This ambiguity is cancelled by an equivalent calculation for 
the eikonal jet function in \eq{FinCondNLO2}, where the hard momentum can be similarly 
rescaled~\cite{Dixon:2008gr,Gardi:2009qi}.

%%%%%%%%%%%%%%%%%%%%%

\subsection{Collinear cancellation}
\label{sec:collcanc}

The cancellation of collinear poles in \eq{FinCondNLO3} has a different subtlety with 
respect to the soft cancellation discussed in \secn{sec:softcanc}. Indeed, \eq{FinCondNLO3} 
understands an explicit mapping to go from the $(n+1)$-particle phase space of the 
real-radiation matrix element to the $n$-particle phase space of the virtual matrix element. 
In particular, we seek an expression that is locally finite in the $n$-body phase space: 
to be precise, we will prove a condition of the form
\beq
\label{eq:JetCompleteConcrete}
  d \Phi(\bar{k}_r) \, d \Phi( k_{12} ) \, J_{f, f_1}^{(1)} \big(  \bar \ell; k_{12} \big)
  +
  d \Phi( k_r) \, d \Phi(k_1) \int d \Phi(k_2) J_{f, f_1 f_2}^{(0)} \big(\ell; k_1, k_2 \big)
  \, = \, \text{finite} \, .
\eeq
This describes the splitting $k_{12} \to k_1 + k_2$, with momentum $k_r$ acting as 
a spectator. Note that all the momenta in \eq{eq:JetCompleteConcrete} are on-shell and 
massless, except the `parent' momenta $\ell$ and $\bar{\ell}$, which are fixed by the 
$\delta$ functions implicit in the jet definitions. The advantage of \eq{eq:JetCompleteConcrete},
as compared to \eq{FinCondNLO3}, is that it is local in $k_{12}$ and $k_1$, {\it i.e.} we 
only integrate the momentum of the radiated parton, $k_2$. We will now show that 
\eq{eq:JetCompleteConcrete} holds, using as an example the splitting $q \to qg$.

We start with the second term, where the integrand (discussed in Appendix \ref{app:jetFuncs})
is given by~\cite{Magnea:2018ebr} 
\beq
\label{eq:Jqqg0}
  J_{q,qg}^{(0)} \big(\ell; k_1, k_2 \big) & = & \frac{4 \pi \al_s C_F}{\kOnekTwo} \, 
  (2 \pi)^d \delta^d( \ell - k_1 - k_2) \, 
  \Bigg\{ \left[1 + \frac{2 \kOnen}{\kTwon} - \frac{\kOnekTwo}{\kTwon^2} \, n^2 \right] 
  \slashed{k}_1
  \nn \\ && \hspace{1cm} + \, \left[1 - \eps + \frac{\kOnen}{\kTwon} \right] \slashed{k}_2 - 
  \frac{\kOnekTwo}{\kTwon} \slashed{n} \Bigg\} \, , 
\eeq
where we keep $n^2 \neq 0$ to regulate spurious collinear singularities. We note that 
the above expression reduces to the usual Altarelli-Parisi splitting kernel $P_{qg}$ 
upon taking the collinear $k_1|| k_2$ limit. Next, we apply the Catani-Seymour 
mapping $(k_1, k_2, k_r) \to (k_{12}, \bar k_r)$~\cite{Catani:1996vz},
\beq
\label{CSmap}
  k_{12} & = &
  k_1 + k_2 - \frac{s_{12}}{s_{1r} + s_{2r}} \, k_r \, ,
  \nn \\
  \bar k_r & = &
  \frac{s_{12} + s_{1r} + s_{2r}}{s_{1r} + s_{2r}} \, k_r \, ,
  \nn \\
  \delta^{(d)} \left(\ell - k_1 - k_2 \right) 
  & = &
  \delta^{(d)} \left(\bar \ell - k_{12} \right) \, ,
\eeq
where $s_{ir} = 2 k_i \cdot k_r$ for $i=1,2$, and $\ell$ is mapped to $\bar \ell$ so 
that the $\delta$ functions align in \eq{eq:JetCompleteConcrete}. The jacobian of this 
change of variables is
\beq
  d \Phi(k_r) \, d \Phi(k_1) \, d \Phi(k_2)
  & = &
  d \Phi(\bar k_r) \, d \Phi(k_{12}) \, d \Phi(k_2)
  \\ && \hspace{-2cm} \times \,
  \bigg( \frac{s_{1r} + s_{2r}}{s_{12} + s_{1r} + s_{2r}} \bigg)^{\! d-3}
  \bigg( \frac{s_{1r} + s_{2r}}{s_{1r}} \bigg) \, 
  \Theta \bigg( \frac{s_{1r}}{s_{1r} + s_{2r}} \bigg) \, 
  \Theta \bigg( \frac{s_{1r} + s_{2r}}{s_{12} + s_{1r} + s_{2r}} \bigg) \, . \nn
\eeq
As the vector $n$ in the jet function definition is arbitrary, we simplify the integration by  
choosing $n = k_{12} + \bar k_r$, so that $n^2 \neq 0$. The integration then depends on 
only two light-like momenta, $\bar k_r$ and $k_{12}$, instead of three. We use them to 
parameterise $k_2$ as
\beq
  k_2 \, = \, z \, k_{12} + y \, (1 - z) \bar k_r + k_{2 \perp} \, .
\eeq
Our phase space is then
\beq
  d \Phi(k_r) \, d \Phi(k_1) \, d \Phi(k_2)
  \, = \,
  d \Phi(\bar k_r) \, d\Phi(k_{12}) \, d\Phi(k_2) \, 
  \frac{(1 - y)^{1 - 2 \eps}}{1 - z} \, \Theta(1 - z) \, \Theta(1-y) \, .
\eeq
Integrating \eq{eq:Jqqg0} over $z$, $y$ and $k_{2\perp}$ we arrive at
\beq
  && d \Phi(k_r) \, d \Phi(k_1) \int d \Phi(k_2) \, J_{f, f_1 f_2}^{(0)} \big(\ell; k_1, k_2 \big)
  \, = \,
  (2 \pi)^d \, \delta^{(d)} (\bar\ell - k_{12}) \, d \Phi(k_{12}) \, d \Phi(\bar k_r)
  \nn \\
  && \qquad \qquad \qquad \qquad
  \times \, \frac{\al_s C_F}{2 \pi}  \Bigg\{ \left[ \frac1{\eps^2} + \frac1{\eps}
  \left(\frac52 - \log \frac{\mu^2}{2 \, k_{12} \cdot \bar k_r} \right) \right] \slashed{k}_{12}
  + \mathcal{O}(\eps^0) \Bigg\} \, .
\eeq
Reinstating the dependence on an arbitrary $n$ by writing $k_{12} \cdot\bar k_r \, = \, 
2 (k_{12} \cdot n)^2/n^2$, we recognise in the second line the structure of the one-loop 
virtual jet function $J_{q,q}^{(1)}$, which is given by~\cite{Dixon:2008gr}
\beq
  J_{q,q}^{(1)} \big(\bar\ell; k_{12} \big)
  \, = \,
  - (2\pi)^d \delta^{(d)} \big( \bar\ell - k_{12} \big) \,
  \frac{\al_s C_F}{2 \pi} \left[\frac1{\eps^2} + \frac1{\eps} 
  \left(\frac52 - \log \frac{n^2 \mu^2}{(2 k_{12} \cdot n)^2} \right) + 
  \mathcal{O}(\eps^0) \right] \slashed{k}_{12} \, , \quad
\eeq
which completes the proof of the cancellation of collinear poles according to 
\eq{eq:JetCompleteConcrete}.

%%%%%%%%%%%%%%%%%%%%%%%%%%%%%%%%%%%%%%%

\section{Tree-level radiative jet functions for different partonic processes}
\label{app:jetFuncs}

In this Appendix we give explicit expressions for tree-level single-radiative jet functions for
QCD partonic processes. In the process, we recover the Altarelli-Parisi splitting functions 
by considering the collinear limit of the results, expressed as the limit $k_\perp\to 0$ in 
the Sudakov parametrisation of radiative momenta,
\beq
  k_1^\mu \, = \, z \,p^\mu + k_\perp^\mu -  
  A \, n^\mu , \qquad 
  k_2^\mu \, = \, (1-z) \, p^\mu -  k_\perp^\mu - B \, n^\mu \, ,
\label{Sudapp}
\eeq
where $p$ is the collinear light-like direction, and
\beq
  A \, = \,
  - \frac{z \, p\cdot n}{n^2}
  \left(
  1 - \sqrt{1-\frac{n^2 \, k_\perp^2}{(z p\cdot n)^2}}
  \, \right) \, ,
  \qquad
  B \, = \, A|_{z \to 1 - z} \, .
\eeq
We have checked that a similar, lengthier calculation for double-radiative process yields the 
Catani-Grazzini~\cite{Catani:1999ss} double splitting kernels for all the relevant partonic 
processes.

For the $q \to qg$ process the calculation yields \eq{eq:Jqqg0}, which we reproduce here 
for convenience:
\beq
\label{J0qg}
  J_{q,qg}^{(0)}(\ell; k_1, k_2) & = &  \frac{4 \pi \al_s C_F}{\kOnekTwo} \, (2\pi)^d
  \delta^d(\ell - k_1 - k_2) \, 
  \Bigg\{ \left[ 1 + \frac{2 \kOnen}{\kTwon} - \frac{\kOnekTwo n^2}{\kTwon^2} \right]
  \slashed{k}_1
  \nn \\
  && \hspace{2cm} + \, \left[1 - \eps + \frac{\kOnen}{\kTwon} \right] \slashed{k}_2 - 
  \frac{\kOnekTwo}{\kTwon} \slashed{n} \Bigg\} \, .
\eeq
In the limit $k_\perp \to 0$ one recovers the $P_{qg}$ splitting function:
\beq
\label{limJ0qg}
  J_{q,qg}^{(0)} (\ell;k_1,k_2) 
  & = &
  \frac{4 \pi \al_s}{\kOnekTwo} \, C_F
  \bigg[ \frac{1 + z^2}{1 - z} - \eps(1 - z) \bigg] \, 
  \slashed{p} \, (2\pi)^d \delta^{(d)}(\ell - p) \, + \, {\cal O}(k_\perp^0)
  \nnb \\ & = &
  \frac{4 \pi \al_s}{\kOnekTwo} \, P_{qg} (z) \, 
  \slashed{p} \, (2\pi)^d \delta^{(d)}(\ell - p) \, + \, {\cal O}(k_\perp^0) \, .
\eeq
For a gluon radiating a quark-antiquark pair, summing over massless flavours, we find
\beq
\label{J0qqbar}
  J_{g,q\bar q}^{(0)} (\ell; k_1, k_2) & = & \frac{4 \pi \al_s n_f T_R}{\kOnekTwo} \, 
  \Bigg[ - g^{\mu\nu} + \frac{2 \kOnen \kTwon - n^2 \kOnekTwo}{\kOnekTwo (\ell \cdot n)^2}
  \left(k_1^\mu k_2^\nu + k_1^\nu k_2^\mu\right) \nn \\
  & & - \, \frac{n^2 \kOnekTwo (k_1^\mu k_1^\nu + k_2^\mu k_2^\nu) + 
  2 \kTwon^2 k_1^\mu k_1^\nu + 2 \kOnen^2 k_2^\mu k_2^\nu}{\kOnekTwo 
  (\ell \cdot n)^2} \nn \\
  & & + \, \frac{1}{(\ell \cdot n)} \left(k_1^\nu n^\mu + k_2^\nu n^\mu + k_1^\mu n^\nu + 
  k_2^\mu n^\nu\right) \Bigg] \, (2\pi)^d \delta^{(d)}(\ell - k_1 - k_2) \, .
\eeq
In the collinear limit, we recover the $P_{q \bar q}$ splitting function:
\beq
\label{limJ0qqbar}
  J_{g,q \bar q}^{(0) \mu \nu} (\ell; k_1, k_2) & = & \frac{4 \pi \al_s}{\kOnekTwo} \, \, d^{\mu}_{\rho} (p,n) \,
  n_f T_R \left( -g^{\rho\sigma} + 4 z (1 - z) \frac{k_\perp^\rho k_\perp^\sigma}{k_\perp^2}
  \right)
  \nnb\\
 && \hspace{45mm}
 \times  \, d_\sigma^{\nu}(p,n) \, (2\pi)^d \delta^{(d)} (\ell - p)  \, + \, {\cal O}(k_\perp^0) \nn \\
  & = & \frac{4 \pi \al_s}{\kOnekTwo} \, d^{\mu}_{\rho} (p,n) \, n_f P_{q \bar q}^{\rho\sigma} (z) \,
  d_\sigma^{\nu}(p,n) \, (2\pi)^d \delta^{(d)} (\ell - p)  \, + \, {\cal O}(k_\perp^0) \, ,
\eeq
where $d^{\mu\nu}(p,n)$ is the gluon polarisation tensor for non-light-like reference 
momenta,
\beq
\label{dmunumass}
  d^{\mu\nu}(p,n) \, = \, - g^{\mu\nu} + \frac{p^\mu n^\nu + p^\nu n^\mu}{p \cdot n} - 
  n^2 \frac{p^\mu p^\nu}{(p\cdot n)^2} \, .
\eeq
For a gluon radiating to two gluons the full result is somewhat lengthier, and we find
\beq
\label{J0gg}
  J_{g,gg}^{(0) \mu \nu} (\ell; k_1, k_2) & = & \frac{4 \pi \al_s C_A}{\kOnekTwo} \,
  \Bigg\{ g^{\mu \nu} \left[ \frac{n^2 \kOnekTwo}{\kTwon^2} - \frac{2 \kOnen}{\kTwon} \right]
  \, - \, \frac{k_1^\mu k_2^\nu}{(\ell \cdot n)^2} \bigg[ \, \frac{2 (1 - \eps) 
  \kOnen \kTwon}{\kOnekTwo} \nn \\
  & & + \, \frac{n^2}{\kOnen \kTwon} \left( n^2 \kOnekTwo + \kOnen^2 + \kTwon^2 
  \right) \bigg] \nn \\
  & & + \, \frac{k_1^\mu k_1^\nu}{(\ell \cdot n)^2} \left[ \,
  \frac{2 (1 - \eps) \kTwon^2}{\kOnekTwo} - 2 n^2 \left( 1+ \frac{\kOnen}{\kTwon} +
  \frac{\kTwon}{\kOnen} \right) + \frac{(n^2)^2 \kOnekTwo}{\kTwon^2} \right] \nn \\
  & & + \, \frac{k_1^\mu n^\nu + k_1^\nu n^\mu}{(\ell \cdot n)}
  \left[ 1 + 2 \frac{\kOnen}{\kTwon} + \frac{\kTwon}{\kOnen} + 
  \frac{n^2 \kOnekTwo \left(\kTwon - \kOnen \right)}{\kOnen \kTwon^2} \right] \nn \\
  & & - \, \frac{\kOnekTwo}{\kOnen \kTwon} \, n^\mu n^\nu +
  \left( k_1 \leftrightarrow k_2 \right) \Bigg\} \, (2\pi)^d \delta^{(d)} (\ell - k_1 - k_2) \, .
\eeq
In the collinear limit, \eq{J0gg} reproduces the $P_{gg}$ splitting function:
\beq
  J_{g,gg}^{(0)\mu\nu}(\ell; k_1, k_2)
  & = &
  \frac{4 \pi \al_s}{\kOnekTwo} \,
  d^{\mu}_{\rho}(p,n) \, 2 C_A
  \Big[
  - g^{\rho\sigma} \Big(\frac z{1 - z} + \frac{1 - z}z \Big)
  - 2 (1 - \eps) z (1 - z) \frac{k_\perp^\rho k_\perp^\sigma}{k_\perp^2}
  \Big]
  \nnb \\ && \hspace{55mm}
  \times \, d_\sigma^{\nu}(p,n) \, (2\pi)^d
  \delta^{(d)} (\ell - p) \, + \, {\cal O}(k_\perp^0)\nnb\\
  & = &
  \frac{4 \pi \al_s}{\kOnekTwo} \,
  d^{\mu}_{\rho}(p,n) \, P_{gg}^{\rho\sigma} (z) \, d_\sigma^{\nu}(p,n) \, (2\pi)^d
  \delta^{(d)} (\ell - p) \, + \, {\cal O}(k_\perp^0) \, .
\eeq
Finally, the single-radiative eikonal jet function is given by
\beq
\label{J0eik}
  J_{\E_f, g}^{(0)} (\beta; k) \, = \, 4 \pi \al_s C_f 
  \left[ \frac{2 \, (\beta \cdot n)}{(\beta \cdot k)(k \cdot n)} - \frac{n^2}{(k \cdot n)^2} \right] \, ,
\eeq
and can be obtained both by direct calculation, or by taking the soft limit of either 
\eq{J0qg} or \eq{J0gg}: as expected, it is spin-independent.

%%%%%%%%%%%%%%%%%%%%%%%%%%%%%%%%%%%%%%%

\section{On the computation of radiative jet functions at one loop}
\label{app:olojetcomp}

In this Appendix we summarise the techniques relevant for the calculation of one-loop 
single-radiative jet functions for the required flavour structures, and we give some details 
about their renormalisation. For all the relevant processes, the underlying integrals 
can be written in terms of the family
\beq
\label{eq:familyDef}
  \Int_{a_1 a_2 a_3 a_4} \, = \, \int\frac{d^dk}{i \pi^{d/2}} \,
  \frac{e^{\eps \gamma_E}}{(k^2)^{a_1} \left( (p_i - k)^2 \right)^{a_2} 
  \left( (p_i + p_j - k)^2 \right)^{a_3} (k\cdot n)^{a_4}} \, ,
\eeq
with $a_i$ being non-negative integers, $p_i^2 = p_j^2 = 0$, and we can set 
$p_i \cdot p_j = - 1$ (a negative value is useful to simplify numerical tests); furthermore, 
as mentioned in the main text, in order to avoid unphysical collinear divergences, we 
set $n^2 = 1$. The numerical values for the invariants $p_i \cdot p_j$ and $n^2$ are 
chosen so as to simplify the calculation: the dependence on such invariants can be 
fully reconstructed using the mass dimension of the integral and its scaling with 
respect to $n$, respectively. In the case of partonic jet functions, the momenta 
$p_i$ and $p_j$ are the outgoing partonic momenta, whereas, for the eikonal, 
jet one of them should be thought of as the light-like Wilson-line velocity. The 
integral family in \eq{eq:familyDef} can be computed from the basis integrals
$\Int_{0011}$, $\Int_{0101}$, $\Int_{1010}$, $\Int_{1011}$ and $\Int_{1111}$. In 
particular, the integrals $\Int_{0011}$, $\Int_{0101}$ and $\Int_{1010}$ can be 
easily evaluated to all orders in $\eps$ by Feynman parameterisation. For instance,
\beq
\label{eq:bubbleInts}
  \Int_{0101} \, = \, 2 \, \left( - 2 p_i \cdot n \right)^{1 - 2 \eps} \, e^{\eps \gamma_E} 
  \, \Gamma (1 - \eps) \Gamma (2 \eps - 1) \, , \qquad 
  \Int_{1010} \, =  \, \frac{2^{- \eps} e^{\eps \gamma_E} \Gamma (1 - \eps)^2 
  \Gamma (\eps)}{\Gamma (2 - 2 \eps)} \, , 
\eeq
and similarly for $\Int_{0011} = \Int_{0101}|_{p_i \to p_i + p_j}$. We note that the 
integral $\Int_{1011}$ is finite in $\eps$ and, therefore, is not needed for the present 
analysis. On the other hand, there is one non-standard  integral we have to evaluate, 
namely the IR divergent box integral $\Int_{1111}$. To tackle this calculation we exploit 
the method of differential equations.

As a first step, we assemble the integral basis
\beq
\label{eq:basis}
  \mathbf{f} & = & \Bigg\{ \frac{\eps (1 - 2 \eps)}{(p_i + p_j) \cdot n} \, \Int_{0011} \, , \, 
  \frac{\eps (1 - 2 \eps)}{p_i \cdot n} \, \Int_{0101} \, , \, 
  \eps (1 - 2 \eps) \, \Int_{1010} \, , \, \eps^2 \sqrt{2 + \left( (p_i + p_j) \cdot n 
  \right)^2} \, \Int_{1011} \, , \nn \\
  & & \qquad \eps^2 (p_i \cdot n) \, \Int_{1111} \Bigg\} \, .
\eeq
Then, we define new variables, $t_1$ and $t_2$, to remove any square roots that may 
appear in the integral basis $\bf f$. They are implicitly defined by 
\beq
\label{DEvar}
  p_i \cdot n \, = \, \frac{2 \sqrt{2} \, t_1 t_2}{t_1^2 - (1 + t_2)^2} \, , \qquad 
  p_j \cdot n \, = \, \frac{2 \sqrt{2} \, t_1}{t_1^2 - (1 + t_2)^2} \, .
\eeq
Using this basis and these variables has the advantage that the differential equations 
take the canonical form~\cite{Henn:2013pwa}
\beq
\label{eq:DE}
  \text{d} \mathbf{f} (\eps, t_1, t_2) \, = \,\eps \left[ \,\, \sum_{l \in \mathcal{A}} A_l \,
  \text{d} \log(l) \right] \cdot \mathbf{f} (\eps, t_1, t_2) \, ,
\eeq
where the $A_l$ are constant $5 \times 5$ matrices, each associated to a letter 
$l$ in the alphabet
\beq
\label{alphabet}
  \mathcal{A} & = & 
  \bigg\{t_1, \, t_2 + 1, \, t_2, \, 
  \frac{t_1 + t_2 + 1}{1 - t_1 + t_2}, \, 
  t_1^2 + t_2^2 + 2 t_2 + 1, \, 
  1 - t_1^2 + t_2^2 + 2 t_2,
  \nn \\ & & \qquad
  t_1^2 - 2 t_1 t_2 + 2 t_1 + t_2^2 + 2 t_2 + 1, \, 
  t_1^2 + 2 t_1 t_2 - 2 t_1 + t_2^2 + 2 t_2 + 1 \bigg\} \, .
\eeq
The overall $\eps$ factor in \eq{eq:basis} is chosen so that $\mathbf{f}$ admits an 
expansion in $\eps$ starting at ${\cal O}(\eps^0)$. We can then integrate \eq{eq:DE} 
in terms of iterated integrals. In principle, one can compute the integrals to arbitrary 
order in $\eps$, assuming the appropriate boundary conditions are known. In our case, 
however, as we are only interested in the poles of the jet functions, one iteration of 
the differential equations is sufficient. This corresponds to functions that have a 
transcendental weight of one, which is the highest possible weight that can appear.
Notice that $f_1$, $f_2$ and $f_3$ are given by \eq{eq:bubbleInts}, and the first 
non-vanishing term of $f_4$ will be of weight two. We can then focus directly on 
$f_5$: upon solving \eq{eq:DE}, without specifying the boundary conditions, we 
get an expression of the form
\beq
\label{eq:f5_solution}
  f_5 \, = \, c_0 + \eps \, \Big[ c_1 + \log g(t_1, t_2) \Big] + \mathcal{O}(\eps^2) \, ,
\eeq
where $g(t_1, t_2)$ is a known polynomial built from the letters in the alphabet 
$\mathcal{A}$, $c_0$ is of weight zero, and $c_1$ is of weight one, ensuring that 
\eq{eq:f5_solution} is of uniform weight. The values of the coefficients can be found 
by computing $f_5$ numerically, for example using \texttt{pySecDec}~\cite{Hahn:2004fe,
Borowka:2017idc}, and then fitting the result: it turns out that $c_0 = - \frac{3}{4}$,
and $c_1 = \frac{7}{4} \log{2}$. Inverting \eq{eq:basis} for $T_{1111}$, and 
reinstating the Mandelstam invariants, we finally find
\beq
\label{eq:J1111}
  \Int_{1111} \, = \, \frac{(- p_i \cdot p_j)^{- 1 - \eps}}{2 p_i \cdot n}
  \left[- \frac{3}{2} \frac{1}{\eps^2} + \frac{1}{2} \frac{1}{\eps}
  \log \left(\frac{- 2 (p_i \cdot n)^2}{(p_i + p_j) \cdot n \sqrt{n^2}
  \sqrt{- p_i \cdot p_j}} \right) + \mathcal{O}(\eps^0) \right] \, .
\eeq
A similar calculation leads to the eikonal version of \eq{eq:J1111}, which we compute 
using $p^2 = \beta^2 = 0$. Up to ${\cal O} (\eps^0)$ corrections, we find
\beq
\label{eq:eikJ1111}
  \int \frac{d^d k}{i \pi^{d/2}} \frac{e^{\eps\gamma_E}}{k^2 (p - k)^2 
  \left( (k - p) \cdot \beta \right) k \cdot n} \, = \, 
  \frac{1}{(p \cdot n)(p \cdot \beta)} \Bigg[ \frac{3}{2} \frac{1}{\eps^2} + 
  \frac{1}{2} \frac{1}{\eps} \log \left(\frac{n^2 (2 \beta \cdot n)^2}{(2 p \cdot n)^4 
  (2 p \cdot \beta)^2} \right) \Bigg] \, . \quad
\eeq
In order to complete our discussion about one-loop radiative jet functions, we report 
below the renormalisation factors in the $\overline{\rm{MS}}$ scheme that were used 
in \secn{sub:radiative_jet_function_calc}. They are given by 
\begin{subequations}
\begin{align}
\label{eq:Zpsi}
  Z_\psi \, = \, 
  & \, 1 - \frac{\al_s}{4 \pi} \frac{1}{\eps} \, C_F + \mathcal{O} (\al_s^2)
  \, , & \text{quark field} \, , \\
\label{eq:ZA} 
  Z_A \, = \, 
  & \, 1 + \frac{\al_s}{4 \pi} \frac{1}{\eps} \left( \frac{5}{3}C_A - \frac{2}{3} n_f \right) 
  + \mathcal{O}(\al_s^2) \, , & \text{gluon field} \, , \\
\label{eq:Zg} 
  Z_g \, = \, 
  & \, 1 - \frac{\al_s}{4 \pi} \frac{1}{\eps} \left(C_F + C_A\right) + \mathcal{O}(\al_s^2)
  \, , & \text{$qqg$ vertex} \, , \\
\label{eq:Zggg} 
  Z_{ggg} \, = \,
  & \, 1 + \frac{\al_s}{4 \pi} \frac{1}{\eps} \, \frac{2}{3} \left(C_A - n_f \right) 
  + \mathcal{O}(\al_s^2) \, , & \text{$ggg$ vertex} \, , \\
\label{eq:ZWi} 
  Z_{W_i} \, = \,
  & \, 1 + \frac{\al_s}{4 \pi} \frac{2}{\eps} \, C_{f_i} + \mathcal{O}(\al_s^2)
  \, , & \text{Wilson line} \, , \\
\label{eq:ZV} 
  Z_{V\psi} \, = \, 
  & \, 1 - \frac{\al_s}{4 \pi} \frac{1}{\eps} \, C_F + \mathcal{O}(\al_s^2)
  \, , &\text{$q$ -- Wilson-line vertex} \, , \\
\label{eq:ZVA} 
  Z_{V \hspace{-1pt} A} \, = \, 
  & \, 1 + \mathcal{O}(\al_s^2)
  \, , & \text{$g$ -- Wilson-line vertex} \, .
\end{align}
\end{subequations}
Eqs.~(\ref{eq:Zpsi})-(\ref{eq:Zggg}) are standard textbook QCD results. Eq.~(\ref{eq:ZWi}) 
can be found for instance in \cite{Dixon:2008gr}. Eqs.~(\ref{eq:ZV}) and (\ref{eq:ZVA}) 
were found by explicit computation of the diagrams involved. We also derive the 
renormalisation factors for the coupling of a Wilson line radiating a gluon, see figure
\ref{fig:jetFunctionRenorm2}: they are given by
\begin{subequations}
\label{eq:ZWilsonGluonCoupling}
\begin{align}
  Z_{W_FA} \, = \, Z_g Z_\psi^{-1} Z_{W_F} \, = \,
  & \, 1 + \frac{\al_s}{2 \pi} \frac{1}{\eps} \left( C_F - \frac{C_A}{2} \right)
  \, , & \text{fundamental Wilson line} \, \\
  Z_{W_AA} \, = \, Z_{ggg} Z_A^{-1} Z_{W_A} \, = \,
  & \,1 + \frac{\al_s}{2 \pi} \frac{1}{\eps} \, \frac{C_A}{2}
  \, , & \text{adjoint Wilson line} \, .
\end{align}
\end{subequations}
Note that both colour factors in \eq{eq:ZWilsonGluonCoupling} are of the form 
$(C_{f_i} - N_c/2)$ expected for vertex graphs. Since also
\beq
\label{jetrenorm2}
  Z_g Z_\psi^{-1} Z_A^{-1/2} \, = \, Z_{ggg} Z_A^{- 3/2} \, \equiv \, Z_{\al_s} \, = \, 
  1 - \frac{\al_s}{4 \pi} \frac{1}{\eps} \frac{b_0}{2} + \mathcal{O}(\al_s^2) \, ,
\eeq
the jets $J_{g,gg}^{\mu\nu} (\ell; k_1, k_2)$ and $J_{g,qq}^{\mu\nu} (\ell;k_1, k_2)$ 
renormalise in the same way.

Finally, the renormalisation factor related to the vertex connecting the Wilson lines 
in directions $\beta_i$ and $n$ is given by the usual cusp renormalisation, involving 
an extra collinear pole. It is given by
\beq
\label{eq:Z_beta_n}
  Z_{\beta_i n} \, = \, 1 - \frac{\al_s}{4 \pi} \, C_{f_i} \, \left( \frac{1}{\eps^2} + 
  \frac{1}{\eps} \log \frac{n^2}{4 ( \beta_i \cdot n)^2} \right) \, .
\eeq

%%%%%%%%%%%%%%%%%%%%%%%%%%%%%%%%%%%%%%%%%%%%%

\section{Consistency relations}
\label{sub:checks}

In Section~\ref{sub:finding_k_12} we mentioned the necessity to verify the relation
$\mathbf{L}_1\big[K_{n+2}^{\two}\big] = \mathbf{L}_1 \big[ K_{n+2}^{\otwo}\big]$ in order 
to confirm the consistency of the strongly-ordered counterterm we derived. Here we report 
the relevant details of the calculation. We note once again that, when hard-collinear 
configurations are involved, factorised matrix elements involve convolutions (which 
are understood in our notation). In these cases the tests we perform below rely on 
the assumption that the necessary phase-space mappings are chosen consistently,
ensuring that they behave properly under iterated limits.

We start by acting on $K_{n+2}^{\two}$ with the soft limit $\SL_1$, {\it i.e.}~considering the
$k_1 \to 0$ limit of each of the constituent counterterms in eqs.~(\ref{CandCountNNLO1})
to (\ref{CandCountNNLO4}). As for the double-soft counterterm in \eq{CandCountNNLO1}, 
accounting for two uniform soft emissions, we get
\beq
\label{eq: initial soft check 12}
  \SL_1\left[ K^{({\bf 2}, \, \rm{2s})}_\npt  \right]
  \, = \,
  {\hard_n^{(0)}}^{\dagger} \,
  S^{(0)}_{n; \, g_2,g_1} \,
  \mathcal{H}_n^{(0)}
  \, = \,
  {\hard_n^{(0)}}^{\dagger} \,
  {{\cal S}^{(0)}_{n,g_2}}^{\!\!\!\dag} \, S^{(0)}_{n+1,g_1} \, {\cal S}^{(0)}_{n,g_2} \,
  \mathcal{H}_n^{(0)}
  \, ,
\eeq
where we have used \eq{softrads.o.}. In order to compare this expression with the soft 
limit of $K^{({\bf 12}, \, \rm{s})}_\npt$, we note that \eq{eq:K12s_2} is written assuming 
$k_2$, as opposed to $k_1$, to be the unresolved momentum, hence a meaningful 
comparison with \eq{eq: initial soft check 12} needs a $1\leftrightarrow2$ relabelling 
of \eq{eq:K12s_2}. We get then
\beq
\label{eq: initial soft check 12 bis}
  \SL_1 \left[ K_{n+2}^{(\mathbf{12}, \, \text{s})} \right]
  & = &
  {\mathcal{H}_n^{(0)}}^{\dagger} 
  \sum_{f_1,f_2}
  \SL_1
  \bigg[
  S^{(0)}_{n;f_{[12]},f_1}
  +
  S^{(0)}_{n,f_{[12]}}
  \Big(
  J_{f_{[12]},f_1f_2}^{(0)}
  -
  J_{\E_{[12]},f_1}^{(0)}
  \Big)
  \bigg] \,
  \mathcal{H}_{n}^{(0)}
  \nnb \\
  & = &
  {\mathcal{H}_n^{(0)}}^{\dagger} 
  S^{(0)}_{n; \, g_2,g_1}
  \mathcal{H}_{n}^{(0)}
  \, = \,
  \SL_1\left[ K^{({\bf 2}, \, \rm{2s})}_\npt  \right]
  \, ,
\eeq
where we have used $\SL_1 \, J_{f_{[12]},f_1f_2}^{(0)} = J_{\E_{[12]},f_1}^{(0)}$.
Next we consider the hard-collinear counterterm associated with a single leg, given by
$K_{n+2, \, i}^{({\bf 2}, \, \rm{2hc})}$ in \eq{CandCountNNLO2}. The single-soft limit 
of the various contributions to this counterterm read
\beq
\label{eq:soft_1}
  \SL_1 \left[J^{(0)}_{f_i,f_1f_2f_3}\right]
  & = &
  J^{(0)}_{f_i,f_2f_3} \, S^{(0)}_{3,f_1}
  \, , \\
  \SL_1 \left[J^{(0)}_{\E_i,f_1f_k}\right]
  & = &
  J^{(0)}_{\E_i,f_k} \, S^{(0)}_{3,f_1}
  \, ,
  \hspace{10mm} k=2,3 \, ,
  \nnb \\
  \SL_1 \left[J^{(0)}_{\E_i,f_2f_3}\right]
  & = & 0
  \, ,
  \nnb \\
  \SL_1 \left[J^{(0)}_{\E_i,f_1} \,
  J_{f_i, f_2 f_3}^{(0),\, \text{hc}}\right]
  & = &
  J^{(0)}_{\E_i,f_1} \,
  J_{f_i, f_2 f_3}^{(0),\, \text{hc}}
  \, ,
  \nnb \\
  \SL_1 \left[J^{(0)}_{\E_i,f_j} \,
  J_{f_i, f_1 f_k}^{(0),\, \text{hc}}\right]
  & = & 0
  \, ,
  \hspace{25mm} j,k=2,3
  \, ,
  \nnb
\eeq
where $S^{(0)}_{3,f_1}$ is the same object appearing in \eq{eq:compl_jet_RV}, namely 
a soft function formed from three Wilson lines in the directions $\beta_2$,  $\beta_3$ 
and $n$ (the latter being the auxiliary vector used in the definition of the jet functions), 
radiating a gluon with momentum $k_1$. Plugging these limits into \eq{CandCountNNLO2}, 
we can assemble
\beq
  \SL_1 \left[ K_{n+2, \, i}^{(\mathbf{2}, \, 2 \text{hc})} \right]
  & = &
  {\hard_n^{(0)}}^{\dagger} \, 
  \sum_{f_1, f_2, f_3}
  J_{f_i, f_2 f_3}^{(0),\, \text{hc}} \,
  \Big(S^{(0)}_{3,f_1} - J^{(0)}_{\E_i,f_1}\Big) \,
  \mathcal{H}_n^{(0)} \, .
\eeq
We can compare this expression with the single-soft limit of \eq{eq:K12hci}. For this purpose, 
we note that in \eq{eq:K12hci} the unresolved momentum is $k_3$, hence a $1\leftrightarrow3$ 
relabelling is necessary. This gives
\beq
\label{eq: soft limit check final}
  \SL_1 \left[K_{n+2, i}^{(\mathbf{12}, \, \text{hc})} \right]
  & = &
  {\hard_n^{(0)}}^{\dagger}
  \!\!\!
  \sum_{f_1,f_2,f_3}
  \bigg[ J_{f_i,f_2f_3}^{(0), \, \text{hc}}
  \Big(S_{3,f_1}^{(0)} - J_{\E_i,f_1}^{(0)}\Big) 
  \nnb \\
  & & \hspace{20mm}
  + \, \, \SL_1 \!\!\! \sum_{kl = \{23,32\}}
  J_{f_i,f_{[1k]} f_l}^{(0), \, \text{hc}}\,
  \Big(
  J_{f_{[1k]},f_1f_k}^{(0)}
  -
  J_{\E_{[1k]},f_1}^{(0)}
  \Big)
  \bigg] \,
  \hard_n^{(0)}
  \nnb \\
  & = &
  {\hard_n^{(0)}}^{\dagger}
  \!\!\! \sum_{f_1,f_2,f_3}
  J_{f_i,f_2f_3}^{(0), \, \text{hc}}
  \Big(S_{3,f_1}^{(0)} - J_{\E_i,f_1}^{(0)}\Big) 
  \, \hard_n^{(0)}
  \, = \,
  \SL_1 \left[ K_{n+2, \, i}^{(\mathbf{2},\, 2\text{hc})} \right]
  \, .
\eeq
One can similarly show that $\SL_1 \, \big[ K^{({\bf  2}, \, {\rm 2hc})}_{\npt, \, ij} \big] = 0$, 
and furthermore one sees that $K_{n+2, i}^{(\mathbf{2}, 1 \text{hc}, 1 \text{s})} = 
K_{n+2,i}^{(\mathbf{12}, \text{1hc}, \text{1s})}$ before taking any limits. Therefore,
the consistency test for the soft limit, $\SL_1 \big[ K_{n+2,i}^{(\mathbf{2})} \big] = \SL_1 
\big[ K_{n+2,i}^{(\mathbf{12})} \big]$, is completed.

Next, we analyse the collinear $\CL_{12}$ limit of the double-unresolved counterterm. 
Beginning, as above, with the double-soft counterterm in \eq{CandCountNNLO1}, we find
\beq
\label{C12K22s}
  \CL_{12} \left[K_{n+2}^{(\mathbf{2}, \, \text{2s})}\right]
  \, = \,
  {\hard_n^{(0)}}^{\dagger} 
  \sum_{f_1,f_2}
  \CL_{12} \, S^{(0)}_{n,f_1f_2} \,
  \mathcal{H}_n^{(0)}
  \, = \,
  {\hard_n^{(0)}}^{\dagger}
  \sum_{f_1,f_2}
  S^{(0)}_{n,f_{[12]}} \,
  J^{(0)}_{f_{[12]},f_1f_2} \, 
  \mathcal{H}_n^{(0)} \, .
\eeq
On the other hand, from \eq{eq:K12s_2} we get
\beq
\label{C12K12s}
  \CL_{12} \left[K_{n+2}^{(\mathbf{12}, \, \text{s})}\right]
  & = &
  {\mathcal{H}_n^{(0)}}^{\dagger} 
  \sum_{f_1,f_2}
  \bigg[
  \CL_{12} \,
  S^{(0)}_{n;f_{[12]},f_2}(k_{[12]},k_2)
  +
  S^{(0)}_{n,f_{[12]}}
  \Big(
  J_{f_{[12]},f_1f_2}^{(0)}
  -
  J_{\E_{[12]},f_2}^{(0)}
  \Big)
  \bigg] \,
  \mathcal{H}_{n}^{(0)}
  \nnb  \\ & = &
  {\mathcal{H}_n^{(0)}}^{\dagger} 
  \sum_{f_1,f_2}
  S^{(0)}_{n,f_{[12]}} \,
  J_{f_{[12]},f_1f_2}^{(0)} \,
  \mathcal{H}_{n}^{(0)}
  \, = \,
  \CL_{12} \left[K_{n+2}^{(\mathbf{2}, \, \text{2s})}\right] \, ,
\eeq
where we have used $\CL_{12} \, S^{(0)}_{n; f_{[12]}, f_2}(k_{[12]}, k_2) = S^{(0)}_{n, f_{[12]}} 
\,  J_{\E_{[12]},f_2}^{(0)}$. Moving on to the contributions of $K^{({\bf 2}, \, {\rm 2hc})}_{\npt, \, i}$ 
in \eq{CandCountNNLO2}, we find
\beq
\label{C12K22hc}
  \CL_{12} \left[J^{(0)}_{f_i, f_1 f_2 f_3}\right]
  & = &
  J^{(0)}_{f_i, f_3 f_{[12]}} \, J^{(0)}_{f_{[12]}, f_1 f_2}
  \, , \\
  \CL_{12} \left[J^{(0)}_{\E_i, f_1 f_2}\right]
  & = &
  J^{(0)}_{\E_{i}, f_{[12]}} \, J^{(0)}_{f_{[12]}, f_1 f_2} \, ,
  \nnb \\
  \CL_{12} \left[J^{(0)}_{\E_i, f_k f_3}\right]
  & = &
  J^{(0)}_{\E_{[12]}, f_{k}} \, J^{(0)}_{\E_{i}, f_3} \, , 
  \hspace{2cm} k=1,2  \, ,
  \nnb \\
  \CL_{12} \left[J^{(0)}_{\E_i,f_j} \,
  J^{(0) , \, \text{hc}}_{f_i, f_k f_3}\right]
  & = &
  0  \, ,
  \hspace{40mm} j,k=1, 2 \, ,
  \nnb \\
  \CL_{12} \left[J^{(0)}_{\E_i, f_3} \,
  J^{(0) , \, \text{hc}}_{f_i, f_1 f_2}\right]
  & = &
  J^{(0)}_{\E_i, f_3} \, J^{(0), \, \text{hc}}_{f_i, f_1 f_2}
  \, = \,
  J^{(0)}_{\E_i, f_3} \, J^{(0), \, \text{hc}}_{f_{[12]}, f_1 f_2} \, , \nnb
\eeq
where spin indices have been understood. Combining these contributions we obtain the 
simple result
\beq
  \CL_{12} \left[K^{({\bf 2}, \, {\rm 2hc})}_{\npt, \, i} \right]
  & = &
  {\mathcal{H}_n^{(0)}}^\dagger \,
  \sum_{f_1, f_2, f_3}
  J^{(0), \, \text{hc}}_{f_i, f_{[12]} f_3} \,
  J^{(0)}_{f_{[12]}, f_1 f_2} \,
  \mathcal{H}_n^{(0)} \, .
\eeq
This expression must be compared with the collinear limit of \eq{eq:K12hci}, after the 
necessary $1 \leftrightarrow3$ relabelling. We get
\beq
\label{eq: collinear limit check final}
  \CL_{12} \left[K_{n+2, i}^{(\mathbf{12}, \, \text{hc})} \right]
  & = &
  {\hard_n^{(0)}}^{\dagger}
  \CL_{12}
  \sum_{f_1, f_2, f_3}
  \bigg[
  J_{f_i, f_2 f_3}^{(0), \, \text{hc}}(\bar k_2,\bar k_3) \, S_{3,f_1}^{(0)} -
  J_{f_i, f_2 f_3}^{(0), \, \text{hc}}(k_2, k_3) \, J_{\E_i, f_1}^{(0)}
  \nnb \\ & & \qquad \qquad \quad
  + \!\!\!\!
  \sum_{kl=\{23,32\}}
  J_{f_i, f_{[1k]} f_l}^{(0), \, \text{hc}} \,
  \Big(
  J_{f_{[1k]},f_1f_k}^{(0)} - J_{\E_{[1k]},f_1}^{(0)}
  \Big)
  \bigg] \,
  \hard_n^{(0)}
  \nnb \\ & = &
  {\hard_n^{(0)}}^{\dagger}
  \CL_{12}
  \sum_{f_1, f_2, f_3}
  J_{f_i, f_{[12]} f_3}^{(0), \, \text{hc}}
  \bigg[
  S_{3, f_1}^{(0)} +
  J_{f_{[12]}, f_1 f_2}^{(0)} -
  J_{\E_{[12]}, f_1}^{(0)}
  \bigg] \,
  \hard_n^{(0)} \, .
\eeq
Noting that $\CL_{12} \, S_{3,f_1}^{(0)}(\beta_{[12]}, \beta_3,n) = J_{\E_{[12]}, f_1}^{(0)}
(\beta_2,n)$, we finally get
\beq
\label{eq: collinear limit check final final}
  \CL_{12} \left[K_{n+2, i}^{(\mathbf{12}, \, \text{hc})} \right]
  & = &
  {\hard_n^{(0)}}^{\dagger}
  \sum_{f_1, f_2, f_3}
  J_{f_i, f_{[12]} f_3}^{(0), \, \text{hc}} \,
  J_{f_{[12]}, f_1 f_2}^{(0)}
  \hard_n^{(0)}
  \, = \,
  \CL_{12} \left[ K^{({\bf 2}, \, {\rm 2hc})}_{\npt, \, i} \right] \, .
\eeq
The remaining counterterms do not cause any difficulties, since $K_{n+2, ij}^{(\mathbf{2},
\, \text{2hc})} = K_{n+2, ij}^{(\mathbf{12}, \, \text{2hc})}$, and similarly $K_{n+2, i}^{(\mathbf{2},
\text{1hc}, \text{1s})} = K_{n+2, i}^{(\mathbf{12}, \text{1hc}, \text{1s})}$. The collinear consistency 
check $\CL_{12} \big[ K_{n+2, i}^{(\mathbf{2})} \big] = \CL_{12} \big[ K_{n+2, i}^{(\mathbf{12})} \big]$ 
is thus completed. This, in turn, concludes the proof that $\mathbf{L}_1\big[K_{n+2}^{\two}\big] 
= \mathbf{L}_1 \big[ K_{n+2}^{\otwo}\big]$.

%%%%%%%%%%%%%%%%%%%%%%%%%%%%%%%%%%%%%%%%%%%%%

%%%%%%%%%%%%%%%%%%%%%%%%%%%%%%%%%%%%%%%%%%%%%

\bibliographystyle{JHEP}

\bibliography{StronglyOrdered}

%%%%%%%%%%%%%%%%%%%%%%%%%%%%%%%%%%%%%%%%%%%%%

%%%%%%%%%%%%%%%%%%%%%%%%%%%%%%%%%%%%%%%%%%%%%

\end{document}